\documentclass[12pt,review]{elsarticle}

\usepackage[colorlinks,citecolor=blue,linktoc=all,linkcolor=cyan]{hyperref}
\usepackage{graphicx}

\usepackage[T1]{fontenc}
\usepackage{dsfont}               
\usepackage{mathrsfs}             
\usepackage{slashed}              
\usepackage{amsmath}
\usepackage{amssymb}
\usepackage{amsbsy}
\usepackage{amsfonts}
\usepackage{ulem}
\usepackage{array} 
\usepackage{makecell}
\usepackage{CJK}
\usepackage{setspace}
\everymath{\displaystyle}

\numberwithin{equation}{section}
\numberwithin{table}{section}
\numberwithin{figure}{section}

\usepackage{tikz,xcolor,hyperref}

\definecolor{lime}{HTML}{A6CE39}
\DeclareRobustCommand{\orcidicon}{
	\begin{tikzpicture}
	\draw[lime, fill=lime] (0,0) 
	circle [radius=0.16] 
	node[white] {{\fontfamily{qag}\selectfont \tiny ID}};
	\draw[white, fill=white] (-0.0625,0.095) 
	circle [radius=0.007];
	\end{tikzpicture}
	\hspace{-2mm}
}
\foreach \x in {A, ..., Z}{%
	\expandafter\xdef\csname orcid\x\endcsname{\noexpand\href{https://orcid.org/\csname orcidauthor\x\endcsname}{\noexpand\orcidicon}}
}

\journal{Progress in Particle and Nuclear Physics}

\usepackage{titlesec}
\usepackage{sectsty}
\titleformat{\section}{\normalfont\Large\bfseries}{\thesection}{1em}{}
\titleformat{\subsection}{\normalfont\large\bfseries}{\thesubsection}{1em}{}
\titleformat{\subsubsection}{\normalfont\normalsize\bfseries}{\thesubsubsection}{1em}{}

\biboptions{sort&compress}

\begin{document}
\begin{CJK*}{UTF8}{gbsn}
	\begin{frontmatter}
		
		\title{Shear viscosity of nucleonic matter}

		\author[mymainaddress,mysecondaryaddress]{Xian-Gai Deng}

		\author[mymainaddress,mysecondaryaddress]{De-Qing Fang}
		
		\author[mymainaddress,mysecondaryaddress]{Yu-Gang Ma\corref{mycorrespondingauthor}}
		\cortext[mycorrespondingauthor]{Corresponding author}
		\ead{mayugang@fudan.edu.cn}

		\address[mymainaddress]{Key Laboratory of Nuclear Physics and Ion-beam Application (MOE), Institute of Modern Physics, Fudan University, Shanghai 200433, China}
		\address[mysecondaryaddress]{Shanghai Research Center for Theoretical Nuclear Physics, NSFC and Fudan University, Shanghai 200438, China}
		
\begin{abstract}
The research status of the shear viscosity of nucleonic matter is reviewed. Some methods to calculate the shear viscosity of nucleonic matter are introduced, including mean free path, Green-Kubo, shear strain rate, Chapman-Enskog and relaxation time approximation. Based on these methods, results for infinite and finite nucleonic matter are discussed, which are attempts to investigate the universality of the ratio of shear viscosity over entropy density and transport characteristics like the liquid-gas phase transition in nucleonic matter. In addition, shear viscosity is also briefly discussed for the quantum chrodynamical matter produced in relativistic heavy-ion collisions. 
				
\end{abstract}
		
		
	\end{frontmatter}
	
	\newpage
	
	\thispagestyle{empty}
	\tableofcontents
	

\newpage
\section{Introduction}\label{intro}


In a non-perfect fluid, viscosity represents internal friction, which causes an irreversible energy dissipation process, and it is a measure of the flow resistance to the liquid, which means the higher value, and the more resistance, and thus the fluid flows more slowly.
Shear viscosity, which describes momentum transfer ability in a flowing fluid, is one of the important transport properties for lots of substances, such as macroscopic matter as well as microscopic and quantum matter, e.g. hot dense quark matter, etc.~\cite{PRAKASH1993321,SThomas2014}. Viscosity, which affects solute dynamics and reaction rates in a solvent, is crucial for understanding biological processes and chemical reactions~\cite{MCKINNIE1981,Kyushiki1990,Malosso2022}. In astrophysics, properties of transport coefficients (bulk viscosity, shear viscosity, and thermal conductivity) have been studied in different layers of neutron stars (NS), e.g. in the crust and in the core~\cite{Shternin2008,Shternin_2013,Shternin_2017}. The transport coefficients were described in the framework of Brueckner-Hartree-Fock (BHF) theory~\cite{BHF23,BHF2}, and comparisons were made using several equations of states (EoSs)~\cite{Shternin_2017}. For viscosity, one of the most important features is the damping of oscillations. So it plays an important role for $r$-modes in rotating pulsars~\cite{KEE2015} and affects the emission and propagation of gravitational waves, which is important for the planning of advanced gravitational wave experiments~\cite{Ofengeim_2015,Bishop2022}. The study on transport coefficients could improve our understanding of the dynamical behaviour of the merging of binary neutron stars through the signal of gravitational waves, which was observed by LIGO and Virgo on August 17, 2017 (GW170817)~\cite{LIGOVirgo_2017}. The most common astrophysical manifestation of the NSs is the radio pulsars, which are driven by the strong magnetic field~\cite{Shternin_2022} as high as 10$^{12}$-10$^{13}$ G~\cite{Manchester_2005}. With such high magnetic fields, it could impact the EoS and have strong effects on the NS matter. How the magnetic field affects the transport coefficients of the magnetized NS should be considered~\cite{Ofengeim_2015}.


Viscous liquids like honey, syrup, and oil flow slowly, and less viscous liquids like water and gasoline flow quickly. The Navier-Stokes (NS) equation can be used to describe the motion of non-perfect fluid in fluid mechanics, where the shear viscosity coefficient and the bulk viscosity coefficient are included~\cite{ZuDN1970,Landau1987}. At the microscopic level, viscosity is caused by the potential and collisions of particles (or molecules). While macroscopically or hydrodynamically, it is determined by a constitutive relationship as the ratio of the shear stress to the shear rate~\cite{HANSEN2013,Haralson2017}. In equilibrium molecular dynamics (EMD), the viscosity $\eta$ is given as an integral of the shear-stress autocorrelation function determined during the simulation~\cite{PGai2006}, which is called Green-Kubo (GK) formula derived from linear response theory~\cite{R_Kubo_1966} or given by Einstein relation ~\cite{Tenney2010}. In the framework of EMD methods, for the convergence of shear-stress autocorrelation function, it would take a long simulation time due to relatively large fluctuations in the shear tensor. In molecular simulations for calculating viscosity, a nonequilibrium molecular dynamics (NEMD) method with a planar Couette flow field at strain rate $\dot{\gamma}=\partial u_{x}/\partial y$ becomes increasingly popular and has been used extensively to predict the rheological properties of real fluids~\cite{PGai2006}. 


In 2005, Kovtun, Son, and Starinets (KSS) \cite{KSS2005} studied the shear viscosity with certain supersymmetric gauge theories with gravity duals and gave a lower limit value for the ratio of shear viscosity to entropy density, $\eta/s$ $\geqslant \hbar/(4\pi k_{B})$ ($\hbar$ is the Planck's constant divided by $4\pi$ and $k_{B}$ is the Boltzmann constant), which is called the KSS bound. It has been conjectured that this ratio is the lowest limit for a large class of quantum field theories~\cite{KSS2005}. A new state of matter, the quark gluon plasma (QGP), is created in ultra-relativistic heavy-ion collisions, which is manifested by the results from the Relativistic Heavy Ion Collider (RHIC) at Brookhaven~\cite{PHENIX2005,PHOBOS2005,STAR2005,Chen_PR,Ko_2023,Rapp,QinGY,ZhangBW} and the Large Hadron Collider (LHC) at CERN~\cite{ATLAS2008,CMS2008,ALICE2008,LHCb2008,wang_2022,zhu_2022}. Based on heavy-ion data from these two colliders and theoretical modeling, lots of efforts are paid to uncover the fundamental properties of the quark gluon plasma~\cite{Nakamura_2005_PRL,Christiansen_2015_PRL,Eskola2019,Annu1,Annu2}. The analysis of the elliptic flow $v_{2}$ in the ultra-relativistic heavy-ion collisions shows that QGP behaves like a near perfect liquid with a lower limit of $\eta/s$~\cite{TSDT2009}. How perfect is the fluid in relativistic heavy-ion collisions~\cite{Romatschke2007}? In explaining the data from RHIC, the results ~\cite{SongHC2008,ShenC2010} from the viscous hydrodynamics model suggest that the ratio of shear viscosity to entropy density has a very small value. It indicates that the quark-gluon plasma (QGP) is nearly an ideal fluid. Moreover, in the analysis by Csernai {\it et al.}~\cite{Csernai2006}, it shows that the transition from hadrons to quarks and gluons has behavior similar to helium, nitrogen, and water at and near their phase transitions in the ratio $\eta/s$. 

Near and below a critical temperature, $T_{c}$, the condensates of bosons or fermion pairs behave like superfluids, which are frictionless~\cite{Joseph2015}. The behavior of shear viscosity versus temperature for bosonic and fermionic system is different below $T_{c}$~\cite{Joseph2015}. For instance, like the bosonic $^{4}$He, viscosity increases as the temperature approaches zero below $T_{c}$~\cite{Woods1963} and in the Fermionic $^{3}$He, the shear viscosity decreases to zero with decreasing of temperature below $T_{c}$~\cite{Alvesalo1974,Alvesalo1975,Wolfle1979}. The shear viscosity for $^{3}$He at the transition temperature $T_{A}$ = 2.6 mK appears a rapid decrease~\cite{Alvesalo1975}. The research on shear viscosity and its ratio over entropy density as well as for a Fermions system with the unitary Fermi gas has attracted lots of attention over the past few years~\cite{Bloch2008,Giorgini2008,Horikoshi2010,Cao2011,Enss2011,Salasnich2011,Guo2011,Joseph2015}. The unitary Fermi gas where particles interact at infinite scattering length is experimentally investigated with ultracold atoms in a balanced mixture of, e.g., the two lowest hyperfine levels of $^{6}$Li at a Feshbach resonance~\cite{Horikoshi2010,Enss2011,Salasnich2011}. It is found that the viscosity of a unitary Fermi gas is closer to that of fermionic $^{3}$He than to that of bosonic $^{4}$He~\cite{Guo2011,Joseph2015}. And both the viscosity and the ratio ($\eta$/s) of the unitary gas are predicted to exhibit a minimum, somewhat below $T_{c}$~\cite{Guo2011}. One found that a minimum value of $\eta/s$$\simeq$ 0.44 (in units of $\hbar$/ k$_{B}$) at the temperature $T/T_{F}$$\simeq$ 0.27, with $T_{F}$ the Fermi temperature~\cite{Salasnich2011}.

And it is also very interesting to see the ratio of shear viscosity to entropy density for nuclear matter, which is composed of protons and neutrons. Experimentally, some nuclear experimental facilities, such as FAIR (the International Accelerator Facility for Antiproton and Ion Research)~\cite{Schoenberg2020}, GANIL (Grand Acc$\acute{\rm e}$l$\acute{\rm e}$rateur National d'Ions Lourds-National Large Heavy Ion Accelerator)~\cite{POUTHAS1995418} in Europe, and High Intensity Accelerator Facility (HIAF) in China \cite{BaiSW,TangMT,YuPY} offer chances to explore phase diagram of nuclear matter in the region of high baryon density as well as transport properties \cite{Luciano2019,LanSW,ZhangY,DingHT,ChenQ,DuYL,SunKJ}.
Theoretically, transport coefficients such as shear viscosity have been studied for a few decades in nuclear physics and attracted lots of attentions~\cite{Bertsch1978,DPawel1984,SLDPawel2003,AM2004,KSS2005,DemirN2009,TSDT2009,LiSX2013,XGDeng2016,Mondal2017,Reichert2021}. The equation of state and transport coefficients are important quantities for the hot and dense hadron gases in nuclear physics~\cite{AM2004}. The behaviors of shear viscosity depend on the nucleon-nucleon cross sections as well as the nuclear equation of state, etc. For infinite nuclear matter, the effects of Pauli blocking, nucleon-nucleon cross section, temperature, density, isospin asymmetry, etc. on shear viscosity are discussed~\cite{DPawel1984,SLDPawel2003,Xujun2011,Xujun2013,XGDeng2022}. With the help of transport models like Boltzmann-Uehling-Uhlenbeck model (BUU) and quantum molecular dynamics model (QMD)~\cite{Bertsch1988,AICHELIN1991,YXZhang2018,Xujun2019,BUUQMD2021,BUUQMD2022}, the relation between shear viscosity and the liquid-gas (LG) phase transitions, or collective flow, or giant dipole resonance (GDR) has been investigated~\cite{LiSX2013,Zhou2013,XGDeng2016,Guo2017}.  

\begin{figure}[htb]
\setlength{\abovecaptionskip}{0pt}
\setlength{\belowcaptionskip}{8pt}
\centering\includegraphics[scale=0.4]{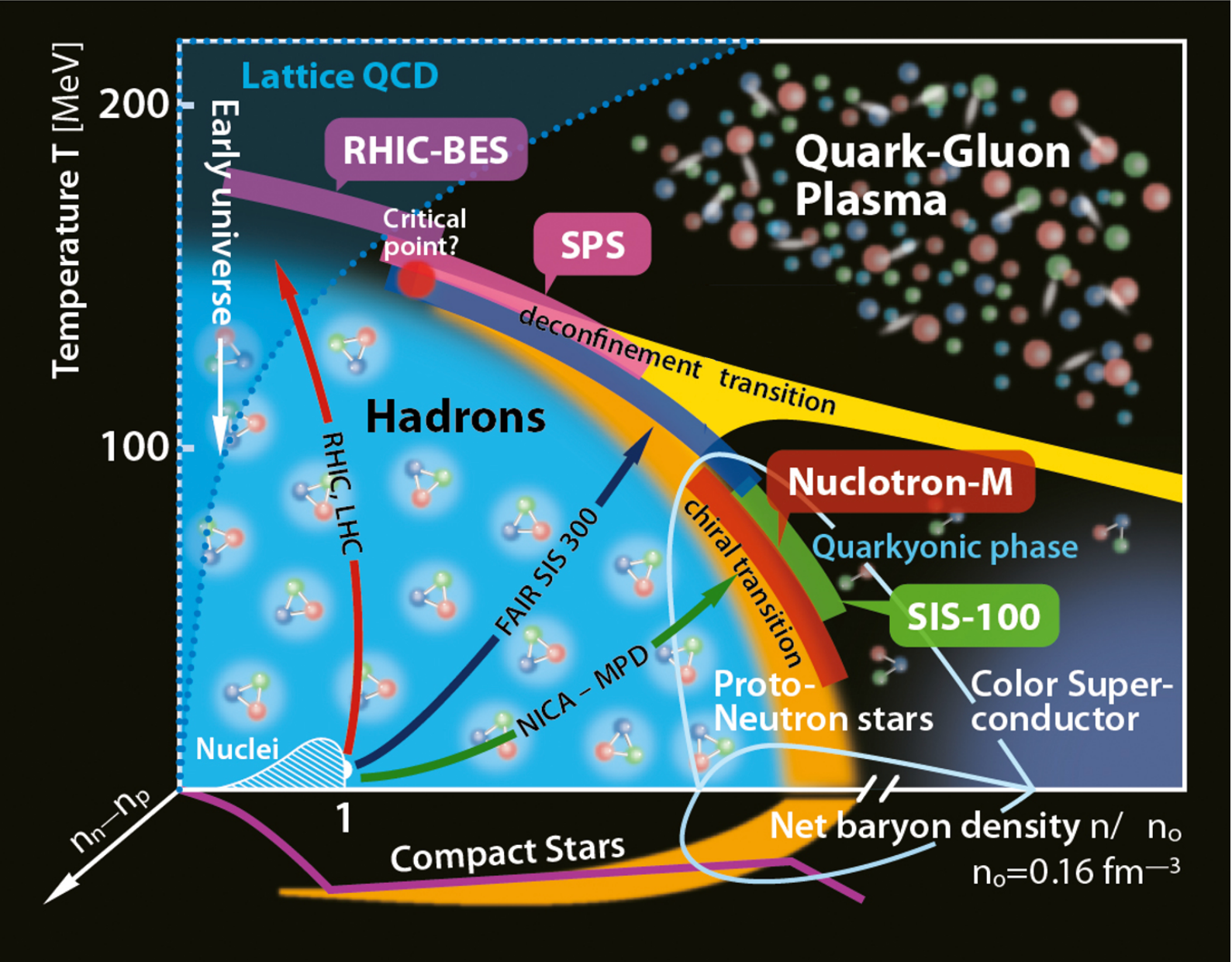}
\caption{(Color online) The QCD diagram. Picture taken from this (CSQCD 2017) conference poster.}
\label{chap0:fig0}
\end{figure}

The nuclear matter driven by the strong interaction acting like the Van der Waals potential for water can undergo the liquid-gas phase transitions in lower energy heavy-ion collisions~\cite{DanielewiczP1979,PochodzallaJ1995,Natowitz2002,YGMa1999,YGMa2005,Borderie2019,WangR2020,LiuC_2022}. The QCD diagram in Fig.~\ref{chap0:fig0}, which presents the relationship among temperature, baryon density (or baryon chemical potential), and nuclear asymmetry (n$_{n}$-n$_{p}$), was explored by the experimental facilities of LHC, RHIC, FAIR , and NICA, etc. The understanding of the properties of the strong interaction phase transition is still rather limited. The exploration of nuclear matter properties and QCD phase diagram would need both new experimental data with extended detection capabilities and theoretical calculations~\cite{CAINES2017121,Ko_2023,Zhang_dc,Ma_2023_1,Jia_jy2023,Ma_2023_2}. As seen from the QCD diagram, the phase transition is not only for nuclear matter but also for the QGP matter. Compilation of data for different substances ~\cite{Csernai2006} demonstrates a drop of the shear viscosity in the region of a phase transition, when the entropy density is employed as a universal reference. Furthermore, from the calculations of shear viscosity of QCD matter in the hadronic phase by the coupled Boltzmann equations of pions and nucleons in low temperatures and low baryon-number densities, one found that $\eta/s$ shows a discontinuity in the bottom of the valley for a first-order phase transition which is similar to water~\cite{ChenJW2007,Meyer_2009_PRD}. The studies on the specific shear viscosity near the liquid$-$gas phase transition may shed new light on the nature of this transition in nucleonic matter~\cite{LiSX2013,Xujun2013}.

\newpage
\section{Analysis methods of shear viscosity}
\label{second}

Shear viscosity is a common property of substances, and there are some methods to calculate it, such as the mean free path, Green-Kubo formula, shear strain rate method, Chapman-Enskog method, and relaxation time method. In this chapter, the details of these methods will be introduced.

\subsection{The definition of shear viscosity}\label{sub1:first-1}
	
\begin{figure}[htb]
\setlength{\abovecaptionskip}{0pt}
\setlength{\belowcaptionskip}{8pt}
\centering\includegraphics[scale=0.6]{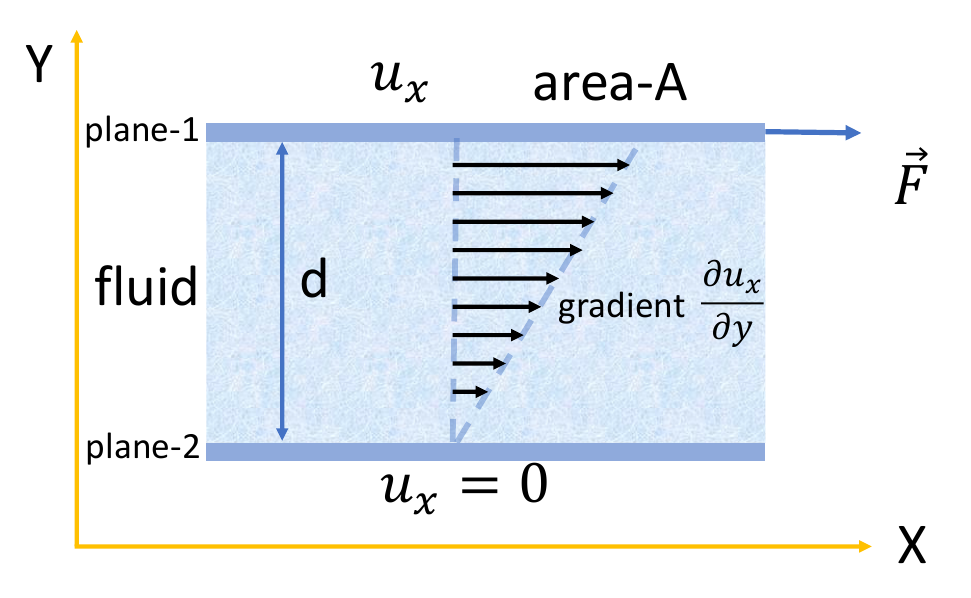}
\caption{(Color online) The illustration of shear stress.}
\label{chap1:fig1}
\end{figure}
Considering two planes `1' and `2' with area $A$ as in Fig.~\ref{chap1:fig1}, a fluid with thickness $d$ is between them. When the plane `2' is fixed, one drags plane `1' with a force $\vec{F}$. Sometime later, the fluid velocity will decrease from the value $u_{x}$ at plane `1' to zero at the stationary plane `2'. The velocity gradient of the fluid is determined by the shear force between flow layers. The shear stress $F/A$ is proportional to $u_{x}/d$ which is called Newton's equation and is written as follows:
\begin{equation}
\begin{split}
\frac{F}{A} = -\eta \frac{u_{x}}{d},
\end{split}                              
\label{chap1:Eqshear1}
\end{equation}
where $\eta$ is called the friction coefficient or the coefficient of shear viscosity. Rewritten Eq.~(\ref{chap1:Eqshear1}) in a differential form, one has 
\begin{equation}
\begin{split}
\frac{F}{A} = -\eta \frac{\partial u_{x}}{\partial y}.
\end{split}                              
\label{chap1:Eqshear2}
\end{equation}
It is seen that the shear viscosity as internal friction causes the irreversible transfer of momentum from points with large velocity to small ones~\cite{Landau1987}. So the physical meaning of shear viscosity is the ability to transfer momentum over a distance of the mean free path, which measures the average distance traveled by a particle between collisions. Hence, a system with lower values of $\eta$ indicates that the constituents of the system interact strongly to transfer momentum easily. On the contrary, a system with a large $\eta$ means that the constituents of the system interact weekly, and it becomes strenuous for the momentum transfer between the constituents~\cite{Ghosh2013}. 

Recalling the fluid mechanics~\cite{Landau1987}, for the ideal fluid, we have,
\begin{equation}
\begin{split}
\frac{\partial p_{\alpha}}{\partial t} =\frac{\partial  \hat{\varrho} u_{\alpha}}{\partial t} = - \frac{\partial T_{\alpha \beta}}{\partial r_{\beta}}, \,\,\,\, \alpha, \beta=x, y, z,
\end{split}                              
\label{chap1:StressTensor1}
\end{equation}
where $p$, $\hat{\varrho}$, and $u$ are the local momentum density, particle mass density and local fluid velocity, respectively. The tensor $T_{\alpha \beta}$ is called momentum flux density tensor which is defined as (for ideal fluid),
 \begin{equation}
\begin{split}
T_{\alpha \beta} = P \delta_{\alpha \beta} + \hat{\varrho} u_{\alpha} u_{\beta},
\end{split}                              
\label{chap1:StressTensor2}
\end{equation}
where $P$ is the pressure and $\delta_{\alpha \beta}$ is equal to 1 (or 0) when $\alpha = \beta$ ($\alpha \neq \beta$). For viscous fluid, an irreversible viscous term-$\Pi_{\alpha \beta}$ is added in momentum flux density tensor,
 \begin{equation}
\begin{split}
T_{\alpha \beta} = \hat{P} \delta_{\alpha \beta} + \hat{\varrho} u_{\alpha} u_{\beta}+\Pi_{\alpha \beta}.
\end{split}                              
\label{chap1:StressTensor3}
\end{equation}
The form of the tensor $\Pi_{\alpha \beta}$ which is called viscous stress tensor, can be written as follows,
 \begin{equation}
\begin{split}
\Pi_{\alpha \beta}= -\eta (\frac{\partial u_{\alpha} }{\partial r_{\beta}}+\frac{\partial u_{\beta} }{\partial r_{\alpha}}-\frac{2}{3}\delta_{\alpha \beta} \frac{\partial u_{\gamma} }{\partial r_{\gamma}})-\xi\delta_{\alpha \beta} \frac{\partial u_{\gamma} }{\partial r_{\gamma}},  \,\,\,\, \alpha, \beta, \gamma=x, y, z,
\end{split}                              
\label{chap1:StressTensor4}
\end{equation}
where $\eta$ and $\xi$ are shear viscosity and bulk viscosity, respectively. In a flat-parallel flow of uniform fluid $u=u_{x}$, the $x,y$ component of $T_{\alpha \beta}$ can be obtained,
\begin{equation}
\begin{split}
T_{xy} = \Pi_{xy}= -\eta \frac{\partial u_{x} }{\partial y}.
\end{split}                              
\label{chap1:StressTensor5}
\end{equation}
In this case, the momentum flux density tensor is equal to the viscous stress tensor. And here the Eq.~(\ref{chap1:StressTensor5}) corresponds to the Eq.~(\ref{chap1:Eqshear2}).

Above, shear viscosity is given by a direct definition or from fluid mechanics. In thermodynamics, a simple estimate of the shear viscosity for a dilute gas can be obtained by the mean free path of particles, which reads as~\cite{TSDT2009,DanielewiczP2009},
\begin{equation}
\begin{split}
\eta = \frac{1}{3}\rho m\bar{v} \lambda,
\end{split}                              
\label{chap2:Eqshear1}
\end{equation}
where $\rho$ is the density, $\bar{v}$ is the average velocity of particles, and $\lambda$ is the mean free path, which can be written as $\lambda$ = $1/(\rho \sigma)$. Here $\sigma$ is the transport cross section.
		
\subsection{Green-Kubo method}\label{sub:second-1}
	
\begin{figure}[htb]
\setlength{\abovecaptionskip}{0pt}
\setlength{\belowcaptionskip}{8pt}
\centering\includegraphics[scale=1.1]{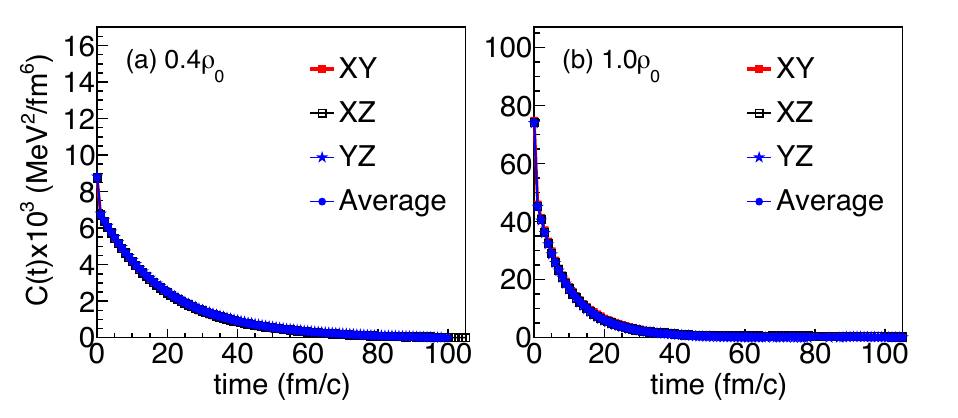}
\caption{(Color online) The evolution of autocorrelation function for different components at $T_{0}$ = 30 MeV and different densities of $0.4\rho_{0}$ (a) or  $1.0\rho_{0}$ (b)  (here $\rho_{0}$ = 0.16 fm$^{-3}$ is the saturation density for nuclear matter). It is worth to mention that the calculations for this figure is without mean field~\cite{XGDeng2021}}.
\label{fig:fig1-0}
\end{figure}

It is a very traditional way~\cite{JR1989} to obtain shear viscosity (or other transport coefficients) for a liquid from pressure or momentum fluctuations with the Green-Kubo relation at equilibrium~\cite{HBerk2002}, which is related to the linear response theory~\cite{R_Kubo_1966}. 
It can be found that the shear viscosity of a gas or liquid system arises from the instantaneous fluctuations due to the thermal motion in the stress tensor. In the classical limit \cite{PKPScott2006}, the standard Green-Kubo formula for the calculation of the shear viscosity coefficient $\eta$ can be expressed by~\cite{R_Kubo_1966,DJE08,AH84,XGDeng2021}:
\begin{equation}
\begin{split}
\eta = \frac{V}{T} \int_{t_{0}}^{\infty} \langle T_{\alpha\beta}(t)T_{\alpha\beta}(t_{0}) \rangle dt;  \,\, \alpha, \beta = x, y, z,  \alpha\neq \beta ,
\end{split}                              
\label{GKubo1}
\end{equation}
where $V$ and $t_{0}$ are system volume and equilibrium time, respectively. And the integral kernel $\langle T_{\alpha\beta}(t)T_{\alpha\beta}(t_{0}) \rangle $ is called the correlation function $C(t)$. The bracket $\langle...\rangle$ denotes average over an equilibrium ensemble of events. In the Green-Kubo method,
the convergence of the correlation function $C(t)$ is important. For the nuclear matter, $C(t)$ was checked in the Improved Quantum Molecular Dynamics (ImQMD) model with mean field~\cite{XGDeng2021} as shown in Fig.~\ref{fig:fig1-0}.
One can see that $C(t)$ of nucleonic matter is convergent. At densities of $0.4\rho_{0}$, the correlation function $C(t)$ tends to zero in a period of time $\sim$ 60 fm/c, and it decreases faster at a higher density of $1.0\rho_{0}$. Since it is symmetric in the simulation box, correlation functions for different components are almost the same.

In Eq.~(\ref{GKubo1}), $T_{\alpha\beta}(t)$ is an off-diagonal element of the stress tensor, which can be given by, 
\begin{equation}
\begin{split}
T_{\alpha\beta}(t) = \frac{1}{V}\int d^{3}r T_{\alpha\beta}(\vec{r}, t),
\end{split}                              
\label{GKubo3}
\end{equation}
where $T_{\alpha\beta}(\vec{r}, t)$ is local tensor which is regarded as the momentum flux density tensor. Let us start from the momentum conservation law~\cite{AAAIMK1959,KAD1963,ZuDN1970}:
\begin{equation}
\begin{split}
 \frac{\partial \textbf{p}(\vec{r}, t)}{\partial t} + \nabla_{r} \cdot \textbf{T} (\vec{r}, t) = 0, 
\end{split}                              
\label{MDC}
\end{equation}
where $\textbf{p}(\vec{r}, t)$ and $\textbf{T}(\vec{r}, t)$ are momentum density and momentum flux density, respectively. One can rewrite Eq.~(\ref{MDC}) as tensor form:
\begin{equation}
\begin{split}
 \frac{\partial p_{\alpha}}{\partial t} + \frac{\partial}{\partial r_{\beta}} T_{\alpha \beta} = 0.
\end{split}                              
\label{MDC-1}
\end{equation}
With a distribution function $f(\vec{r},\vec{p},t)$, the local momentum density can be defined as~\cite{AI2009}:
\begin{equation}
\begin{split}
\textbf{p}(\vec{r}, t)= \int \vec{p} f(\vec{r},\vec{p},t) d^{3}p. 
\end{split}                              
\label{MDC-2}
\end{equation}
In statistical mechanics, a system is no longer considered as a continuous fluid, but rather treated an ensemble consisted of particles with the number $A$. In a many-particle system, the distribution function $f(\vec{r},\vec{p},t)$ could have Dirac's form~\cite{Bertsch1988}:
\begin{equation}
\begin{split}
f(\vec{r},\vec{p},t) = \sum_{i}^{A} \delta (\vec{r}- \vec{R}_{i}(t))\delta (\vec{p}- \vec{P}_{i}(t)),
\end{split}                              
\label{MDC-3}
\end{equation}
where $\vec{R}_{i}(t)$ and $\vec{P}_{i}(t)$ are the time-dependent coordinate and momentum of particle-$i$, or Gauss form~\cite{AICHELIN1991}:
\begin{equation}
\begin{split}
f(\vec{r},\vec{p},t) = \frac{1}{(\pi \hbar)^{3}}\sum_{i}^{A} {\rm exp}\Big{[}-\frac{(\vec{r}-\vec{R}_{i}(t))^{2}}{2L}\Big{]}{\rm exp}\Big{[}-\frac{2L(\vec{p}-\vec{P}_{i}(t))^{2}}{\hbar^{2}}\Big{]},
\end{split}                              
\label{MDC-4}
\end{equation}
where $L$ is square of width of Gauss wave packet in unit of $\rm fm^{2}$. With distribution function $f(\vec{r},\vec{p},t)$, one can have following integral relations:
\begin{equation}
\begin{split}
\int f(\vec{r},\vec{p},t) d^{3}p = \rho (\vec{r}) \, ,
\end{split}                              
\label{MDC-6-0}
\end{equation}
\begin{equation}
\begin{split}
\int f(\vec{r},\vec{p},t) d^{3}r = g (\vec{p}) \, .
\end{split}                              
\label{MDC-6-1}
\end{equation} 
where $\rho (\vec{r})$ and $g (\vec{p})$ are the densities in coordinate and momentum spaces, respectively. Substituted $f(\vec{r},\vec{p},t)$ into Eq.(\ref{MDC-2}), one can get:
\begin{equation}
\begin{split}
\textbf{p}(\vec{r}, t)= \sum_{i}^{A} \vec{P}_{i}\rho_{i}, 
\end{split}                              
\label{MDC-7}
\end{equation} 
where the $\rho_{i}$ is density of particle $i$ which can be expressed as:
\begin{equation}
\begin{split}
\rho_{i}(\vec{r})= \delta(\vec{r}-\vec{R}_{i}), 
\end{split}                              
\label{MDC-8}
\end{equation} 
or
\begin{equation}
\begin{split}
\rho_{i}(\vec{r}) = \frac{1}{(2\pi L)^{3/2}} \exp \Big[-\frac{(\vec{r}-\vec{R}_{i})^{2}}{2L} \Big] \, .
\end{split}                              
\label{MDC-9}
\end{equation}
Taking derivative of Eq.~(\ref{MDC-7}) with respect to time:
\begin{equation}
\begin{split}
\frac{\partial\textbf{p}}{\partial t}= \sum_{i}^{A} \frac{\partial \vec{P}_{i}}{\partial t} \rho_{i} +\sum_{i}^{A} \vec{P}_{i}\frac{\partial \rho_{i}}{\partial t} .
\end{split}                               
\label{MDC-10}
\end{equation}
For the derivative of $\rho_{i}$ with respect to time:
\begin{equation}
\begin{split}
\frac{\partial \rho_{i}}{\partial t}= \frac{\partial \rho_{i}}{\partial \vec{R}_{i}} \frac{\partial \vec{R}_{i}}{\partial t}=- \frac{\partial \rho_{i}}{\partial \vec{r}} \frac{\partial \vec{R}_{i}}{\partial t} \,.
\end{split}                               
\label{MDC-11-0}
\end{equation}
Thus the second term of right hand side of Eq.~(\ref{MDC-10}) can reach:
\begin{equation}
\begin{split}
\sum_{i}^{A} \frac{\partial \rho_{i}}{\partial t} P_{i\alpha}= - \frac{\partial}{ \partial r_{\beta}} \sum_{i}^{A} \Big{(}\frac{P_{i\alpha}P_{i\beta}}{m} \Big{)} \rho_{i} \, .
\end{split}                               
\label{MDC-11}
\end{equation}
And for the first term of right hand side of Eq.~(\ref{MDC-10}):
\begin{equation}
\begin{split}
\sum_{i}^{A} \frac{\partial \vec{P}_{i}}{\partial t} \rho_{i} = \sum_{i}^{A} \vec{F}_{i} \rho_{i}  \, .
\end{split}                               
\label{MDC-12}
\end{equation}
If the interaction is only two-body, we have
\begin{equation}
\begin{split}
\vec{F}_{i} =\sum_{i \neq j} \vec{F}_{ij}\, .
\end{split}                               
\label{MDC-13}
\end{equation}
Thus,
\begin{equation}
\begin{split}
\sum_{i}^{A}\vec{F}_{i}\rho_{i} = \frac{1}{2} \Big{(}\sum_{i}^{A}\vec{F}_{i}\rho_{i}+\sum_{j}^{A}\vec{F}_{j}\rho_{j}\Big{)} = \frac{1}{2} \Big{(}\sum_{i}\sum_{i \neq j} \vec{F}_{ij}\rho_{i} +\sum_{j}\sum_{j \neq i} \vec{F}_{ji}\rho_{j} \Big{)}  \, ,
\end{split}                               
\label{MDC-14}
\end{equation}
where $\vec{F}_{ij}$ is force exerted on particle-$i$ by particle-$j$. And  $\vec{F}_{ij}$ = $-\vec{F}_{ji}$, then
\begin{equation}
\begin{split}
\sum_{i}^{A}\vec{F}_{i}\rho_{i} =  \frac{1}{2} \Big{[} \sum_{i}\sum_{i \neq j} \vec{F}_{ij}\Big{(}\rho_{i}-\rho_{j}  \Big{)}\Big{]} \, .
\end{split}                               
\label{MDC-15}
\end{equation}
With Taylor series (due to $R_{ij\beta} = R_{i\beta}- R_{j\beta}$ which is relative position of particle $i$ and $j$ is finite):
\begin{equation}
\begin{split}
\rho_{i}(r_{\beta}-R_{i\beta})&=\rho_{i}[r_{\beta}-R_{j\beta} + (R_{j\beta}-R_{i\beta})]   \\
                                            &=\rho_{i}[r_{\beta}-R_{j\beta} +  R_{ji\beta}]     \\   
                                            &= \sum_{n=0}^{N} \frac{(R_{ij\beta})^{n}}{n!} \rho_{j}^{(n)}(r_{\beta}-R_{j\beta})\\
                                            &\approx  \rho_{j}(r_{\beta}-R_{j\beta})+ R_{ji\beta} \frac{\partial}{\partial r_{\beta}}\rho_{j}(r_{\beta}-R_{j\beta})\, .   
\end{split}                              
\label{MDC-16}
\end{equation}
 Then $R_{ij\beta} = -R_{ji\beta} $, 
\begin{equation}
\begin{split}
\rho_{i}(r_{\beta}-R_{i\beta})-\rho_{j}(r_{\beta}-R_{j\beta})= -R_{ij\beta} \frac{\partial}{\partial r_{\beta}}\rho_{j}(r_{\beta}-R_{j\beta}) .   
\end{split}                              
\label{MDC-17}
\end{equation}
According to Eq.~(\ref{MDC-12}), Eq.~(\ref{MDC-15}), and Eq.~(\ref{MDC-17}), 
\begin{equation}
\begin{split}
\sum_{i}^{A} \frac{\partial \vec{P}_{i}}{\partial t} \rho_{i}  = - \frac{1}{2} \sum_{i}\sum_{i \neq j}  F_{ij\alpha} R_{ij\beta}\frac{\partial}{\partial r_{\beta}}\rho_{j}(r_{\beta}-R_{j\beta}) =
 \frac{\partial}{\partial r_{\beta}} \Big{(}- \frac{1}{2} \sum_{i}\sum_{i \neq j}  F_{ij\alpha} R_{ij\beta} \rho_{j}\Big{)} \, .
\end{split}                               
\label{MDC-18}
\end{equation}
If interaction includes three-body component, in the same way, one can get
\begin{equation}
\begin{split}
\frac{1}{2} \sum_{i} \sum_{i\neq j}\sum_{i\neq j \neq k} F_{ijk\alpha}\rho_{i}= \frac{1}{6}\sum_{i}\sum_{i\neq j}\sum_{i\neq j \neq k} (F_{ijk\alpha}\rho_{i}+F_{jki\alpha}\rho_{j} +F_{kij\alpha}\rho_{k})  \, .
\end{split}                               
\label{MDC-19}
\end{equation}
With the condition,
\begin{equation}
\begin{split}
 F_{ijk\alpha}+F_{jki\alpha} +F_{kij\alpha}=0\, .
\end{split}                               
\label{MDC-20}
\end{equation}
Then we obtain,
\begin{equation}
\begin{split}
\frac{1}{2} \sum_{i} \sum_{i\neq j}\sum_{i\neq j \neq k} F_{ijk\alpha}\rho_{i}\approx \frac{\partial}{\partial r_{\beta}} \Big{[}- \frac{1}{6}\sum_{i}\sum_{i\neq j}\sum_{i\neq j \neq k} (F_{ijk\alpha}R_{ik\beta}+F_{jki\alpha}R_{jk\beta} )\rho_{k} \Big{]} \, .
\end{split}                               
\label{MDC-21}
\end{equation}
Finally, according to Eqs.(\ref{MDC-1}), (\ref{MDC-11}), (\ref{MDC-12}), (\ref{MDC-19}), and (\ref{MDC-21}), the momentum flux density tensor can be given,
\begin{equation}
\begin{split}
&T_{\alpha\beta}(\vec{r},t) = \sum_{i}^{A} \frac{P_{{i}\alpha}P_{{i}\beta}}{m_{i}}\rho_{i}(\vec{r},t)   \\
& + \frac{1}{2} \sum_{i}^{A} \sum_{i\neq j}^{A} F_{ij\alpha}R_{ij\beta} \rho_{j}(\vec{r},t)             \\   
 & + \frac{1}{6} \sum_{i}^{A} \sum_{i\neq j}^{A}\sum_{i\neq j \neq k}^{A} (F_{ijk\alpha}R_{ik\beta}+ F_{jki\alpha}R_{jk\beta} )\rho_{k}(\vec{r},t) \\
&+\cdots,
\end{split}                              
\label{MDC-22}
\end{equation}
The first term on the right-hand side is the momentum term, the second is the two-body interaction term, and the third is the three-body interaction term. 

The $\eta$ in Eq.~(\ref{GKubo1}) is a normal (or traditional) form of the Green-Kubo formula. In the last few decades, it has been extensively used in molecular dynamics simulations. In Ref.~\cite{XGDeng2021}, a more general form of the Green-Kubo formula is given for the calculation of the shear viscosity for both Fermi-Dirac and Boltzmann distribution systems, which is read as,
\begin{equation}
\begin{split}
\eta_{new} = \frac{mVA}{\langle \frac{1}{3}\sum_{i}^{A}p_{i}^{2} \rangle} \int_{t_{0}}^{\infty}  \langle T_{\alpha\beta}(t) T_{\alpha\beta}(t_{0}) \rangle  dt,
\end{split}                              
\label{GKubo2}
\end{equation}
where $m$ is the particle mass. The derivation can be found in the Ref.~\cite{XGDeng2021}. It is obvious, of course, in  the classic limit, one has,
\begin{align}
\langle \frac{1}{3}\sum_{i}^{A}p_{i}^{2} \rangle = Amk_{B}T, 
\label{MDC-23} 
\end{align}
where $k_{B}$ is the Boltzmann constant (here $k_{B}$ set to 1). Substituting Eq.~(\ref{MDC-23}) into Eq.~(\ref{GKubo2}), we can get Eq.~(\ref{GKubo1}). To calculate shear viscosity directly from the Green-Kubo relation, one simply performs an equilibrium simulation of an $A$-particle system.

\subsection{Shear strain rate method}\label{sub:second-2}
\begin{figure}[htb]
\setlength{\abovecaptionskip}{0pt}
\setlength{\belowcaptionskip}{8pt}
\centering\includegraphics[scale=1.3]{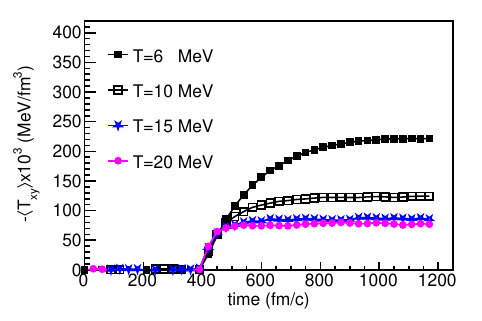}
\caption{(Color online) Stress tensor as a function of time at different temperatures and density of 0.4$\rho_{0}$ ($\rho_{0}$ = 0.16 fm$^{3}$ called saturation density of nuclear matter) without the mean field~\cite{XGDeng2021}.}
\label{fig:fig1}
\end{figure}

Instead of measuring the intrinsic fluctuations of the system, the viscosity can also be obtained using the non-equilibrium molecular dynamics (NEMD) calculation by imposing a Couette flow. This method was proposed by Evans and Morriss \cite{PGai2006} and named the SLLOD algorithm, which has been extensively applied to predict the rheological properties of real fluids \cite{PGai2006}. The SLLOD algorithm is related to the dynamics with strain rate $\dot{\gamma}$ associated with the Lees-Edwards periodic boundary conditions ~\cite{evans_morriss_2008} and it is for shear viscosity with a planar Couette flow field at shear rate $\dot{\gamma}$ = $\partial v_{x}/\partial y$ which is the change in streaming velocity $v_{x}$ in the $x$-direction with vertical position $y$. We can obtain shear viscosity by
\begin{equation}
\begin{split}
\eta = -\frac{\langle T_{xy} \rangle }{\dot{\gamma}}.
\end{split}                              
\label{GKubo6}
\end{equation}
In Ref.~\cite{XGDeng2021}, the strain rate was implanted in the framework of an improved quantum molecular dynamic (ImQMD) model without mean field, which is the cascade mode. As in Fig.~\ref{fig:fig1}~\cite{XGDeng2021}, it shows the stress tensor as a function of time at different temperatures, which increases when the shear rate is $\dot{\gamma}$ at 400 fm/c and reaches a constant value after a time equals to the calculated shear viscosity.

\begin{figure}[htb]
\setlength{\abovecaptionskip}{0pt}
\setlength{\belowcaptionskip}{8pt}
\centering\includegraphics[scale=1.3]{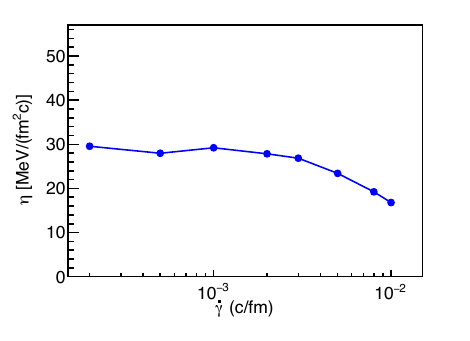}
\caption{(Color online) Deduced shear viscosity as a function of the shear rate $\dot{\gamma}$ applied within the SLLOD algorithm, in the system at the density of $0.2 \, \rho_{0}$ and temperature $T = 10 \, \text{MeV}$, when no mean field is employed.}
\label{fig:fig2}
\end{figure}

The effect of the shear rate on the shear viscosity has been checked. As in Fig.~\ref{fig:fig2}, one can only notice in some ranges the shear viscosity keeps constant.
So in the simulations, the shear rate cannot be too large or small. If it is small, flow fields cannot be implanted in the system. On the other hand, when shear rate is too large, energy can not be kept conservative when the dynamical equations of motion are solved with a finite time step. 

By adding the shear rate, the dynamical equations of motion for the system can be rewritten as
\begin{align}
&\frac{d\vec{r}_{i}}{dt} = \frac{\vec{p}_{i}}{m_{i}} + \dot{\gamma}y_{i} \hat{x}               
\label{GKubo77}  \\
&\frac{d\vec{p}_{i}}{dt} = \vec{F}_{i}-\dot{\gamma}p_{yi} \hat{x}-a \vec{p}_{i}.        
\label{GKubo88}
\end{align}
Eq.~(\ref{GKubo77})  and Eq.~(\ref{GKubo88}) are called the SLLOD equations. In order to keep the kinetic energy conservation, one needs a `thermostat'. Thus a multiplier is applied to motion equations. Considering conservation of kinetic energy, one can get
\begin{equation}
\begin{split}
a = \frac{\sum_{i}(\vec{F}_{i}\cdot \vec{p}_{i}/m_{i}-\dot{\gamma}p_{xi}p_{yi}/m_{i})}{\sum_{i}p_{i}^{2}/m_{i}}.
\end{split}                              
\label{GKubo9}
\end{equation}

\subsection{Chapman-Enskog method}\label{sub:second-3}
	
Chapman and Enskog (CE) have extracted the shear viscosity $\eta$ in a gas of nonrelativistic particles by using the Boltzmann kinetic equation for the phase-space distribution function~\cite{Chamman-52}. One should notice that originally the CE method was for Maxwell-Boltzmann statistics, but here it is for the Fermi system with the same strategy~\cite{SLDPawel2003}. In the CE method, one can solve the Boltzmann equation by applying a series expansion to the distribution function~\cite{MNMLinZW2022}. In Ref.~\cite{DPawel1984}, one considered Boltzmann-Uehling-Uhlenbeck equation without potential $U$~\cite{Bertsch1988}, 
\begin{eqnarray}
\setlength{\abovedisplayskip}{-3cm}
\setlength{\belowdisplayskip}{-3cm}
\frac{\partial}{\partial t}{f_{1}(\vec{r},\vec{p},t)} +\frac{\vec{p}}{m}\cdot\nabla_{\vec{r}}{f_{1}(\vec{r},\vec{p},t)}=I_{coll.}
\label{BUUequation}
\end{eqnarray}%
The right-hand side (rhs) of Eq.(\ref{BUUequation}) is the collision term ($I_{coll.}$), which denotes collision process of 1+2$\rightarrow$3+4, and it can be read as~\cite{YXZhang2018}:
\begin{eqnarray}
I_{coll.}=\big{(}\frac{\partial {f_{1}}}{\partial t} \big{)}_{coll.}=\frac{\rm g}{(2\pi \hbar)^{3}}\int d^{3}p_{2}d\Omega\frac{d\sigma_{NN}}{d\Omega}|\vec{\upsilon}_{1}-\vec{\upsilon}_{2}|{\times}\mathcal{P}{\times}       
\delta^{3}(\vec{p}_{1}+\vec{p}_{2}\!-\!\vec{p}_{3}\!-\!\vec{p}_{4}).
\label{ICOLLequation}
\end{eqnarray}%
where $\rm g=4$ is the spin and isospin degeneracy factor, and $\sigma_{NN}$ is the nucleon-nucleon cross section. And the $\delta^{3}(\vec{p}_{1}+\vec{p}_{2}\!-\!\vec{p}_{3}\!-\!\vec{p}_{4})$ term is for keeping momentum conservation in the collisions. In Eq.(\ref{ICOLLequation}), $\mathcal{P}$ denotes the Pauli blocking factor, which reads,
\begin{eqnarray}
\mathcal{P}=f_{3}f_{4}(1-f_{1})(1\!-\!f_{2})\!-\!f_{1}f_{2}(1\!-\!f_{3})(1\!-\!f_{4})
\label{Pauliequation}
\end{eqnarray}%
where the first term of the right-hand side is the gain term (3+4$\rightarrow$1+2) and the second term of the right-hand side is the loss term (1+2$\rightarrow$3+4). 

Allowing the system to be slightly out of equilibrium, the first order of $f$ can be written as~\cite{DPawel1984,Uehling1933,DaD1996},
\begin{eqnarray}
f=f_{eq}+\delta f= f^{0}+f^{0}(1-f^{0})\Phi \, ,
\label{CEequation1}
\end{eqnarray}%
where $f_{eq}=f^{0}$. Substituting Eq.(\ref{CEequation1}) into the collision integral Eq.(\ref{ICOLLequation}) and neglecting terms of quadratic and higher order in $\Phi$~\cite{DPawel1984,DaD1996}, one can obtain
\begin{eqnarray}
\frac{\partial}{\partial t}{f_{1}^{0}} +\frac{\vec{p}}{m}\cdot\nabla_{\vec{r}}{f_{1}^{0}}&=&\frac{\rm g}{(2\pi \hbar)^{3}}\int d^{3}p_{2}d\Omega\frac{d\sigma_{NN}}{d\Omega}|\vec{\upsilon}_{1}-\vec{\upsilon}_{2}|{\times}f^{0}_{1}f^{0}_{2}(1-f_{3}^{0})(1-f^{0}_{4})       \notag \\
&{\times}& (\Phi_{3}+\Phi_{4}-\Phi_{1}-\Phi_{2}) {\times} \delta^{3}(\vec{p}_{1}+\vec{p}_{2}\!-\!\vec{p}_{3}\!-\!\vec{p}_{4}).
\label{CEequation2}
\end{eqnarray}%
In the first-order approximation, the contribution of $\delta f$ on the left side can be neglected, and $f_{1}$ is replaced by $f_{1}^{0}$. Considering the existence of a motion with a slightly inhomogeneous velocity $\vec{u}$ in the liquid~\cite{AAAIMK1959}, we can write the distribution function $f^{0}$ as,
\begin{eqnarray}
f^{0}= \frac{1}{e^{(E-\vec{p}\cdot\vec{u}-\mu)/T}+1},
\label{CEequation3}
\end{eqnarray}%
where $\mu$ is chemical potential. By substituting (\ref{CEequation3}) to the left-hand side of (\ref{BUUequation}), expanding in terms of gradients of temperature and fluid velocity and assuming $\bigtriangledown$$\cdot$$\vec{u}=0$ for an incompressible fluid that drives bulk viscosity and is neglected here~\cite{AAAIMK1959,DPawel1984,PChakraborty2011}, we have
\begin{eqnarray}
\frac{\partial f^{0}}{\partial t} +\frac{\vec{p}}{m}\cdot\nabla_{\vec{r}}{f^{0}}&=&\frac{1}{2mT}f^{0}f^{0}\Big{[}(p_{\alpha}p_{\beta}
-\frac{1}{3}\delta_{\alpha\beta}p^{2})(\frac{\partial u_{\alpha}}{\partial r_{\beta}}+\frac{\partial u_{\beta}}{\partial r_{\alpha}}-\frac{2}{3}\delta_{\alpha\beta}\bigtriangledown\cdot\textbf{u})+\frac{1}{mT}(p^{2}-\frac{5}{3}\langle{p^{2}}\rangle)p_{\alpha}\frac{\partial{T}}{\partial{r_{\alpha}}}\Big{]} \notag \\
&=&\frac{1}{2mT}f^{0}f^{0}\Big{[}(p_{\alpha}p_{\beta}
-\frac{1}{3}\delta_{\alpha\beta}p^{2})(\frac{\partial u_{\alpha}}{\partial r_{\beta}}+\frac{\partial u_{\beta}}{\partial r_{\alpha}})+\frac{1}{mT}(p^{2}-\frac{5}{3}\langle{p^{2}}\rangle)p_{\alpha}\frac{\partial{T}}{\partial{r_{\alpha}}}\Big{]}.
\label{CEequation4}
\end{eqnarray}%
where $\langle{p^{2}}\rangle$ is an average over distribution. In Eq.(\ref{CEequation4}), we only pick up the terms which are relevant to shear viscosity and heat conduction coefficient. For symmetry consideration between the right-hand side of Eq.(\ref{CEequation2}) and the right-hand side of Eq.(\ref{CEequation4}), it can be found that $\Phi$ has the form
\begin{eqnarray}
\Phi=q^{(1)}(p_{\alpha}p_{\beta}
-\frac{1}{3}\delta_{\alpha\beta}p^{2})(\frac{\partial u_{\alpha}}{\partial r_{\beta}}+\frac{\partial u_{\beta}}{\partial r_{\alpha}})+q^{(2)}(p^{2}-\frac{5}{3}\langle{p^{2}}\rangle)p_{\alpha}\frac{\partial{T}}{\partial{r_{\alpha}}},
\label{CEequation6}
\end{eqnarray}%
where the $q^{(1)}$ and $q^{(2)}$ are constant factors~\cite{AAAIMK1959,DPawel1984}. And by multiplying both sides of Eq.(\ref{CEequation2}) by $\Phi_{1}$ and integrating over the momenta $d^{3}p_{1}$, one can obtain the constants. 

\begin{figure}[htb]
\setlength{\abovecaptionskip}{0pt}
\setlength{\belowcaptionskip}{8pt}
\centering\includegraphics[scale=0.9]{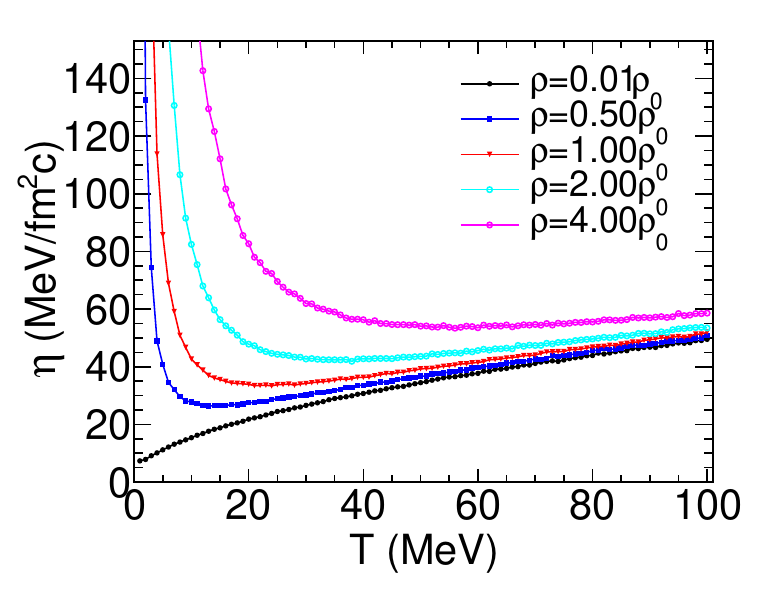}
\caption{(Color online) Shear viscosity as a function of temperature at different nuclear matter densities.}
\label{fig:fig3}
\end{figure}

Shear viscosity ($\eta$) as the coefficient proportional to the shear momentum flux tensor (one also can see from Eq.~\ref{chap1:StressTensor4}) without bulk viscosity,
\begin{eqnarray}
\Pi_{\alpha\beta}&=&-\eta (\frac{\partial u_{\alpha}}{\partial r_{\beta}}+\frac{\partial u_{\beta}}{\partial r_{\alpha}}-\frac{2}{3}\delta_{\alpha\beta}\bigtriangledown\cdot\vec{u})  \notag \\
 &=&-\eta (\frac{\partial u_{\alpha}}{\partial r_{\beta}}+\frac{\partial u_{\beta}}{\partial r_{\alpha}})
\label{CEequation7}
\end{eqnarray}%
where the $\bigtriangledown$$\cdot$$\vec{u}$ is assumed to be zero. A slight out of equilibrium produces a change in the dissipative stress tensor; at the lowest order in $\delta f$, we can obtain,
\begin{eqnarray}
\Pi_{\alpha\beta}&=&\frac{\rm g}{(2\pi\hbar)^{3}}\int d^{3}p \frac{p_{\alpha}p_{\beta}}{m} f= \frac{\rm g}{(2\pi\hbar)^{3}}\int d^{3}p \frac{p_{\alpha}p_{\beta}}{m}f_{0}+\frac{\rm g}{(2\pi\hbar)^{3}}\int d^{3}p \frac{p_{\alpha}p_{\beta}}{m} \delta f  \notag \\
                          &=& \frac{\rm g}{(2\pi\hbar)^{3}}\int d^{3}p \frac{p_{\alpha}p_{\beta}}{m} \delta f            \notag \\
                          &=&\frac{\rm g}{(2\pi\hbar)^{3}}\int d^{3}p \frac{p_{\alpha}p_{\beta}}{m} f^{0}(1-f^{0})\Phi .
\label{CEequation8}
\end{eqnarray}%
The equilibrium term with $f_{0}$ in $\Pi_{\alpha\beta}$ is zero. Substituting (\ref{CEequation6}) into (\ref{CEequation8}), one can get the expression for shear viscosity~\cite{DPawel1984,BBDPawel2019}:
\begin{equation}
\begin{split}
\eta& = \frac{5T}{9}\frac{\Big{(}\int d^{3}p_{1}{\,\,}p^{2}_{1}f_{1}\Big{)}^{2}}{\displaystyle \int d^{3}p_{1}d^{3}p_{2}d\Omega {\,} v_{12}q_{12}^{4}sin^{2}{\theta}\frac{d\sigma_{NN}}{d\Omega} f_{1}f_{2}\tilde{f}_{3}\tilde{f}_{4}}, 
\end{split}                              
\label{CEequation9}
\end{equation}
where $v_{12} = |\vec{v}_{1}-\vec{v}_{2}|$, $q_{12} = |\vec{p}_{1}-\vec{p}_{2}|/2$ and $\theta$ are relative velocity, relative momentum, and scattering angle between the total momentum $\vec{p}_{t}=\vec{p}_{1}+\vec{p}_{2}=\vec{p}_{3}+\vec{p}_{4}$ and the relative momentum of the final state $\vec{p}_{r}=\vec{p}_{3}-\vec{p}_{4}$, respectively. $\sigma_{NN}$ is the total nucleon-nucleon cross section, and $\tilde{f}_{3}$$\tilde{f}_{4}$ = [1-$f(p_{3})$][1-$f(p_{4})$] is the Pauli blocking term. In Eq.(\ref{CEequation9}), the conservations of momentum and energy between $p_{1}, p_{2}$ and $p_{3}, p_{4}$ are needed to be taken into account in the collisions. 

By solving Eq.(\ref{CEequation9}) with the Monte-Carlo approach, one obtains the shear viscosity as shown in Fig.\ref{fig:fig3}. Here the $\frac{d\sigma_{NN}}{d\Omega}$ is set to 40 \rm{mb}/4$\pi$. It is seen that the shear viscosity increases as temperature increases for very low density (0.01$\rho_0$) which can be treated as Fermion gas. For the Fermion gas, the Pauli blocking effect is weak, and so the behavior for shear viscosity acts like that of classic gas. For higher densities, however, shear viscosity decreases with increasing of temperature in low-temperature region since Pauli blocking effect is strong. As temperature increases again, particle velocity becomes dominative leading to the increasing of shear viscosity.

\subsection{Relaxation time method}\label{sub:second-4}

In the relaxation time approximation method, one assumes that the effect of collisions is always to bring the perturbed distribution function close to the equilibrium distribution function~\cite{WAnton2012,Plumari_2012_PRC},  $f=f_{eq}+\delta f $ which is as Eq.(\ref{CEequation1}). Then one writes the collision term exponentially with a relaxation time of $\tau$, 
\begin{eqnarray}
\mathcal{D} f= -\frac{f-f_{eq}}{\tau} = -\frac{\delta f}{\tau}
\label{Rtimeequation1}
\end{eqnarray}%
where $\mathcal{D} = \frac{\partial}{\partial t} +\vec{v}\cdot\nabla_{\vec{r}}-\nabla_{\vec{r}}U\cdot\nabla_{\vec{p}}$. In the relaxation time approximation, one starts from shear force between flow layers per unit area and obtains the shear viscosity~\cite{Xujun2011}
\begin{eqnarray}
\eta = -\sum_{\tau} \frac{\rm g}{(2\pi)^{3}} \int {\tau}_{\tau}(p) \frac{p_{z}^{2}p_{x}^{2}}{pm^{\star}_{\tau}} \frac{df_{\tau}^{0}}{dp}d^{3}p \,,
\label{Rtimeequation2}
\end{eqnarray}%
where the subscript $\tau$ is the isospin factor, which is set to 1/2 ($-$1/2) for neutrons (protons). So here the degeneracy factor $\rm g$=2. And $m^{\star}_{\tau}$ is the effective mass, which reads as,
\begin{eqnarray}
\frac{1}{m^{\star}_{\tau}} = \frac{1}{m}+\frac{1}{p}\frac{dU_{\tau}}{dp} \, ,
\label{Rtimeequation3}
\end{eqnarray}%
where $U_{\tau}$ is the single-particle potential for a nucleon with momentum $\vec{p}$ and isospin $\tau$ in a nuclear matter medium, which can be found in Ref.~\cite{Xujun2011}. In Eq.~(\ref{Rtimeequation2}), $f_{\tau}$ is the local momentum distribution, which is written as,
\begin{eqnarray}
f_{\tau}= \frac{1}{{\rm exp}{\big{[}\frac{p^{2}}{2m}-U_{\tau}(\vec{p})-\mu_{\tau}\big{]}\big{/}T}+1} \,.
\label{Rtimeequation4}
\end{eqnarray}%
And in Eq.~(\ref{Rtimeequation2}), $\tau_{\tau}(p)$  is the relaxation time which can be expressed as,
\begin{eqnarray}
\frac{1}{\tau_{\tau}(p_{1})} = \frac{1}{\tau_{\tau}^{\rm same}(p_{1})}  +\frac{1}{\tau_{\tau}^{\rm diff}(p_{1})} 
\label{Rtimeequation5}
\end{eqnarray}%
where $\tau_{\tau}^{\rm same (diff)}(p_{1})$ is the average collision time for a nucleon with isospin $\tau$ and momentum $p_{1}$ when colliding with other
nucleons of same (different) isospin, and they can be calculated from \cite{Xujun2011},
\begin{eqnarray}
\frac{1}{\tau_{\tau}^{\rm same}(p_{1})} &=& \big{(}{\rm g} - \frac{1}{2}\big{)}\frac{1}{2\pi} \int p_{2}^{2}dp_{2} \,d{\rm cos}\,\theta_{12} \, d\, {\rm cos} \theta \frac{d\sigma_{\tau,\tau}}{d\Omega}   \Big{|} \frac{\vec{p}_{1} }{m_{\tau}^{\star}}-\frac{{\vec{p}}_{2}}{m_{\tau}^{\star}} \Big{|}  \notag \\
&\times& \big{[} f^{0}_{\tau}(p_{2})- f^{0}_{\tau}(p_{2})f^{0}_{\tau}(p_{3})- f^{0}_{\tau}(p_{2})f^{0}_{\tau}(p_{4})+ f^{0}_{\tau}(p_{3})f^{0}_{\tau}(p_{4})\big{]} \,, \,\,\,\,\,\,\,\,\,\,\,\,\,\,\,\,\,\,\,
\label{Rtimeequation6-0}
\end{eqnarray}%
\begin{eqnarray}
\frac{1}{\tau_{\tau}^{\rm diff}(p_{1})} &=& \frac{{\rm g}}{2\pi} \int p_{2}^{2}dp_{2} \,d{\rm cos}\,\theta_{12} \, d\, {\rm cos} \theta \frac{d\sigma_{\tau,\tau}}{d\Omega}   \Big{|} \frac{\vec{p}_{1} }{m_{\tau}^{\star}}-\frac{{\vec{p}}_{2}}{m_{-\tau}^{\star}} \Big{|}  \notag \\
&\times& \big{[} f^{0}_{-\tau}(p_{2})- f^{0}_{-\tau}(p_{2})f^{0}_{\tau}(p_{3})- f^{0}_{-\tau}(p_{2})f^{0}_{-\tau}(p_{4})+ f^{0}_{\tau}(p_{3})f^{0}_{-\tau}(p_{4})\big{]} \,,
\label{Rtimeequation6-1}
\end{eqnarray}%
where $\theta_{12}$ is the angel between $\vec{p}_{1}$ and $\vec{p}_{2}$.

In this chapter, the definition of shear viscosity is given from two planes, and some methods to estimate shear viscosity are given. There are more methods for calculating shear viscosity as in Refs.~\cite{Wesp_2011_PRC,Plumari_2012_PRC,Rose_2018_PRC}.

\newpage
\section{Theoretical aspects of shear viscosity in nucleonic matter}\label{third}

\subsection{Infinite nuclear matter}\label{sub:first-1}

\begin{figure}[htb]
\setlength{\abovecaptionskip}{0pt}
\setlength{\belowcaptionskip}{8pt}
\centering\includegraphics[scale=1.0]{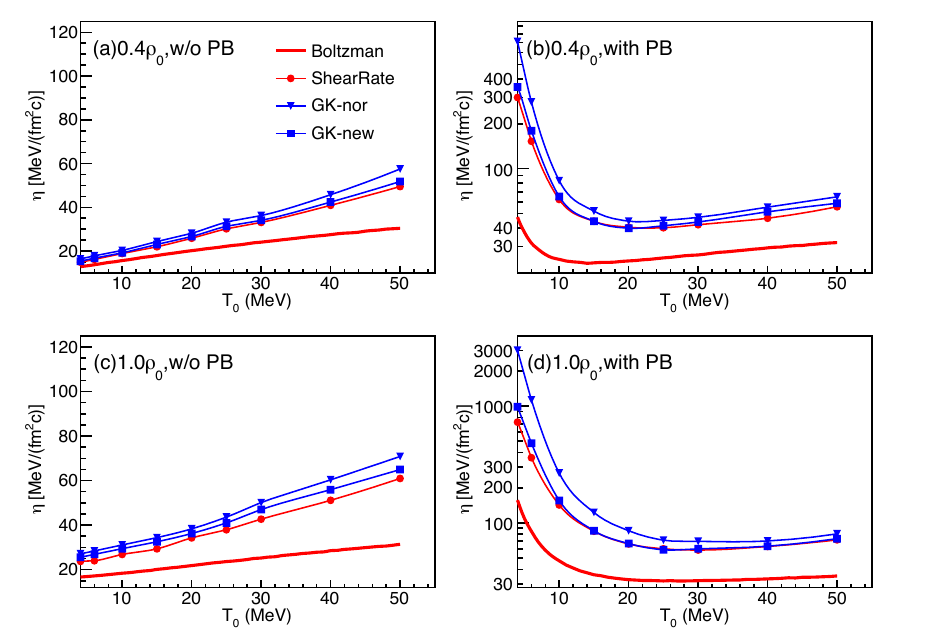}
\caption{(Color online) Different calculations of shear viscosity as a function of temperature in cases of w/ or w/o Pauli blocking (PB) at  0.4 $\rho_0$ and 1.0 $\rho_0$ and the nucleon-nucleon cross section is fixed at 40 mb~\cite{XGDeng2021}.}
\label{fig:fig4}
\end{figure}

In the last section, some approaches to extracting the shear viscosity of nuclear matter or a Fermi system are introduced. In the previous work~\cite{XGDeng2021}, comparisons have been done among methods of Chapman-Enskog (for the Fermi system), shear strain rate, and Greek-Kubo in a periodic box by using an improved quantum molecular dynamic (ImQMD) model without mean field as displayed in Fig.~\ref{fig:fig4}. It should be mentioned that `Boltzman' denotes the Chapman-Enskog method as the Eq.~(\ref{CEequation9}). The GK-nor and GK-new are the normal Green-Kubo formula as Eq.~(\ref{GKubo1}) and new form of the Green-Kubo formula as Eq.~(\ref{GKubo2}), respectively. 
As shown in Fig.~\ref{fig:fig4}, the Pauli blocking effect is studied in the calculation. It is worth to mention that for a Fermi system Pauli blocking should be taken into account. It can be seen that in Fig.~\ref{fig:fig4}(a) and Fig.~\ref{fig:fig4}(c) without the Pauli blocking, shear viscosity increases with temperature at both densities of $0.4\rho_{0}$ and $1.0\rho_{0}$. Seeing from Fig.~\ref{fig:fig4} (b) and Fig.~\ref{fig:fig4}(d) with the Pauli blocking, unlike Fig.~\ref{fig:fig4}(a) and Fig.~\ref{fig:fig4}(c) in the low-temperature region, shear viscosity increases with the decreasing of temperature. It is due to a stronger Pauli blocking effect in the low temperature region. At lower temperature and with stronger Pauli blocking effect, it indicates that a good Fermi sphere is formed and thus the elastic scattering rate is greatly suppressed due to the Pauli blocking. Hence, the system in lower temperature region has larger shear viscosity. However, an uncertainty in extracting the value of the shear viscosity at T = 0 is given here for the small or finite values of shear viscosity being given in the Refs.~\cite{Auerbach1975,Davies_1976_PRC,Davies_1977_PRC,Nix_1980_PRC,DinhDang2011}, such an uncertainty is explained by different dissipation mechanisms~\cite{DinhDang2011}. As temperature increases, the Pauli blocking effect becomes weaker and collision increases which makes the momentum transfer easier and shear viscosity decreasing. As the temperature increases again, even though the collision rate increases, velocity becomes dominant. Thus, shear viscosity increases with temperature. It is concluded that the naive kinetic approach is not valid for the description of the unitary Fermi gas at low temperature, where the system becomes strongly correlated~\cite{YK2016}. For the dilute gas case as black dot-line in Fig.~\ref{fig:fig3}, it increases with temperature since it is dominated by the term of temperature (or particle velocity). 

\begin{figure}
\setlength{\abovecaptionskip}{0pt}
\setlength{\belowcaptionskip}{8pt}
\centering\includegraphics[scale=1.0]{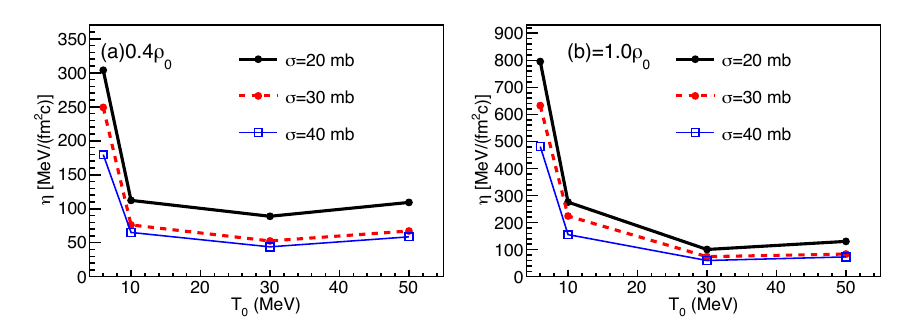}
\caption{(Color online) Shear viscosity as a function of temperature ($T_{0}$) at densities 0.4$\rho_0$ (a) and 1.0$\rho_0$ (b) for different cross sections with Green-Kubo formula as Eq.~(\ref{GKubo2}) in ImQMD model.}
\label{fig:fig5}
\end{figure}

Besides, one can find that, among these three approaches, shear viscosity determined by the Boltzmann type equation of Eq.~(\ref{CEequation9}) is lower than that calculated by the shear rate and the Green-Kubo method. At very low densities such as 0.05$\rho_0$, however, we found that the result by the Boltzmann type equation (Eq.~\ref{CEequation9}) is consistent with other methods, which indicates that a low density approximation for Eq.~\ref{CEequation9} could be valid. Then one would expect that shear viscosities determined by the shear rate and the Green-Kubo approaches are the same. In Ref.~\cite{XGDeng2021}, we re-derived and gave a new form of the Green-Kubo formula for the calculation of shear viscosity as Eq.~(\ref{GKubo2}). 
In the Boltzmann statistical analysis without Pauli blocking, we found that shear viscosities from the shear rate method, the normal form of the Green-Kubo formula, and new form of the Green-Kubo formula are close. As the Pauli blocking is taken into account as shown in Fig.~\ref{fig:fig4} (b) and Fig.~\ref{fig:fig4} (d), however, one sees that in the low temperature region, the shear viscosity from the normal form of the Green-Kubo formula (\ref{GKubo1}) is higher than those from both the shear strain rate method and the new form of the Green-Kubo formula. However, shear viscosities from the shear rate method and the new form of the Green-Kubo formula are almost the same. At high temperatures, shear viscosities from the shear rate method, the normal form of the Green-Kubo formula, and the new form of the Green-Kubo formula are to be the same since particle collisions become dominant and Pauli blocking can be ignored as temperature increases. So one should be careful when applying the normal form of the Green-Kubo formula to extract shear viscosity of Fermi system at low-temperature region.

\begin{figure}
\setlength{\abovecaptionskip}{0pt}
\setlength{\belowcaptionskip}{8pt}
\centering\includegraphics[scale=0.5]{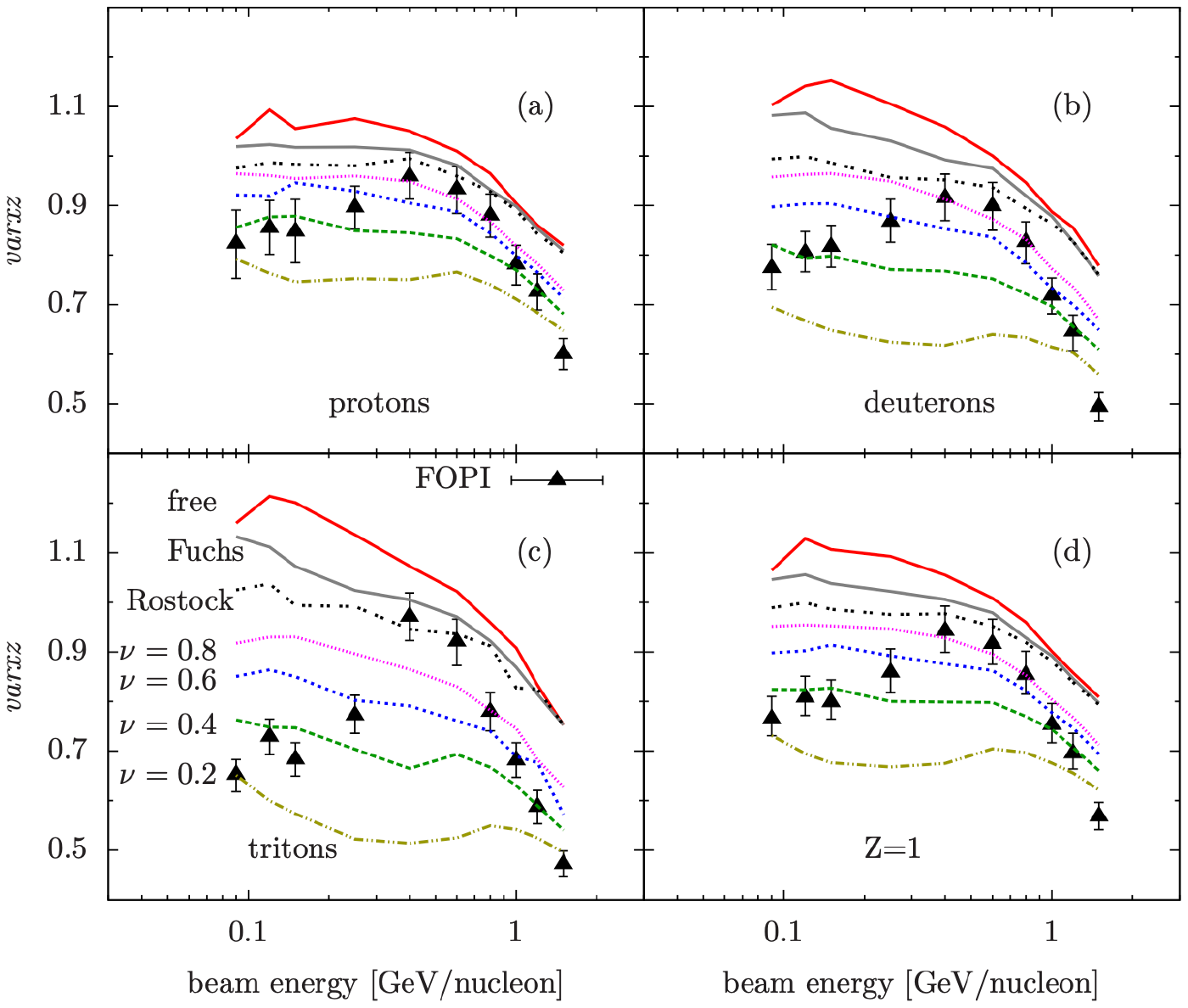}
\caption{(Color online) Stopping observable $varxz$ for protons (a), deuterons (b), tritons (c), and particles with Z=1 (d) in Au + Au collisions at different beam energies at bred $<$ 0.15. Lines show the effects of different in-medium reductions of the NN cross section~\cite{BBDPawel2019}. The “tempered” cross-section reductions are marked with their tunable parameter $\nu$. Symbols are experimental data from the FOPI Collaboration~\cite{REISDORF2010}.}
\label{fig:fig5-1}
\end{figure}

\begin{figure}
\setlength{\abovecaptionskip}{0pt}
\setlength{\belowcaptionskip}{8pt}
\centering\includegraphics[scale=0.5]{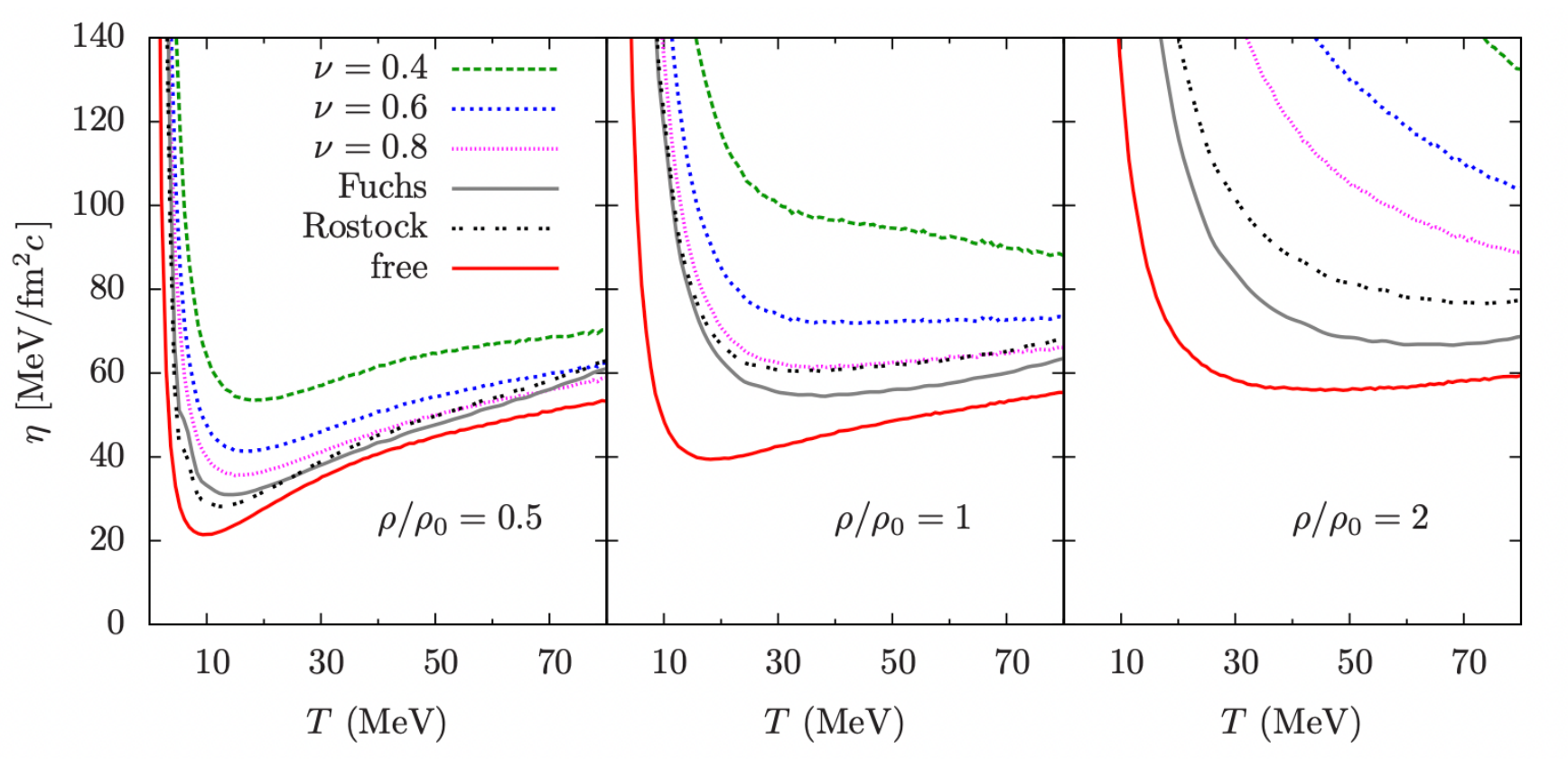}
\caption{(Color online) Shear viscosity for symmetric nuclear matter at different densities and temperatures, deduced from Boltzmann equation with in-medium cross sections~\cite{BBDPawel2019}.}
\label{fig:fig5-2}
\end{figure}

From both the Chapman-Enskog method and relaxation time approximation method, we can see that the collision has an important role for the shear viscosity. For the results in Fig.~\ref{fig:fig4}, the nucleon-nucleon cross section is fixed at 40 mb. When one increases the cross section, the shear viscosity would reduce, as shown in Fig.~\ref{fig:fig5}. It is obvious that as the cross-section increases, momentum transfer becomes quite easier, which would lead to lower shear viscosity. It can be seen that the shear viscosity is sensitive to the nucleon-nucleon cross section. Thus, one adjusts the in-medium nucleon-nucleon (NN) cross section in a transport model which is named pBUU~\cite{Danielewicz1991,Danielewicz1992,PanQB1993,Danielewicz2000} to reproduce the nuclear stopping data and use the fitted cross sections to calculate the shear viscosity self-consistently~\cite{BBDPawel2019}. The NN cross-section should also be affected by the surrounding medium. The authors discussed three kinds of media cross-sections with nuclear stopping. One is 
\begin{eqnarray}
\sigma_{NN}^{\rm med}= \sigma_{0} {\rm tanh} \Big{(} \frac{\sigma_{NN}^{\rm free}}{\sigma_{0}} \Big{)},
\label{NNcross-section-1}
\end{eqnarray}%
where $\sigma_{0} = \nu \rho^{-2/3}$, $\nu$ and $\rho$ are a parameter in order of 1 and medium of number density, respectively. And $\sigma_{NN}^{\rm free}$ is free NN cross section. A in-medium cross section carried out at the University of Rostock~\cite{Alm1994} within a thermodynamic T-matrix approach is given by a parametrization of the cross-section reduction~\cite{BBDPawel2019},
\begin{eqnarray}
\sigma_{NN}^{\rm med}= \sigma_{NN}^{\rm free}{\rm exp}  \Big{(} -0.6 \frac{\rho/\rho_{0}}{1+[T_{c.m.}/150 \, {\rm MeV}]^{2}} \Big{)},
\label{NNcross-section-2-0}
\end{eqnarray}%
where $T_{c.m.}$ is the total kinetic energy of the two interacting particles in the frame where the local medium is at rest. The last one is derived by Fuchs et al. ~\cite{Fuchs2001} which is parameterized with,
\begin{eqnarray}
\sigma_{nn}^{\rm med}= \sigma_{nn}^{\rm free}{\rm exp}  \Big{(} -1.7 \frac{\rho/\rho_{0}}{1+[T_{c.m.}/12 \, {\rm MeV}]^{3/2}} \Big{)}, 
\label{NNcross-section-3} \\
\sigma_{np}^{\rm med}= \sigma_{np}^{\rm free}{\rm exp}  \Big{(} -1.4 \frac{\rho/\rho_{0}}{1+[T_{c.m.}/33 \, {\rm MeV}]} \Big{)}.  \quad\,\,\,
\label{NNcross-section-4}
\end{eqnarray}%
 With the help of the different cross-section reductions above, the authors can compare predictions from pBUU to stopping data, as shown in Fig.~\ref{fig:fig5-1}. The stopping observable $varxz$ is defined as~\cite{REISDORF2007},
 \begin{eqnarray}
varxz= \frac{\triangle y_{x}}{\triangle y_{z}},
\label{NNcross-section-2-1}
\end{eqnarray}%
where $\triangle y_{x}$ is the variance of particle rapidity along a randomly chosen direction that is transverse to the beam and $\triangle y_{z}$ is the variance of the standard particle longitudinal rapidity. From Fig.~\ref{fig:fig5-1}, it is seen that around the peak region in $varxz$ versus energy with the Rostock cross section or the tempered cross section with $\nu$ = 0.8, the results from pBUU can describe the data well. Then, by combining the constrained in-media cross-sections $\sigma_{NN}^{\rm med}$ and the Eq.~(\ref{CEequation9}), one can obtain shear viscosity as shown in Fig.~\ref{fig:fig5-2}. In such a way, the shear viscosity is hopefully of greater generality than even the transport model itself~\cite{BBDPawel2019}. Shear viscosity versus temperature is sensitive to different kinds of in-media cross-sections $\sigma_{NN}^{\rm med}$. But it can be seen that a uniform factor could not describe the data for the whole energy region, which indicates that the factor for in-media cross-sections should be energy dependence. Here, the behavior of shear viscosity versus temperature is no more different from that in Fig.~\ref{fig:fig3}. In Ref.~\cite{BBDPawel2019}, it only considers the elastic collision cross section. Like comments in Ref.~\cite{BBDPawel2019}, however, as collision energy or temperature increases to a high value, meson effects and modifications of inelastic processes need to be explored in the simulations.

\begin{figure}
\setlength{\abovecaptionskip}{0pt}
\setlength{\belowcaptionskip}{8pt}
\centering\includegraphics[scale=0.8]{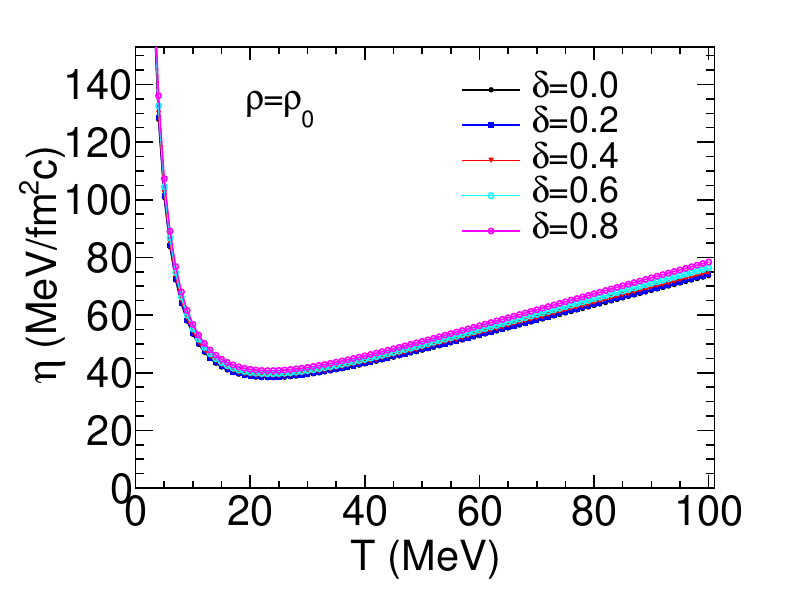}
\caption{(Color online) Shear viscosity as a function of temperature at different isospin asymmetries.}
\label{fig:fig6}
\end{figure}

\begin{figure}
\setlength{\abovecaptionskip}{0pt}
\setlength{\belowcaptionskip}{8pt}
\centering\includegraphics[scale=0.6]{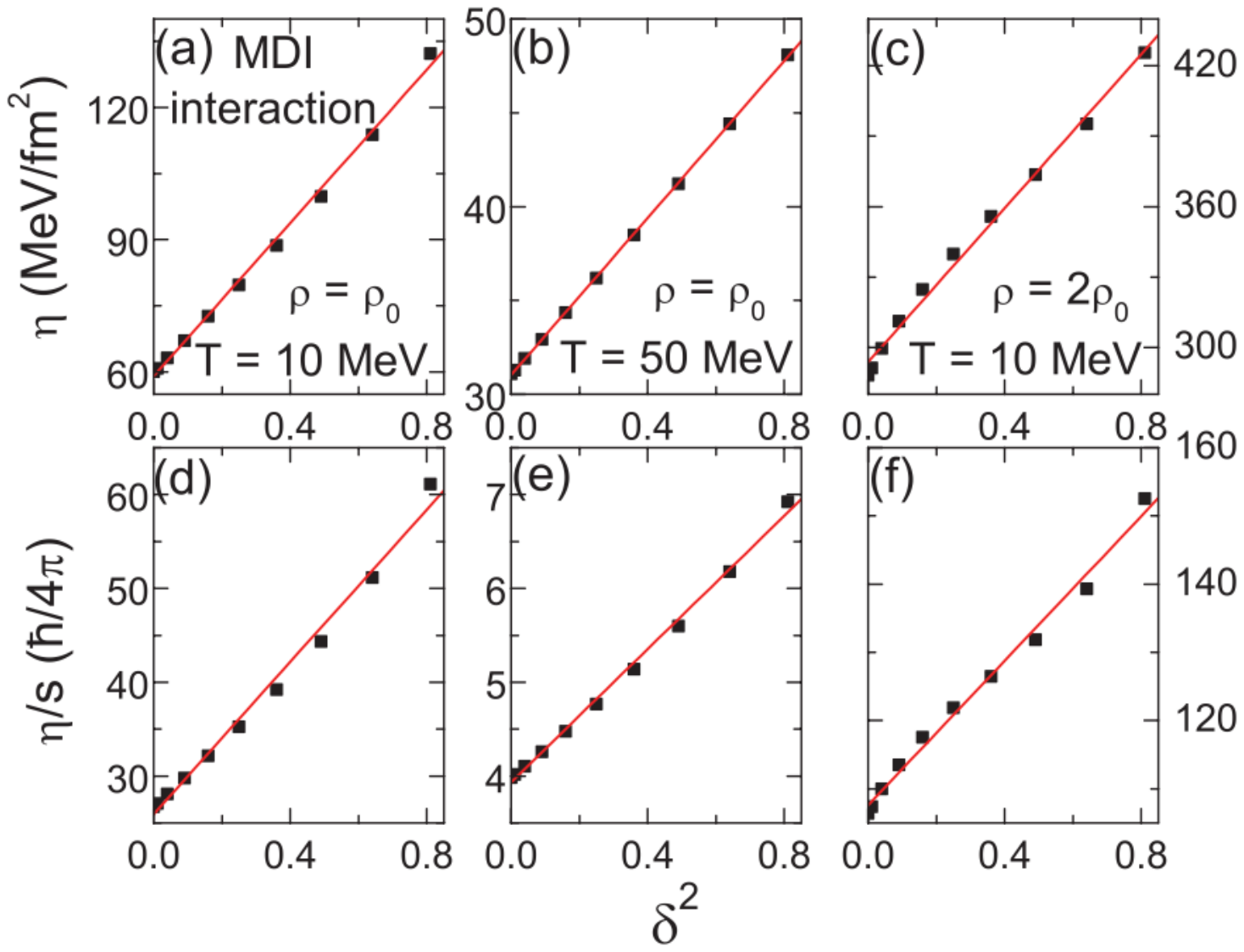}
\caption{(Color online) Shear viscosity $\eta$ and specific viscosity $\eta/s$ (s is entropy density) as functions of isospin asymmetry $\delta^{2}$ for different densities and temperatures. The solid lines are from the linear fit~\cite{Xujun2011}.}
\label{fig:fig7}
\end{figure}

In a more general sense, Shi and Danielewicz~\cite{SLDPawel2003} solved the set of Boltzmann equations for a binary system of fermions to find thermodynamic fluxes driven by specific thermodynamic forces and to find general expressions for the shear viscosity and other transport coefficients. The transport coefficients are calculated numerically for nuclear matter using experimental nucleon-nucleon cross sections. 
The numerical results for the coefficients have been obtained,
\begin{eqnarray}
\eta&=&\Big{(}1+0.10\delta^{2}\Big{)}\Big{[}\frac{856}{T^{1.10}}(\frac{\rho}{\rho_0})^{1.81}-\frac{240.9}{T^{0.95}}(\frac{\rho}{\rho_0})^{2.12} +2.154T^{0.75}\Big{]},                                                                    \label{coefficientA} \\[3mm]
D_{I}&=&\Big{(}1-0.19\delta^{2}\Big{)}\Big{[}\frac{11.34}{T^{2.38}}(\frac{\rho}{\rho_0})^{1.54}+\frac{1.746}{T}(\frac{\rho}{\rho_0})^{0.56} +0.00585T^{0.913}(\frac{\rho}{\rho_0})\Big{]},                                             \label{coefficientB}\\[3mm]
\kappa&=&\Big{(}1+0.10\delta^{2}\Big{)}\Big{[}\frac{0.235}{T^{0.755}}(\frac{\rho}{\rho_0})^{0.951}-0.0582(\frac{\rho}{\rho_0})^{0.0816}+0.0238T^{0.5627}(\frac{\rho}{\rho_0})^{0.0171}\Big{]}.                                    
\label{coefficientC}
\end{eqnarray}%
where $\rm T$ is temperature in MeV, $\eta$ is shear viscosity in MeV/(fm$^{2}$c), $D_{I}$ is isospin diffusivity in fm$\cdot$c, heat conductivity $\kappa$ is in c/fm$^{2}$; and $\delta=\frac{\rho_{n}-\rho_{p}}{\rho}$ is isospin asymmetry. From these Eqs. (\ref{coefficientA})$-$(\ref{coefficientC}), it is found that the shear viscosity diffusion coefficient and heat-conduction coefficient weakly depend on isospin asymmetry. One can also see from Fig.~\ref{fig:fig6} in which shear viscosity as a function of temperature at different isospin asymmetries is displayed. 

On the other hand, one derives the shear viscosity of neutron-rich nuclear matter from the relaxation time approximation approach using an isospin- and momentum-dependent interaction and the nucleon-nucleon cross sections taken as those from the experimental data modified by the in-medium effective masses as used in the isospin-dependent Boltzmann-Uehling-Uhlenbeck (IBUU) transport model calculations~\cite{Xujun2011}. The expression for shear viscosity is shown by Eq.~(\ref{Rtimeequation2}).

As shown in Fig.~\ref{fig:fig7}, shear viscosity $\eta$ and specific viscosity $\eta/s$ (s is entropy density) as functions of isospin asymmetry $\delta^{2}$ for different densities and temperatures are displayed. Here, the entropy density $s$ can be obtained by~\cite{Xujun2011},
\begin{eqnarray}
s=-\sum_{\tau} \frac{{\rm g}}{(2\pi\hbar)^{3}} \int \Big{[}f_{\tau}{\rm \ln} f_{\tau} +(1-f_{\tau}){\rm \ln}(1-f_{\tau}) \Big{]} d^{3}p.
\label{Rtimeequation6}
\end{eqnarray}%
With $\delta^{2}=0$ for the symmetry nuclear matter in Fig.~\ref{fig:fig7} (a), a simple comparison is shown for shear viscosity from the Boltzmann type equation of Eq.~(\ref{CEequation9}), shear rate, and the Green-Kubo methods. We can see that the shear viscosity by relaxation time approximation approach is close to the one by Boltzmann type equation. However, it is less than those from the shear rate and the Green-Kubo methods. And $\eta$ shows parabolic approximation relation to isospin asymmetry $\delta^{2}$. It presents that isospin asymmetry has a large effect on shear viscosity, which could be different from the results obtained by solving the set of Boltzmann equations as mentioned above. The reason here could be due to the isospin interaction used in the calculations. The specific viscosity $\eta/s$  which is a hot agenda and relates to a bound value of $\hbar/4\pi$ is shown in Fig.~\ref{fig:fig7}. It can be seen that specific viscosity $\eta/s$ also shows parabolic approximation relation to isospin asymmetry $\delta^{2}$. From Fig.~\ref{fig:fig7} (d), (e), and (f), we can see that at T= 10 MeV and $\rho=\rho_{0}$, $\eta/s$ is larger than 25 times of  $\hbar/4\pi$. When T=50 MeV, the specific viscosity is about 4$\sim$5 $\hbar/4\pi$, which is already close to KSS bound. At a higher density at $\rho=2\rho_{0}$, specific viscosity becomes larger with the increase in shear viscosity. 

\begin{figure}[htb]
\setlength{\abovecaptionskip}{0pt}
\setlength{\belowcaptionskip}{8pt}
\centering\includegraphics[scale=0.4]{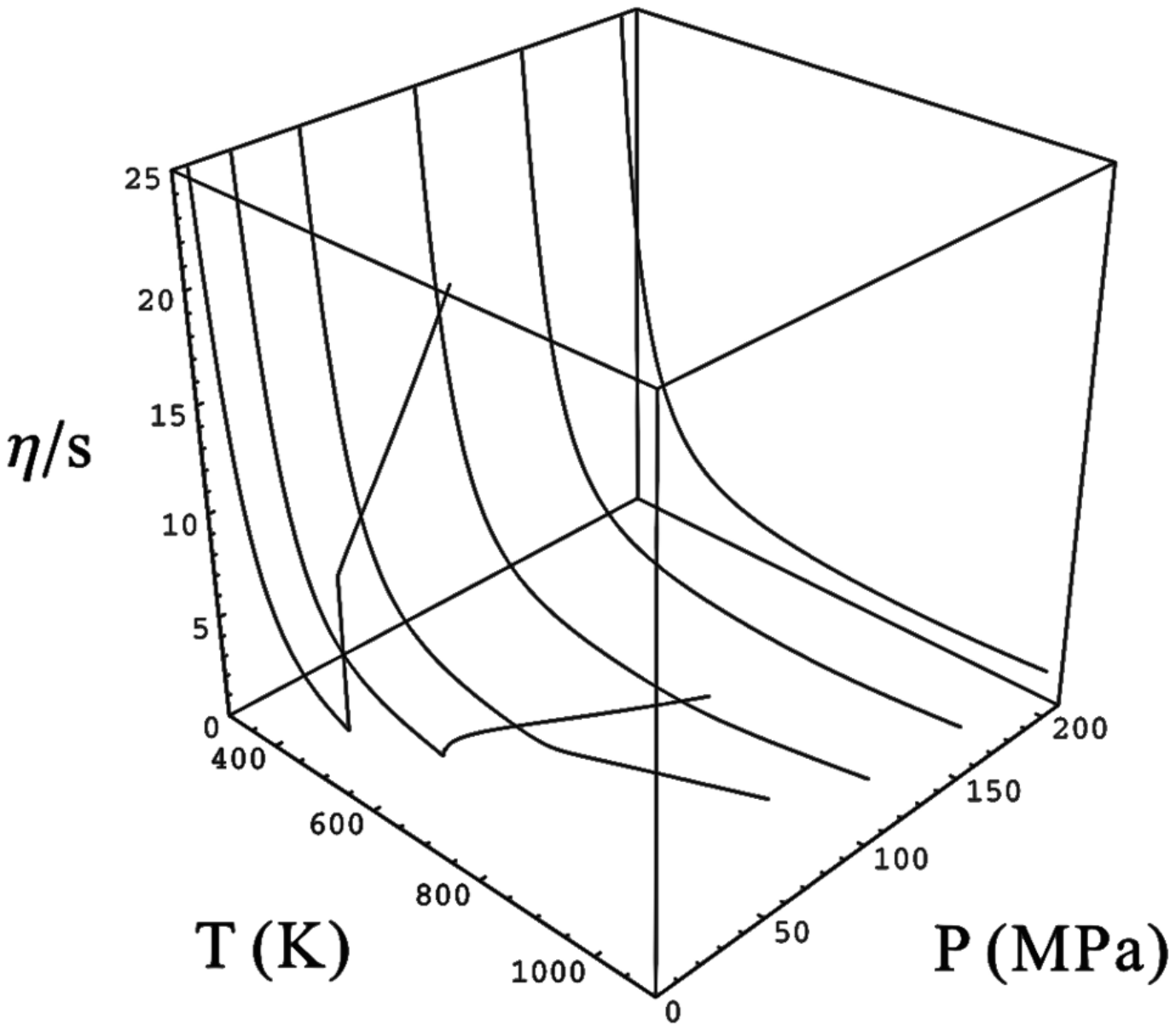}
\caption{(Color online) Specific shear viscosity $\eta/s$ of water as a function of temperature in different pressures~\cite{ChenJW2007}.}
\label{fig:fig7-1}
\end{figure}

For common matter like water, as shown in Fig.~\ref{fig:fig7-1}, it can be seen that below the critical pressure of 22.06 MPa, $\eta/s$ shows a discontinuity at the minimum. Above the critical pressure, $\eta/s$ becomes smooth~\cite{ChenJW2007}. In Ref.~\cite{ChenJW2007}, from the results computed for the shear viscosity of QCD in the hadronic phase by the coupled Boltzmann equations of pions and nucleons in low temperatures and low baryon-number densities and combined common matter such as Ar, CO, CO$_{2}$, H, He, H$_{2}$S, Kr, N, NH$_{3}$, Ne, O, and Xe, the authors suspect that the general feature for a first-order phase transition is that $\eta/s$ has a discontinuity in the bottom of the $\eta/s$ valley. And the valley gradually disappears when the phase transition turns into a crossover. Near the critical point, there is a smooth valley in the crossover, which is similar to the confinement-deconfinement crossover of QCD at $\mu=$ 0. And as the pressure is far away from the critical point, the valley shape will disappear. The authors argue that these are general features of first-order phase transitions which indicates that specific viscosity $\eta/s$ may be used as a probe to study the phenomenon of liquid$-$gas phase transition. Also, in Fig.~\ref{fig:fig7-1}, the $\eta/s$ of water shows a positive jump across the phase transition line from the gas to liquid phase. The sign of the $\eta/s$ jump of water for first-order phase transitions is similar to the materials including Ar, CO$_{2}$, H, He, H$_{2}$S, Kr, N, NH$_{3}$, Ne, O, and Xe. But it might not be universal, as the jump for QCD in the limit of a large number of colors is negative from the low to high temperature phases \cite{Csernai2006} and the behavior of nucleonic matter~\cite{Xujun2013}.

The specific viscosity ($\eta/s$) of neutron-rich nucleonic matter near its liquid$-$gas phase transition was studied with the relaxation time approximation approach~\cite{Xujun2013}. In this framework, one treats the phase coexistence region with a volume fraction $\xi$, and the average number and entropy densities are given by,
\begin{eqnarray}
\rho&=&\xi\rho_{l}+(1-\xi)\rho_{g},                                                                    
\label{InfinitEq-1} \\[3mm]
s&=&{\xi}s_{l}+(1-\xi)s_{g}.                                    
\label{InfinitEq-2}
\end{eqnarray}%
where $\rho_{l(g)}$ and $s_{l(g)}$ are the number and entropy densities of the liquid (gas) phase, respectively. To construct a phase coexistence state for nuclear matter, the Gibbs conditions are used. For phase coexistence, it is governed by the Gibbs conditions. For asymmetric nuclear matter, two-phase coexistence equations are~\cite{Xujun2007},
\begin{eqnarray}
\mu_{i}^{l}(T,\rho^{l}_{i})&=&\mu_{i}^{g}(T,\rho^{g}_{i}) \,\,\,\, (i= n \,{\rm and} \,p) ,                                                       
\label{InfinitEq-2-1} \\[3mm]
P_{i}^{l}(T,\rho^{l}_{i})&=&P_{i}^{g}(T,\rho^{g}_{i}) \,\,\,\, (i= n \,{\rm or} \,p) ,                                   
\label{InfinitEq-2-2}
\end{eqnarray}%
where P is the pressure. To determine shear viscosity, one considers a stationary flow field in the $z$ direction and the shear force on the particles for a single phase of gas or liquid~\cite{Xujun2011,Xujun2013},
\begin{eqnarray}
F_{i}=\sum\langle (p_{z}-m_{\tau}u_{z})\rho_{\tau}v_{x} \rangle_{i},                           
\label{InfinitEq-3}
\end{eqnarray}%
where $\tau$ refers to a neutron or proton, $i = l$ is for the liquid phase and $g$ is for the gas phase, and $m_{\tau}$ is the nucleon mass. And $\langle\cdots\rangle$ denotes the thermal average of the product of the flux $\rho_{\tau}v_{x}-$ in the $x$ direction and the momentum transfer $p_{z}-m_{\tau}u_{z}$ in the $z$ direction. The shear viscosity is determined by,
\begin{eqnarray}
F_{l(g)}=-\eta_{l(g)}\frac{\partial u_{z}}{\partial x}.                           
\label{InfinitEq-4}
\end{eqnarray}%
In the phase coexistence region, one can imagine gas bubbles in a liquid or liquid droplets in a gas. Thus, finally, the average shear viscosity of the mixed phase can be expressed as,
\begin{eqnarray}
\eta&=&{\xi}\eta_{l}+(1-\xi)\eta_{g}.                                    
\label{InfinitEq-5}
\end{eqnarray}%
Here, $\eta_{l}$ and $\eta_{g}$ can be separately calculated owing to the density is uniform in each phase. And the strategy is as in Ref.\cite{Xujun2011}. 

\begin{figure}[htb]
\setlength{\abovecaptionskip}{0pt}
\setlength{\belowcaptionskip}{8pt}
\centering\includegraphics[scale=0.22]{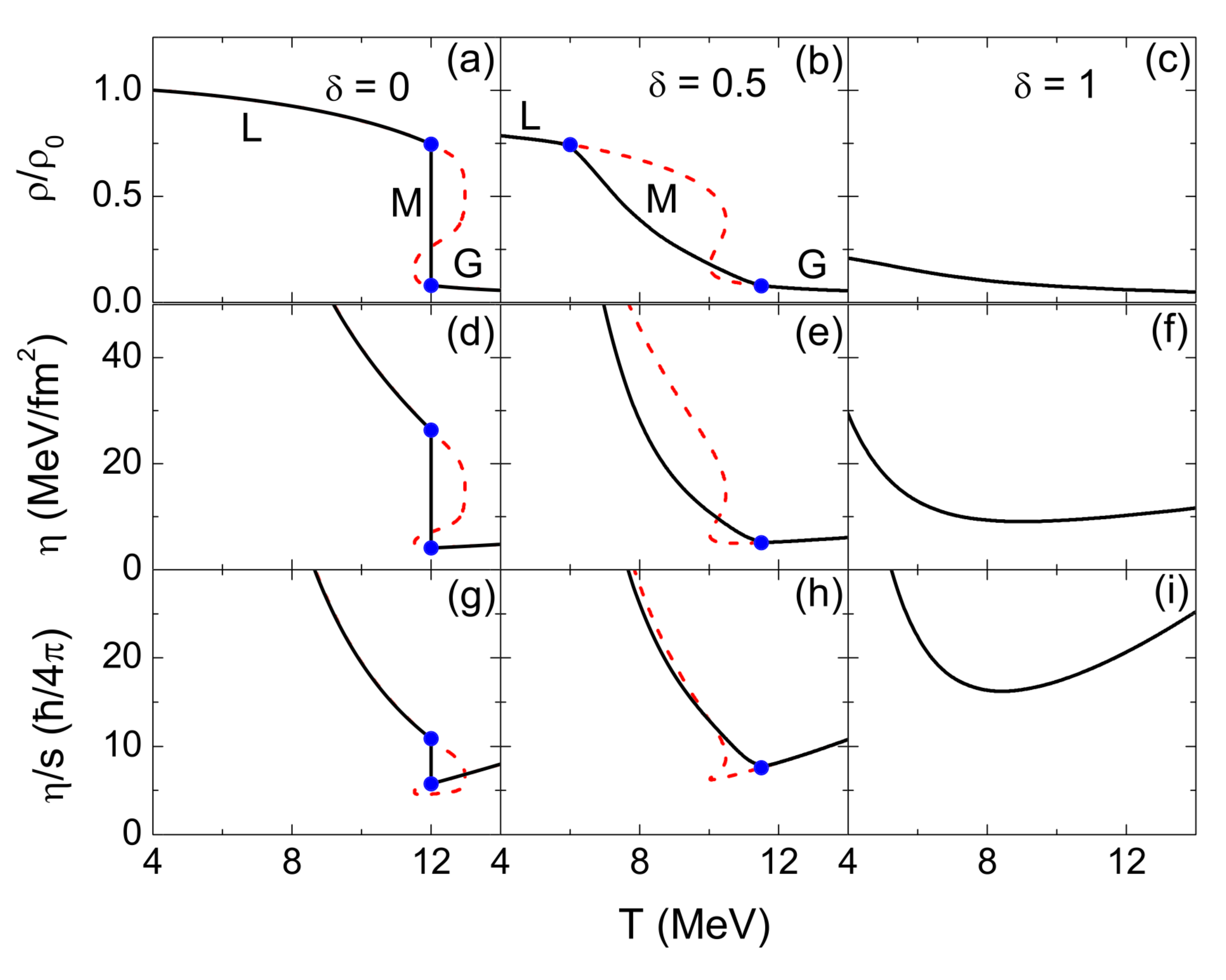}
\caption{(Color online) Density (top panels), shear viscosity (middle panels), and specific shear viscosity (bottom panels) as functions of  temperature at the fixed pressure of P = 0.1 MeV/fm$^{3}$ for isospin symmetric matter ($\delta$ = 0) (left column), neutron-rich matter ($\delta$ = 0.5) (middle column), and pure neutron matter ($\delta$ = 1) (right column) with the stiffer symmetry energy x = 0. Solid lines are results including the liquid–gas phase transition with `L', `M', and `G' denoting the liquid phase, the mixed phase, and the gas phase, respectively. Dashed lines are results obtained by assuming the liquid$-$gas phase transition does not happen inside the binodal surface~\cite{Xujun2013}.}
\label{fig:fig8}
\end{figure}

In Fig.\ref{fig:fig8}, density (top panels), shear viscosity (middle panels), and specific shear viscosity (bottom panels) as functions of temperature at the fixed pressure of P = 0.1 MeV/fm$^{3}$ for different isospin symmetries with the stiffer symmetry energy ${\rm x}$= 0 are shown. For a given pressure, the shear viscosity of the liquid phase is greater than that of the gas phase. And with $\delta$ = 0, there is a sudden jump for density, shear viscosity, and specific shear viscosity at a temperature T$\sim$12 MeV, which is due to liquid$-$gas phase transition of first-order of symmetric nucleonic matter~\cite{Xujun2013}. The behavior here for shear viscosity versus temperature is similar to that of $^{3}He$~\cite{Alvesalo1975}. From Fig.\ref{fig:fig8}, it can be seen that the minimum of the specific shear viscosity is exactly located at the critical temperature when a first-order phase transition happens, while it is located at the boundary of the gas phase if the phase transition is a second-order~\cite{Xujun2013}. In a Fermi system, as we know, there is competition between the Pauli blocking and collision among particles with changes of temperature. So even for pure neutron matter without a liquid-gas phase transition, it is also seen that there is still a valley-shape around phase transition region. In Fig.\ref{fig:fig8} (a), (b), (c), the jumps show a negative sign and differ from water as in Fig.~\ref{fig:fig7-1}.

\begin{figure}[htb]
\setlength{\abovecaptionskip}{0pt}
\setlength{\belowcaptionskip}{8pt}
\centering\includegraphics[scale=0.5]{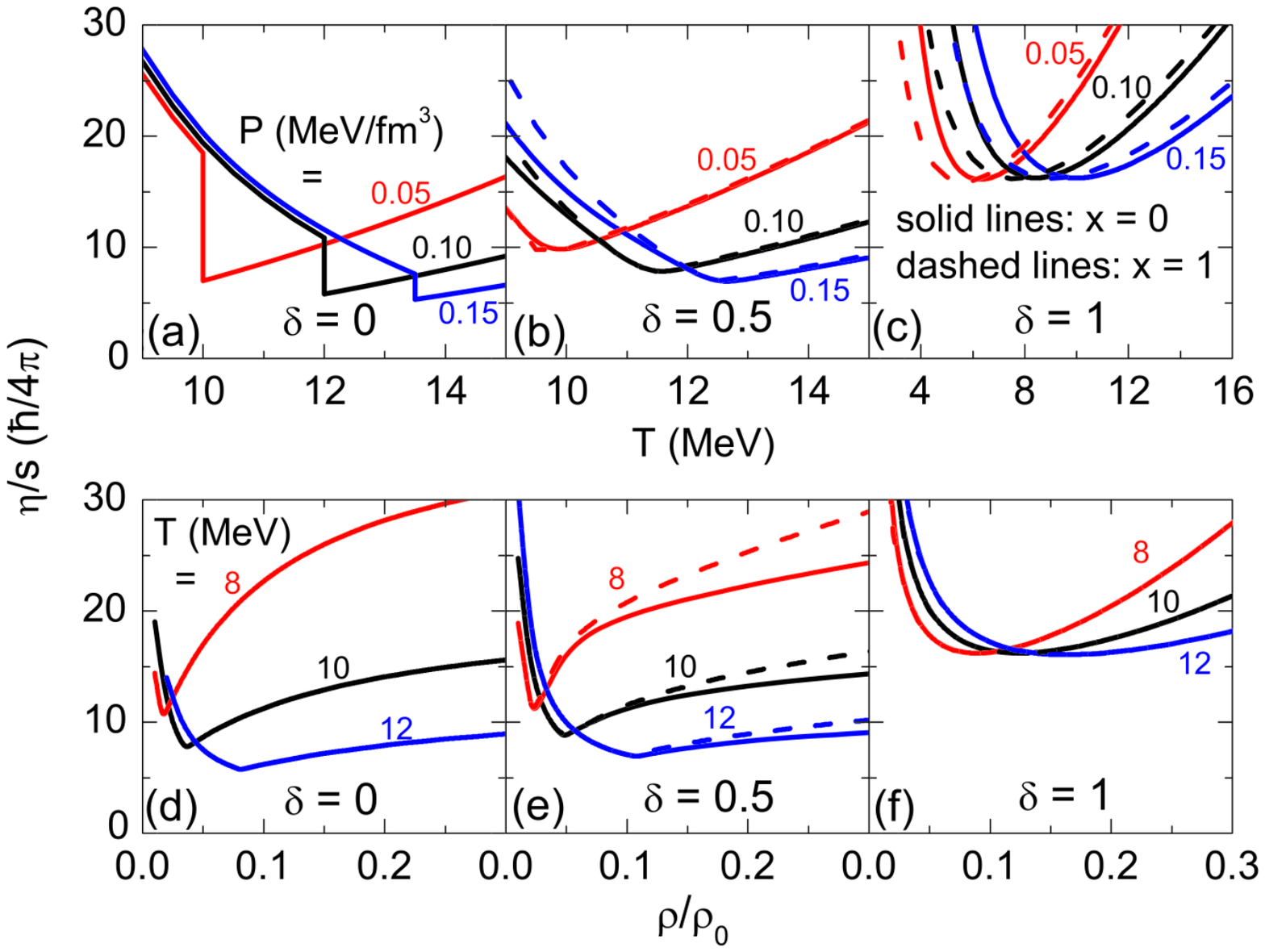}
\caption{(Color online) The specific shear viscosity as functions of  temperature (upper panels) and density (lower panels) at different fixed pressures and temperatures, respectively, in isospin symmetric matter ($\delta$ = 0), neutron-rich matter ($\delta$= 0.5), and pure neutron matter ($\delta$= 1) for both symmetry energies {\rm x}  = 0 and {\rm x} = 1~\cite{Xujun2013}.}
\label{fig:fig9}
\end{figure}

Furthermore, one can get more results for specific shear viscosity as functions of temperature and density from Ref. ~\cite{Xujun2013}, as shown in Fig.\ref{fig:fig9}. The results show that the specific shear viscosity of nucleonic matter is about 4$-$5 $\hbar/4\pi$ for isospin-symmetric nucleonic matter and is generally smaller than that in neutron-rich nucleonic matter. It is obvious that there are valley shapes for symmetric nuclear matter and pure neutron matter. The symmetry energy effect is investigated as the solid and dashed lines display. In the phase coexistence region, the softer (x = 1) symmetry energy would give a larger specific shear viscosity than that with the stiffer (x = 0) symmetry energy for the liquid-like region, as in Fig.\ref{fig:fig9} (b) and Fig.\ref{fig:fig9} (e). And for pure neutron matter without the liquid-gas phase transition under a fixed pressure, specific shear viscosity for x = 1 is similar to that for x = 0, as in Fig.\ref{fig:fig9}(c). At fixed temperature, specific shear viscosities from different symmetry energies are the same for pure neutron matter, as shown in Fig.\ref{fig:fig9} (f).

\begin{figure}
\setlength{\abovecaptionskip}{0pt}
\setlength{\belowcaptionskip}{8pt}
\centering\includegraphics[scale=1.2]{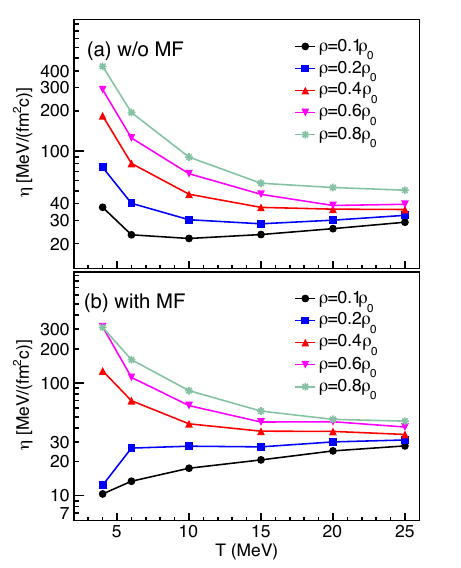}
\caption{(Color online) Shear viscosity as a function of temperature at different densities in the calculations without (a) and with (b) mean field.~\cite{XGDeng2022}.}
\label{fig:fig10}
\end{figure}
\begin{figure}
\setlength{\abovecaptionskip}{0pt}
\setlength{\belowcaptionskip}{8pt}
\centering\includegraphics[scale=1.2]{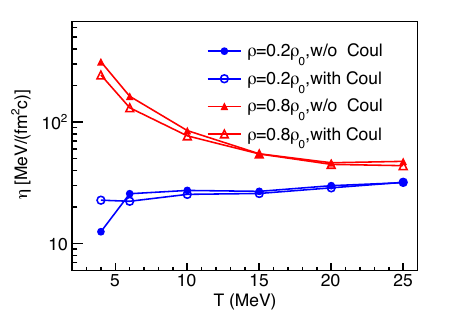}
\caption{(Color online) Shear viscosity as a function of temperature in the calculations without and with the Coulomb interaction~\cite{XGDeng2022}.}
\label{fig:fig11}
\end{figure}

From the above discussions, one can get that Pauli blocking, intermediate cross-section, the asymmetry of nucleonic matters have an effect on the shear viscosity. Both of the Pauli blocking and inter-mediate cross-section affect nucleon-nucleon collisions and then the shear viscosity. A system with different asymmetries of nucleonic matters would have different nucleon-nucleon potentials and different nucleon-nucleon cross-sections which would make shear viscosity different. However, one can notice that these calculations are with uniform nucleonic matter. There are no fragment effects considered in those calculations. The nuclear matter in the gas phase is different from the gas of classical fluids (as the water vapour), which is composed of not only single nucleons, neutrons, and protons, but also complex particles like alpha-particles and light fragments depending on the temperature conditions~\cite{Bugaev_2001,Borderie_2019}. In this case, one extracts the shear viscosity coefficient by using the shear strain rate method and setting up a periodic cubic box within the improved quantum molecular dynamic (ImQMD) model~\cite{XGDeng2022}. In the framework of ImQMD, potential energy density~\cite{ZhangYX2006,WangN2016} without the spin-orbit term can be written as,
\begin{equation}
\begin{split}
V_{loc}&= \frac{\alpha}{2}\frac{\rho^{2}}{\rho_{0}} + \frac{\beta}{\gamma+1}\frac{\rho^{\gamma+1}}{\rho_{0}^{\gamma}} +\frac{g_{sur}}{2\rho_{0}} \, (\bigtriangledown\rho)^{2}     \\
           &+g_{\tau} \frac{\rho^{\eta +1}}{\rho_{0}^{\eta}} +\frac{g_{sur,iso}}{\rho_{0}}\,[\nabla(\rho_{n}-\rho_{p})]^2+\frac{C_s}{2\rho_{0}}\rho^{2} \, \delta^{2} \,,
\end{split}                              
\label{QMDpotential}
\end{equation}
where $\rho = \rho_\text{n} + \rho_\text{p}$ is the net nucleon density, $\rho_\text{n}$ and $\rho_\text{p}$ are the proton and neutron densities, respectively, and $\delta = (\rho_{n}-\rho_{p})/\rho$ is the normalized asymmetry of neutron and proton. The Skyrme density functional with interactions such as Eq.~\eqref{QMDpotential}, as well as other approaches for nuclear systems, yield the liquid-gas phase transition at the subnormal densities for the near-symmetric nuclear matter. By switching on and off the mean field and thus inducing the phase transition, it can be able to observe the impact of clumping on viscosity in the phase transition region. The calculations are only for the isospin-symmetric nucleonic matters with the nucleon-nucleon cross section fixed at 40 mb. 

In the case without mean field and in cascade mode, there is no phase transition and fragment only when mean field is taken into account. However, when the mean field is turned on, the nucleon potential and kinetic energy will change due to the fragment formation which is one of the challenges for simulations. To avoid the system being heated up, it needs to adjust this transient state temperature to the initial temperature by introducing the cooling factor `h', which can be nominally positive and less than `1'. 

\begin{figure}
\setlength{\abovecaptionskip}{0pt}
\setlength{\belowcaptionskip}{8pt}
\centering\includegraphics[scale=1.2]{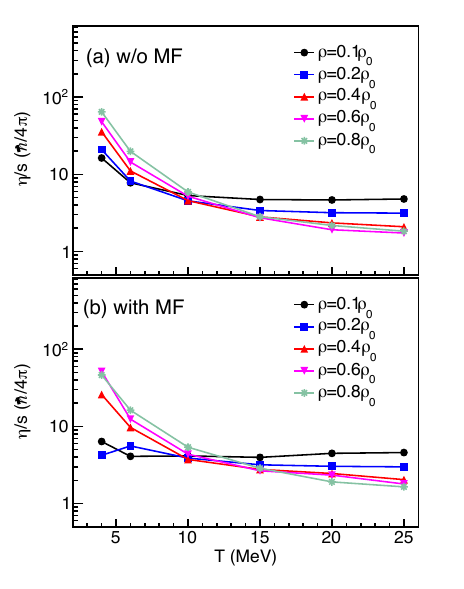}
\caption{(Color online) The specify shear viscosity over entropy density $\eta/s$ as a function of temperature at different densities in the calculations without (a) and with (b) mean field~\cite{XGDeng2022}.}
\label{fig:fig12}
\end{figure}

Without mean field and in cascade mode, the viscosity is that of a gas of nucleons, as shown in Fig.\ref{fig:fig10} (a) and it is similar to that in Fig.\ref{fig:fig4} (b) and (d). From the results with mean field in Ref.~\cite{XGDeng2022} as shown in Fig.\ref{fig:fig10} (b), it is seen that the shear viscosities are similar to those without mean field at high temperatures and densities. However, at densities below $0.4 \, \rho_0$ and temperatures less than $10 \, \text{MeV}$, it is found that the shear viscosity has a strong reduction when the mean field is switched on. The reasons could be understood in the following way: in the region of the phase transition, separated fragments, which move slowly, absorb most of the mass for the box system and stall the transport of momentum. Since those fragments are moving slowly, they play only a passive role in transporting the momentum. In the phase coexistence state, except for liquid-like fragments, there are gas phase nucleons in the system, which could become the primarily transported part. These gas phase nucleons rather collide with the fragments than with each other, which shortens their mean free path. In addition, with most mass in the fragments, the role of the Pauli principle in the gas is reduced which decreases the path further. Thus, in these cases, shear viscosity would be reduced. As density increases, the fragments would connect to each other, and the low-density gas phase is contained in voids in the liquid. Then momentum transport can progress, and the Pauli principle can be restored. 

Moreover, when the fragments form, the long-range Coulomb interaction may play a role for shear viscosity~\cite{XGDeng2022}. It is worth to mentioning that in most of the calculations for the infinite and symmetric nuclear-matter systems, the Coulomb interactions are switched off, and the near-symmetric matter has comparable numbers of neutrons and protons. In Ref. \cite{XGDeng2022}, the effect of Coulomb interaction on the shear viscosity in coexistence nucleonic matter has been investigated. From the results with and without Coulomb interaction, as shown in Fig.~\ref{fig:fig11}, it is found that shear viscosity is reduced by the Coulomb interaction in the region of low temperature. For an understanding of this phenomena, one can imagine that two fluid layers pass from one to another. However, the repulsive Coulomb interaction among protons will reduce momentum transformation or momentum flux between two layers, which will decrease shear viscosity. In low-temperature region, the Coulomb interaction could be important in comparison with thermal motion. With the increasing of temperature, the Coulomb interaction effect becomes smaller since nuclear matter at higher temperature is more uniform and the effect of Coulomb interaction becomes negligible. Based on the above arguments, we can explain the temperature dependence of Coulomb correction on shear viscosity.

\renewcommand\arraystretch{1.8}
\begin{table}[htp]
\setlength{\belowcaptionskip}{0.2cm}
\centering  
\caption{$\eta/s$ of different atomic and molecular systems:  Helium, Nitrogen and H$_{2}$O in certain ranges of  temperature and pressure \cite{Csernai2006,XGDeng2022}. } 
\label{ETAOS1}
\begin{tabular}{p{4cm}|p{3.0cm}<{\centering}|p{3.0cm}<{\centering}|p{3.0cm}<{\centering}}
\hline
Systems [Ref.]  &  T (K)    & Pressure (MPa) & $\eta/s~(\hbar/4\pi)$  \\ %
\Xhline{1pt}
Helium \cite{Csernai2006}  &  2$-$20 & 0.1$-$1 & 8.8$-$126  \\ %

Nitrogen \cite{Csernai2006}  & 50$-$600& 0.1$-$10 & 11.9$-$7000  \\ %

H$_{2}$O \cite{Csernai2006}  &  300$-$1200 & 10$-$100 & 25.5$-$377  \\ %
\hline
\end{tabular}
\end{table}

\renewcommand\arraystretch{1.8}
\begin{table}[htp]
\setlength{\belowcaptionskip}{0.2cm}
\centering  
\caption{$\eta/s$ of different quark matter (QGP) and finite nucleonic  matter (FNM) as well as infinite nucleonic matter (INM)  in certain ranges of  temperature and density. Here T$_{c}$$\approx$ 170 GeV is transition temperature from the hadronic phase to QGP phase \cite{DeyJ2021,XGDeng2022}. } \label{ETAOS2}
\begin{tabular}{p{4cm}|p{3.0cm}<{\centering}|p{3.0cm}<{\centering}|p{3.0cm}<{\centering}}
\hline
Systems [Refs.]  &  T (MeV)  &  $\rho$/$\rho_{0}$  &  $\eta/s~(\hbar/4\pi)$  \\ %
\Xhline{1pt}
QGP1~\cite{DeyJ2021,SASAKI2010,Marty2013,Deb2016}  &  $\leqslant$$\rm T_{c}$ & --- & 3.77$-$25.1    \\ %

QGP2 \cite{DeyJ2021,SASAKI2010,Marty2013,Deb2016,Dias2012} &  $>$$\rm T_{c}$ & --- & 1.00$-$6.91    \\ %

FNM1 \cite{Guo2017,Mondal2017}  &  0.8$-$2.1  & $\sim$1.0 &   2.5$-$6.5  \\ 

FNM2 \cite{Fang2014} &  3.5$-$16  & 0.9$-$1.25 &   3.0$-$70.0  \\ %

FNM3 \cite{Zhou2013}            &  6$-$16  & 0.2$-$0.3 &   9.5$-$20.0  \\

INM1 \cite{Xujun2013}      &  8$-$14 &  0.01$-$0.3 & 4$-$30 \\ %

INM2 \cite{XGDeng2022} &  4$-$25 & 0.1$-$0.8 & 2$-$55   \\ %
\hline
\end{tabular}
\end{table}

For the simulation system investigated in Ref.~\cite{XGDeng2022}, it is easy to extract the entropy density by Eq.(\ref{Rtimeequation6}). Then one can obtain the specific shear viscosity $\eta/s$ which is an interesting topic for a bound value of $\hbar/4\pi$, known as the Kovtun-Son-Starinets (KSS) bound \cite{KSS2005} as mentioned above. To briefly summarize the values of ratios of shear viscosity over entropy density for different systems, including atomic and molecular systems, QGP matter, finite nucleonic matter (FNM), as well as infinite nuclear matter (INM), are summarized in Tables \ref{ETAOS1} and \ref{ETAOS2} (The specific shear viscosities for QGP matter and finite nucleonic matter are discussed in later sections). It is worth to mentioning that the FNM is matter in a system with finite size like a single nucleus or a heavy-ion collision system. The hot quark gluon plasma (QGP) has a very low $\eta/s$ which is very close to $\hbar/4\pi$ and behaves as a nearly perfect fluid~\cite{TSDT2009}. When the QGP cools down and coalesces into a relativistic hadron gas state, specific shear viscosity $\eta/s$ increases and becomes significantly higher than the KSS bound~\cite{ItakuraK2008,DemirN2009}. When the temperature goes down, we would get some hot finite nucleonic matter; the $\eta/s$ is around 3.0 - 70 times of $\hbar/4\pi$ for the density of 0.2 - 1.25 $\rho_0$ \cite{Zhou2013,Fang2014}, in which the minimum $\eta/s$ value corresponds to the liquid-gas phase transition as discussed in the previous model calculations. And here the value of $\eta/s$ is larger than 2 times of $\hbar/4\pi$ for the infinite nuclear matter, which is consistent with the results in Ref.~\cite{Xujun2013,Zhou2013,LiSX2013,XGDeng2016}. For the finitely colder nuclear matter around the saturation density at much lower temperature (0.8$-$2.1 MeV), it is 2.5$-$6.5 times of $\hbar/4\pi$~\cite{Mondal2017,Guo2017}. From the two tables, one can see that the QGP matter at around 170 MeV has the lowest $\eta/s$ which is close to the KSS bound ($\hbar/4\pi$). However, the nuclear matter at around a few to a few tens MeV, which is in a liquid gas phase transition range, has also relative low $\eta/s$, i.e., about several times the KSS bound. The above low $\eta/s$ may reflect the universal property of strong interaction matter regardless of partonic and nucleonic level. On the contrary, for atomic and molecular substances of He, Ni, and water, they have relatively large $\eta/s$ which are dominated by electromagnetic interaction~\cite{XGDeng2022}.

\subsection{Finite nuclear matter}\label{sub:first-2}

A nucleus is a many-body quantum system in which the constituents with finite nucleons are governed by strong interaction. A useful way to investigate the shear viscosity of finite nuclear matter is through heavy-ion collisions. In the past decades, different model-dependent calculations for shear viscosity, especially for the specific shear viscosity $\eta/s$, have been studied in intermediate-energy heavy-ion collisions.

Within the framework of the Boltzmann-Uehling-Uhlenbeck model, Li $et.\,al.$~\cite{LiSX2013} calculated the shear viscosity in Au + Au collisions below 100 MeV/A by using the Green-Kubo formula as Eq.~(\ref{GKubo1}). In their calculations, an isolated volume is chosen from a spherical space with a radius of 5 fm. For simplicity, the correlation functions can be rewritten in exponential form~\cite{AM2004},
\begin{equation}
\begin{split}
\langle T_{\alpha\beta}(t_{0})T_{\alpha\beta}(t_{0}+t^{\prime}) \rangle \propto {\rm exp} (-\frac{t^{\prime}}{\tau_{\pi}});  \,\, \alpha, \beta=x, y, z,  \alpha\neq \beta .
\end{split}                              
\label{FiniteNM-Eq1}
\end{equation}
The calculations of the stress tensor here only include the momentum part, and the mean field part is neglected. And the shear viscosity coefficient can be rewritten in a simple form:
\begin{equation}
\begin{split}
\eta= \frac{V}{T} \langle T_{\alpha\beta}(t_{0})T_{\alpha\beta}(t_{0}) \rangle {\tau_{\pi}};  \,\, \alpha, \beta=x, y, z,  \alpha\neq \beta .
\end{split}                              
\label{FiniteNM-Eq2}
\end{equation}
where $\tau_{\pi}$ is the relaxation time of the shear flux. With the phase space information, one could get the time evolution of the correlation function. By fitting with ${\rm exp} (-\frac{t^{\prime}}{\tau_{\pi}})$, $\tau_{\pi}$ and the shear viscosity can be obtained. And the entropy density can be calculated by the Gibbs formula,
\begin{equation}
\begin{split}
s= \frac{\epsilon+P-\mu_{n}\rho}{T},
\end{split}                              
\label{FiniteNM-Eq3}
\end{equation}
where $\epsilon$ is energy density inside the volume with 5 fm radius. And $\mu_{n}$ is the nucleon chemical potential, $\rho$ is nucleon density of system within the given sphere. 
\begin{figure}
\setlength{\abovecaptionskip}{0pt}
\setlength{\belowcaptionskip}{8pt}
\centering\includegraphics[scale=0.38]{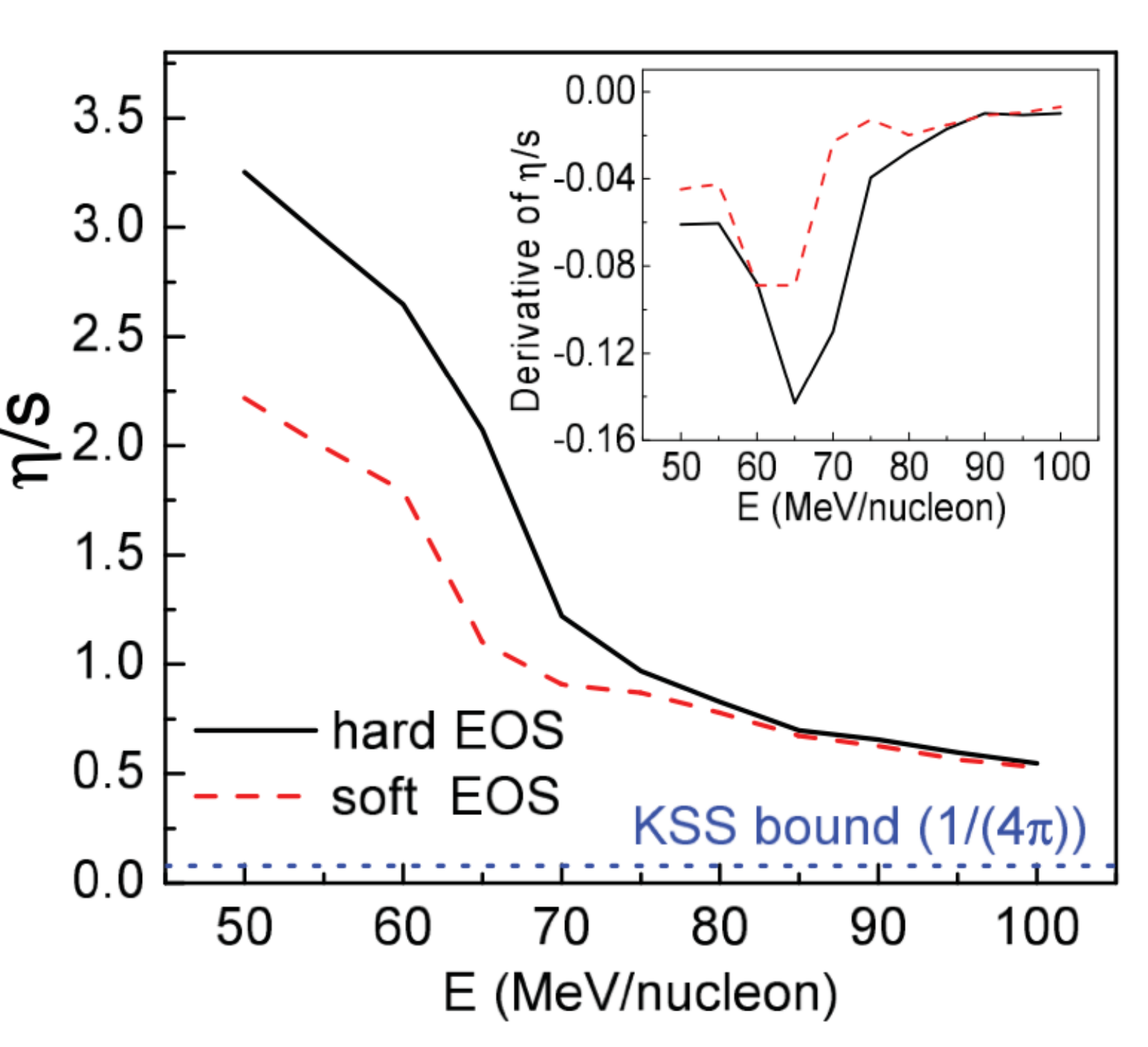}
\caption{(Color online) $\eta/s$ (here $\eta/s$ is divided by $\hbar$ and thus without unit) as a function of beam energy for the central Au + Au collision in a spherical volume with radius of 5 fm. The inset shows the derivative of $\eta/s$ versus beam energy~\cite{LiSX2013}.}
\label{fig:fig13}
\end{figure}
As shown in Fig.\ref{fig:fig13}, the results for $\eta/s$ show a rapid fall with the increasing of incident energy up to E $<$ 70 MeV/A and then drop slowly to a value close to 0.5 when E $>$ 70 MeV/A. Here it is worth to mentioning that $\eta/s$ is a dimensionless quantity since it has been divided by $\hbar$ in some figures blow. 
The results show a strong dependence of $\eta/s$ on the equation of state (EoS) and a harder EoS would give a larger $\eta/s$.  In the inset of Fig.\ref{fig:fig13}, there seems to be a turning point around E$\sim$65 MeV/A. This turning point may indicate a change in the dynamical behavior of the system that is not caused by a phase transition. Since the BUU model is based on one body theory, there is no fluctuation, correlation, or the phase transition behavior. Compared to the KSS bound, here $\eta/s$ is larger than 6 times of $1/4\pi$ (here $\hbar$ is set to `1') in the energy region of E $<$ 100 MeV/A.

\begin{figure}
\setlength{\abovecaptionskip}{0pt}
\setlength{\belowcaptionskip}{8pt}
\centering\includegraphics[scale=0.28]{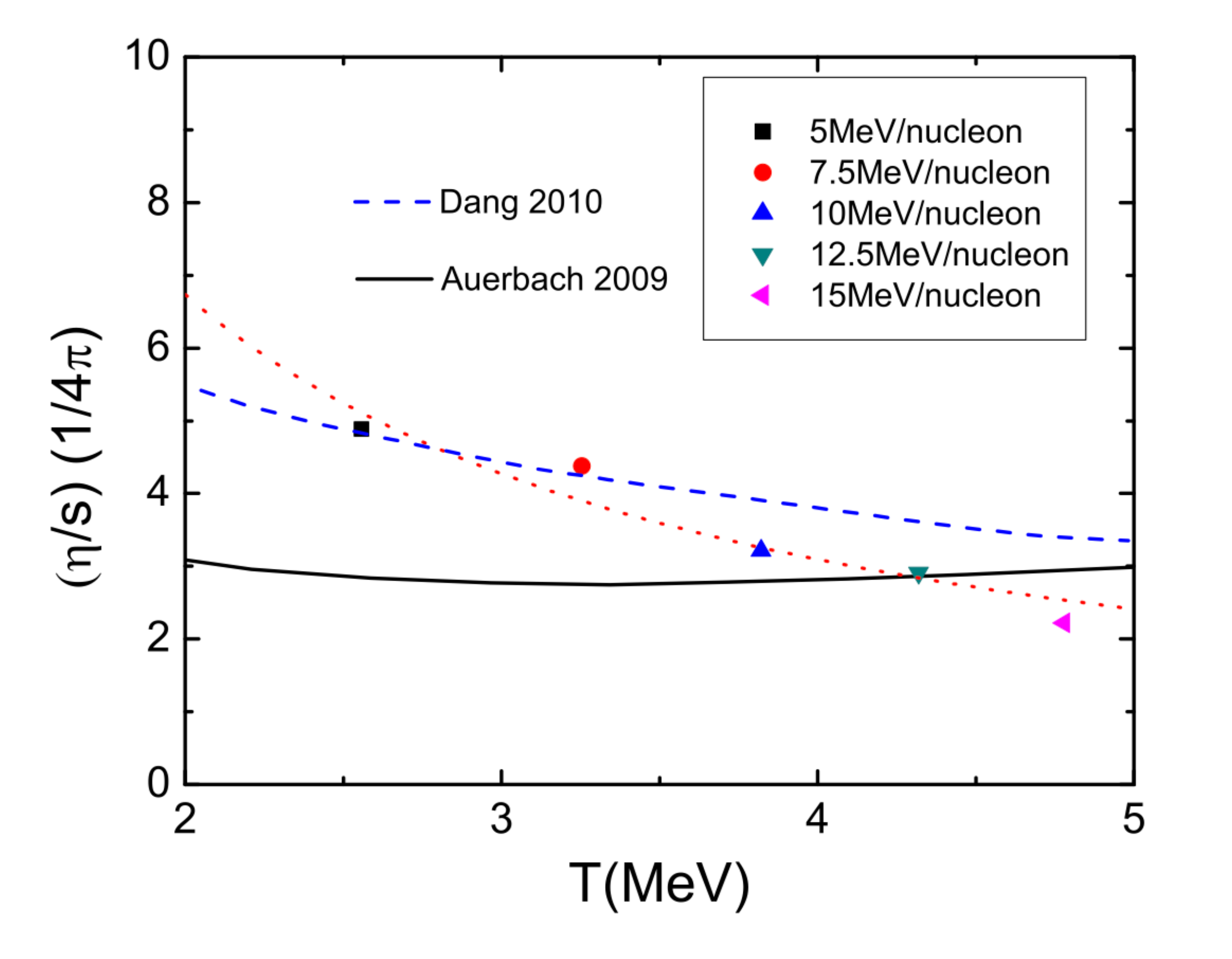}
\caption{(Color online) $\eta/s$ as a function of temperature in a central region for head on $^{40}$Ca + $^{100}$Mo collisions~\cite{Guo2017}. The solid line (Auerbach and Shlomo) is taken from Ref. ~\cite{AuerbachN2009}, and the dashed line (Dinh Dang) is an extrapolation of the phonon-damping model prediction for $^{208}$Pb in Ref.~\cite{DinhDang2011}.}
\label{fig:fig13-1}
\end{figure}

With the same method of Green-Kubo formulas, Guo $et al.$ investigated the specific shear viscosity in low-energy $^{40}$Ca + $^{100}$Mo heavy-ion collision by using an extended quantum molecular dynamics model~\cite{Guo2017}. As shown in Fig.~\ref{fig:fig13-1}, the ratio of shear viscosity to entropy density of the nuclear fireball shows a slight drop as temperature or beam energy increases. The results are consistent with those in Ref.~\cite{AuerbachN2009,DinhDang2011}. Here the specific shear viscosity shows 2 times $1/4\pi$ ($\hbar$=1) which is very close to the KSS limit. 

Later, Zhou et al.~\cite{Zhou2013}, investigated the thermodynamic and transport properties of nuclear fireballs created in the central region of heavy-ion collisions below 400 MeV/A within the isospin-dependent quantum molecular dynamic (IQMD) model. By fitting the results from Eq.(\ref{CEequation9}), one can get shear viscosity in a numerical form~\cite{DPawel1984},
\begin{eqnarray}
\eta= \frac{1700}{T^{2}} \Big{(}\frac{\rho}{\rho_{0}}\Big{)}^{2} + \frac{2}{1+0.001T^{2}} \Big{(}\frac{\rho}{\rho_{0}}\Big{)}^{0.7}+ \frac{5.8\sqrt{T}}{1+160T^{-2}}                  
\label{FiniteNM-Eq4-0}
\end{eqnarray}%
where $\eta$ is in unit of MeV/(fm$^{2}$c), and $T$ is in unit of MeV. Combined temperature and density in the collision process, one can obtain shear viscosity. In the simulations, temperature and entropy density are calculated by the generalized hot Thomas Fermi formalism (GHTFF) ~\cite{DaoTK1992}.
\begin{figure}
\setlength{\abovecaptionskip}{0pt}
\setlength{\belowcaptionskip}{8pt}
\centering\includegraphics[scale=0.19]{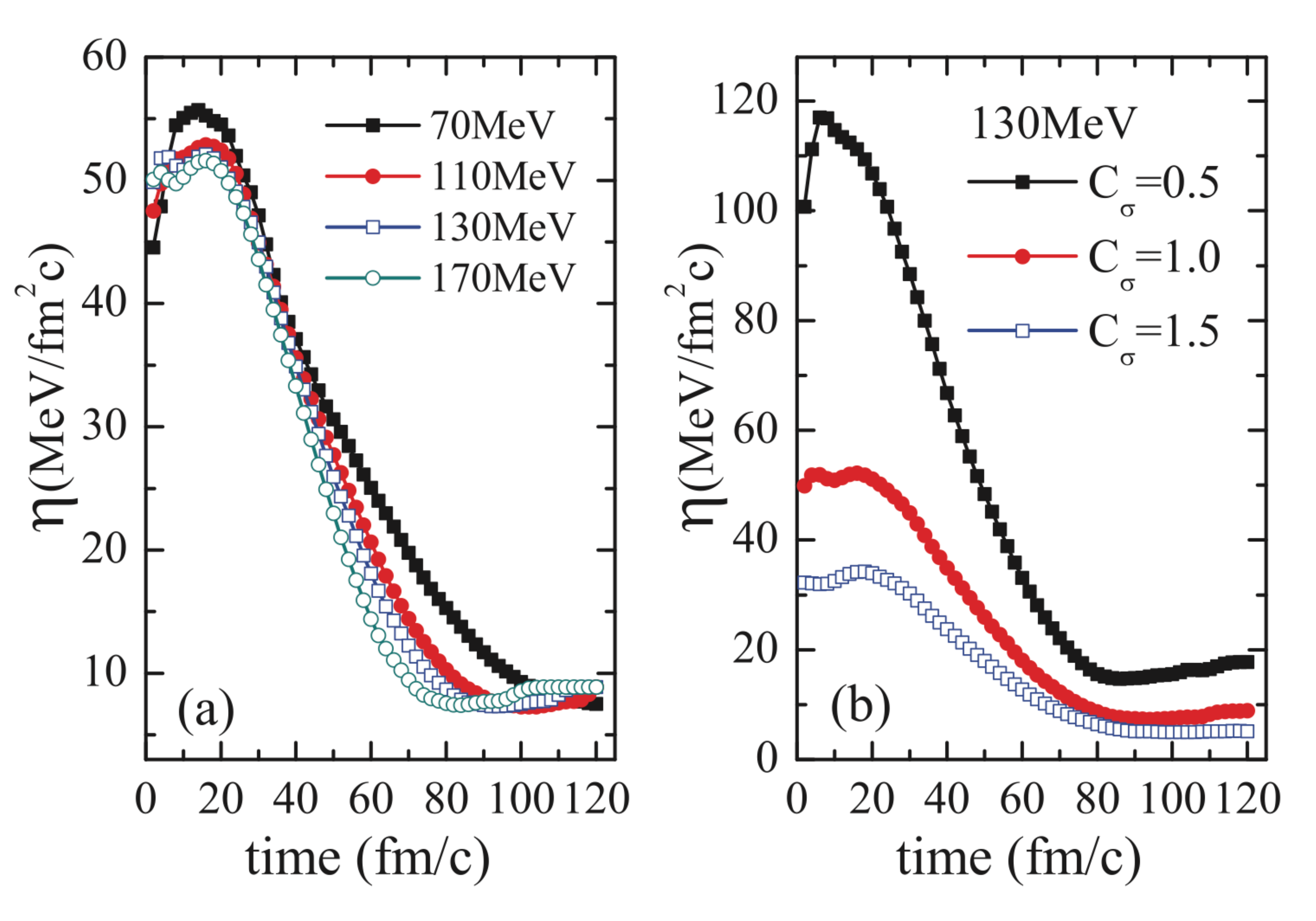}
\caption{(Color online) Time evolution of shear viscosity at different beam energies and cross section factors~\cite{Zhou2013}.}
\label{fig:fig14}
\end{figure}
\begin{figure}
\setlength{\abovecaptionskip}{0pt}
\setlength{\belowcaptionskip}{8pt}
\centering\includegraphics[scale=0.22]{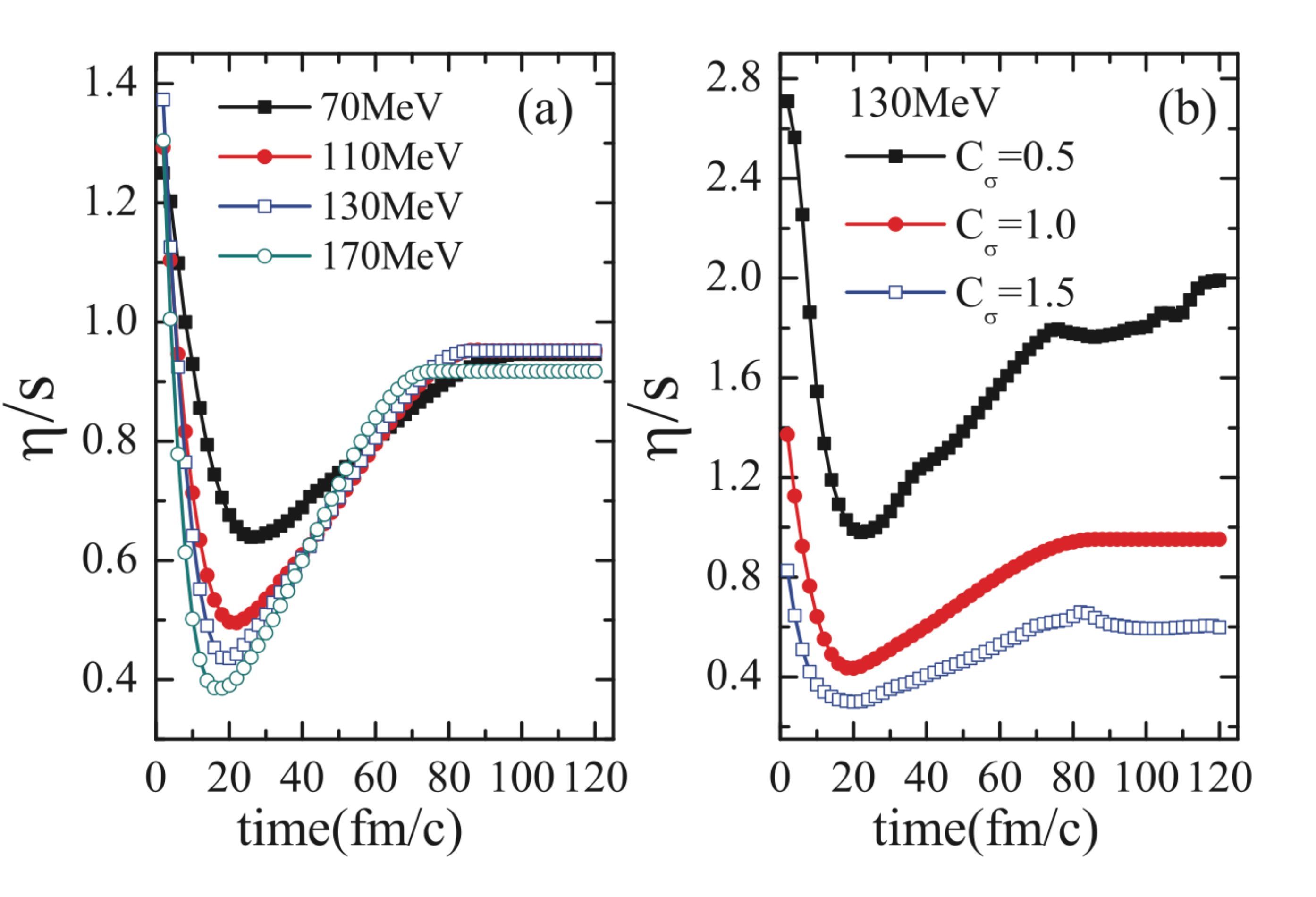}
\caption{(Color online) Time evolution of specific shear viscosity at different beam energies and cross section factors~\cite{Zhou2013}.}
\label{fig:fig15}
\end{figure}
\begin{figure}
\setlength{\abovecaptionskip}{0pt}
\setlength{\belowcaptionskip}{8pt}
\centering\includegraphics[scale=0.25]{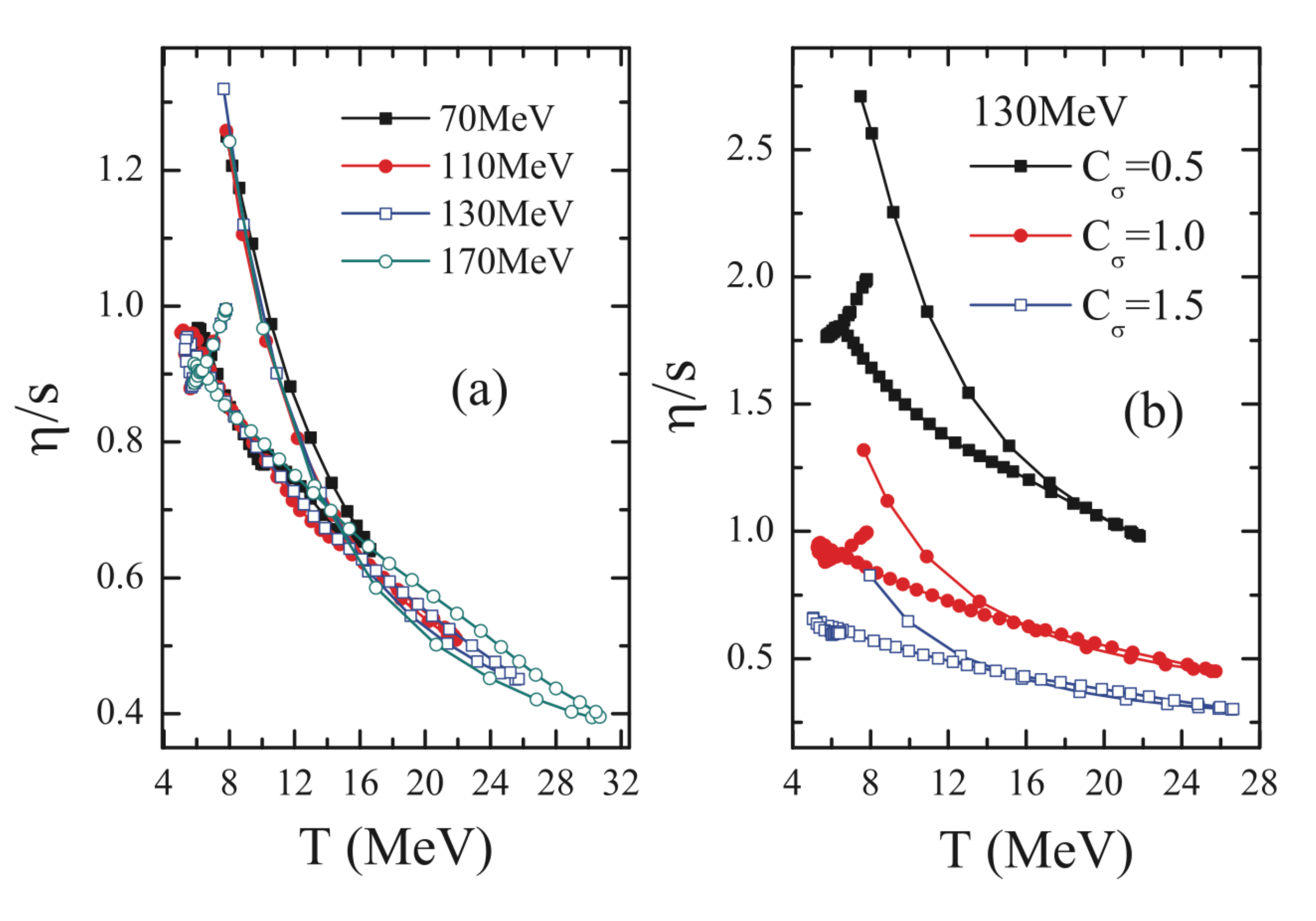}
\caption{(Color online) Specific shear viscosity as a function of temperature at different beam energies and cross section factors~\cite{Zhou2013}.}
\label{fig:fig16}
\end{figure}
In order to study the effect of cross section on shear viscosity, the nucleon-nucleon (NN) cross section is multiplied by a coefficient $C_{\sigma}$, $\sigma_{\rm NN}= C_{\sigma}\sigma_{\rm NN}^{free}$~\cite{Zhou2013}.
In this case, the expression for shear viscosity changes to,
\begin{eqnarray}
\eta(T,\rho,C_{\sigma})= \frac{\eta(T,\rho)}{C_{\sigma}},               
\label{FiniteNM-Eq4-1}
\end{eqnarray}%
The coefficient $C_{\sigma}$ is set to 0.5, 1.0, or 1.5 in the simulations. As shown in Fig.\ref{fig:fig14}, shear viscosity reaches the maximum value at the compression stage during the collision process. 
But specific shear viscosity shows the minimum value as in Fig.\ref{fig:fig16}. As shown in Fig.\ref{fig:fig16}, specific shear viscosity shows a transient minimal $\eta/s$ = (5-10)1/4π in the largest compression stage. It can be noticed that the shear viscosity of nuclear matter in heavy-ion collisions is time-dependent. 

\begin{figure}
\setlength{\abovecaptionskip}{0pt}
\setlength{\belowcaptionskip}{8pt}
\centering\includegraphics[scale=0.38]{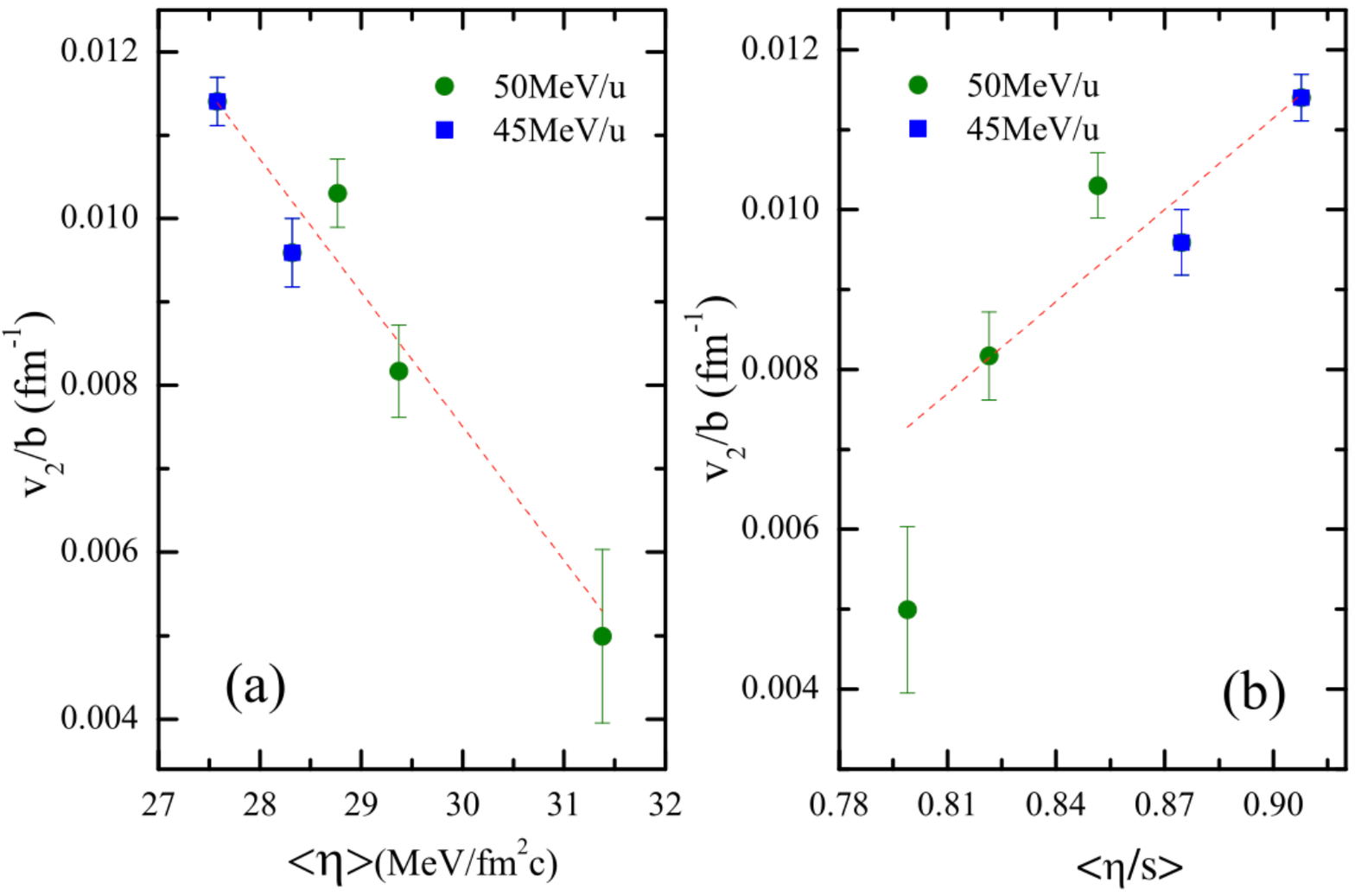}
\caption{(Color online) The scaled elliptic flow as a function of the average shear viscosity and specific viscosity at 45 MeV/u and 50 MeV/u (Here MeV/u is the same with MeV/A) and corresponding impact parameters~\cite{Zhou2014}. The red-dash lines are linear fits.}
\label{fig:fig17}
\end{figure}

In a fluid, shear viscosity plays a resistance role to the flow. As we know, in non-central heavy-ion collisions, particles have collective behavior. Indeed, in a heavy-ion collision, there are two distinct ways for shear viscosity to influence the final emitted particles~\cite{Heinz_2013}. One is that it increases the magnitude and decreases the anisotropies of the hydrodynamically generated
transverse flow which reflects the amount of shear viscosity over the entire expansion history of the fireball. Another is that it causes a deviation, $\delta f$, of the final phase-space distribution $f(x, p)$ from its isotropic local equilibrium form, $f(x, p) = f_{0}(x, p)+\delta f(x, p)$. In heavy-ion collisions, anisotropic flow  can be defined as the orders of harmonic coefficients of Fourier expansion of the particle azimuthal distribution, such as the second-order elliptic flow $v_{2}$,
\begin{eqnarray}
v_{2}= \langle {\rm cos} (2\phi) \rangle= \Big{\langle} \frac{p_{x}^{2}+p_{y}^{2}}{p_{x}^{2}-p_{y}^{2}} \Big{\rangle},              
\label{FiniteNM-Eq5}
\end{eqnarray}%
where $\phi$ is the azimuthal angle, $p_{x}$ and $p_{y}$ are the momenta of the particle at the x and y axes, respectively. For the further work from Zhou {\it et al.}~\cite{Zhou2014}, they consider the correlation between the elliptic flow $v_{2}$ scaled by the impact parameter b and the shear viscosity $\eta$ as well as the specific viscosity $\eta$/s. Here $v_{2}$ is reduced by impact parameter b for reducing influence from b. As shown in Fig.~\ref{fig:fig16}, it is seen that $v_{2}$/b decreases linearly with increasing of average shear viscosity $\langle \eta \rangle$. It is also found that $v_{2}$/b mostly increases with increasing of average specific viscosity $\langle \eta/s \rangle$. So it presents a strong correlation between the elliptic flow and the shear viscosity or specific viscosity $\eta$/s. Somehow, larger shear viscosities quickly reduce collective flow. In relativistic heavy-ion collisions, the large elliptic flow observed at RHIC energies provides compelling evidence for strongly interacting matter that behaves as an almost perfect liquid~\cite{PHENIX2005,STAR2005,PHOBOS2005,ARSENE20051,GYULASSY200530,Heinz_2013}. Here for the $v_{2}$/b versus $\langle \eta \rangle$ consistent with the findings in relativistic heavy-ion collisions, and it is opposite for the $v_{2}$/b versus specific viscosity $\eta$/s. For the later one, the authors argued that it is due to the stronger dissipation in the hadronic phase than in the partonic phase~\cite{Zhou2014}.

\begin{figure}
\setlength{\abovecaptionskip}{0pt}
\setlength{\belowcaptionskip}{8pt}
\centering\includegraphics[scale=0.55]{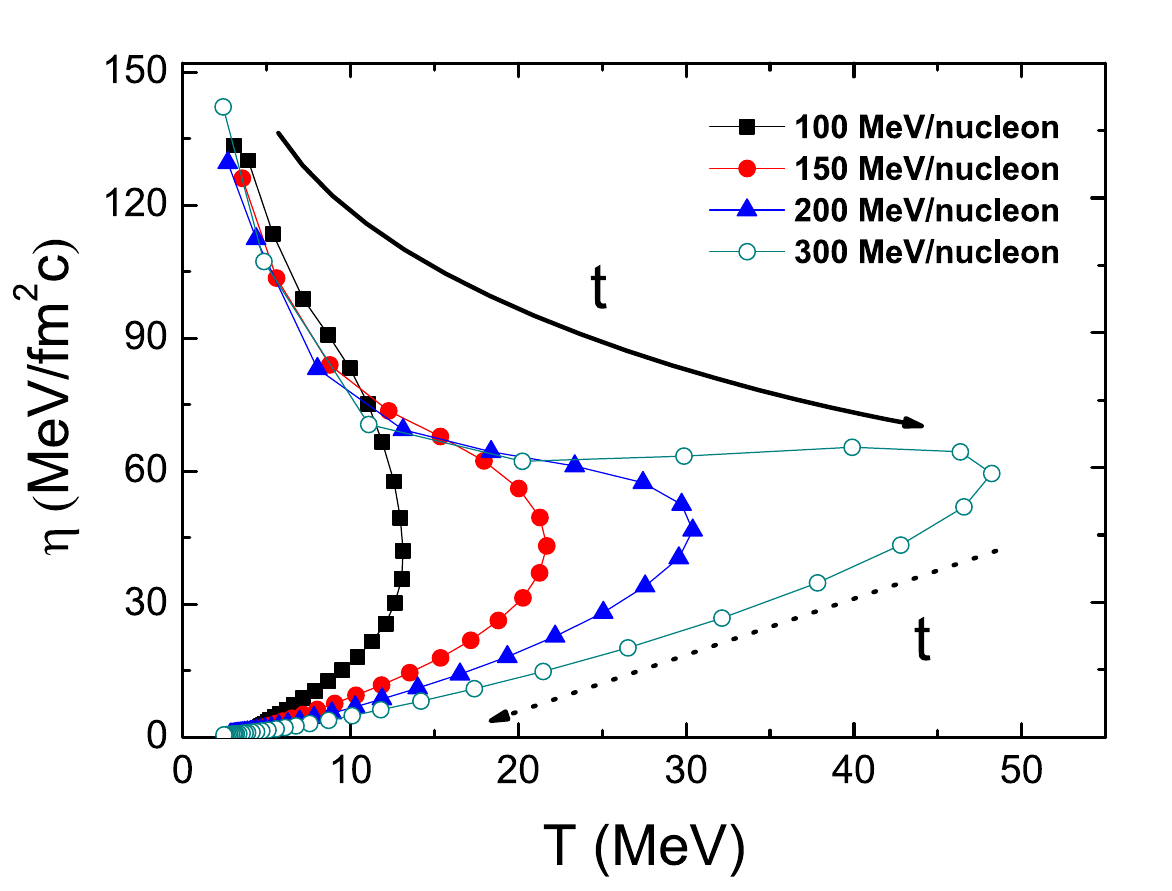}
\caption{(Color online) Time evolution of specific shear viscosity at different beam energies for the soft EoS in the central region of X[−5,5], Y[−5,5], and Z[−5,5]~\cite{XGDeng2016}. Solid and dotted arrows indicate the compression process and expansion process, respectively.}
\label{fig:fig18}
\end{figure}

A similar work is done in Ref.~\cite{XGDeng2016} but with an extension model of Boltzmann-Uehling-Uhlenbeck, namely the Van der Waals Boltzmann-Uehling-Uhlenbeck (VdWBUU) model. In this model, its cross section within the nucleonic medium is derived from the mean field, which is given,
\begin{eqnarray}
\sigma=(\frac{9\pi}{16})^{1/3}b^{\prime{2/3}},
\label{CROSS}
\end{eqnarray}%
where $b^{\prime}$ denotes the proper volume of the constituent particle, which can be related geometrically to its cross section for interaction with other particles and reads as,
\begin{eqnarray}
a^{\prime}&=&-a,                                                        \label{AAA}\\[3mm]
b^{\prime}&=&\frac{b\kappa\rho^{\kappa}+2\gamma{a_{s}}(\frac{\rho}{\rho_{0}})^{1+\gamma}\tau_{z}I}
{(\frac{f_{5/2}(z)}{f_{3/2}(z)})\rho{T}+b\kappa\rho^{1+\kappa}
+2\gamma{a_{s}}\rho_{0}(\frac{\rho}{\rho_{0}})^{1+\gamma}\tau_{z}I}.    \label{BBB}
\end{eqnarray}%
\begin{figure}
\setlength{\abovecaptionskip}{0pt}
\setlength{\belowcaptionskip}{8pt}
\centering\includegraphics[scale=0.55]{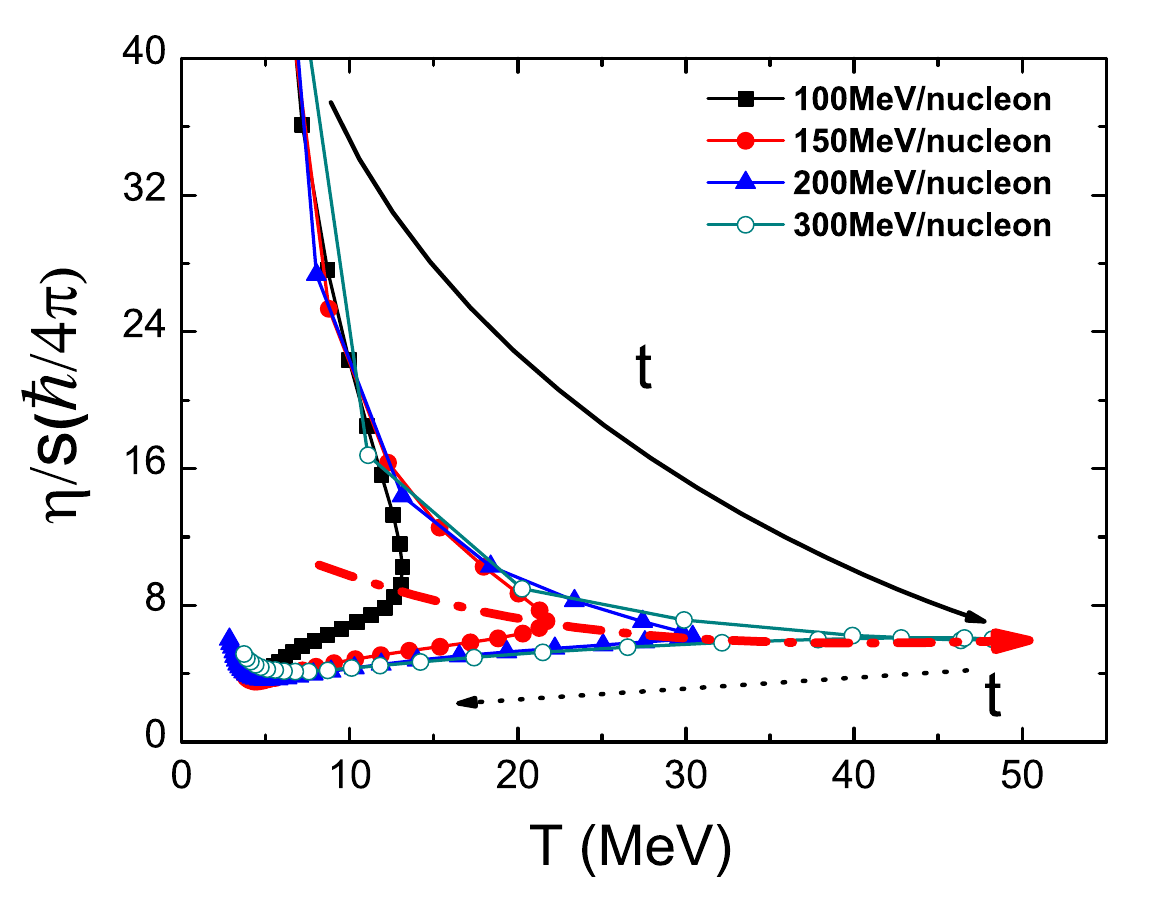}
\caption{(Color online) The same as in Fig.~\ref{fig:fig18} but for the correlation between specific shear viscosity (in units of $\hbar/4\pi$) and temperature~\cite{XGDeng2016}.}
\label{fig:fig19}
\end{figure}
where $a^{\prime}$ is related to attractive interaction among particles; $z = \mu/T$ is the fugacity value of nucleons and $f_{5/2 (3/2)}$ are Fermi integrals; $a_{s}$ is coefficient of the symmetry energy term; $\gamma$ describes the density dependence; and $a, b,$ and $\kappa$ are parameters for the nuclear equation of state (EoS). The more detail can be seen in Ref.~\cite{XGDeng2016}.
In simulations, shear viscosity is calculated in a central region which is defined as a $[-5,5]^3$ fm$^3$ or $[-3,3]^3$ fm$^3$ box with its center located in the c.m. with a numerical expression with isospin asymmetry taken into account as Eq.~(\ref{coefficientA}). And temperature for shear viscosity calculation is extracted by a quantum Fermion fluctuation approach, which is proposed in Ref.~\cite{ZhengH2011}. The entropy density can be determined by the density and temperature~\cite{ZhengH2012}:
\begin{eqnarray}
s=\frac{U-A}{T}\frac{1}{V}=[\frac{5}{2}\frac{f_{5/2}(z)}{f_{3/2}(z)}-\ln{z}]\rho.
\label{entropy}
\end{eqnarray}%
where $f_{m}(z)=\frac{1}{\Gamma(m)}\int_{0}^{\infty}\frac{x^{m-1}}{z^{-1}e^{x}+1}dx$, and $z=e^{\frac{\mu}{T}}$ is fugacity. $U$ is the internal energy and $A$ is Helmholtz free energy~\cite{LIFSHITZ1980}. 

From the results of Ref.~\cite{XGDeng2016} as shown in Fig.~\ref{fig:fig18}, it can be seen that in the compression process, shear viscosity decreases with increasing temperature. While in the expansion process, the behavior is inverted. It is also noticed that in the classical liquid, as the temperature goes higher, the particles become separated over larger distances, they lose local orientation order, and their interactions become weaker, hence the decrease in shear viscosity. A more microcosmic interpretation is the competition between Pauli blocking and collision as for higher density cases as in Fig.~\ref{fig:fig3}. On the contrary, the shear viscosity of gas increases with increasing of temperature due to the momentum of nucleons increasing. Thus, it is assumed that nuclear matter is in the liquid-like phase state during the compression process and in a gas-like phase during the later process. Around the inflection point, it should be in mixed-like phase~\cite{XGDeng2016}. But we cannot say there is a phase transition since the one-body theory for BUU model used in that work ~\cite{XGDeng2016}. In Fig.~\ref{fig:fig19}, specific shear viscosity at a maximum compress stage tends to remain constant as beam energy increases, as the red-dot arrow shows. It is different from the shear viscosity versus T, as in Fig.~\ref{fig:fig18}. 

\begin{figure}
\setlength{\abovecaptionskip}{0pt}
\setlength{\belowcaptionskip}{8pt}
\centering\includegraphics[scale=0.28]{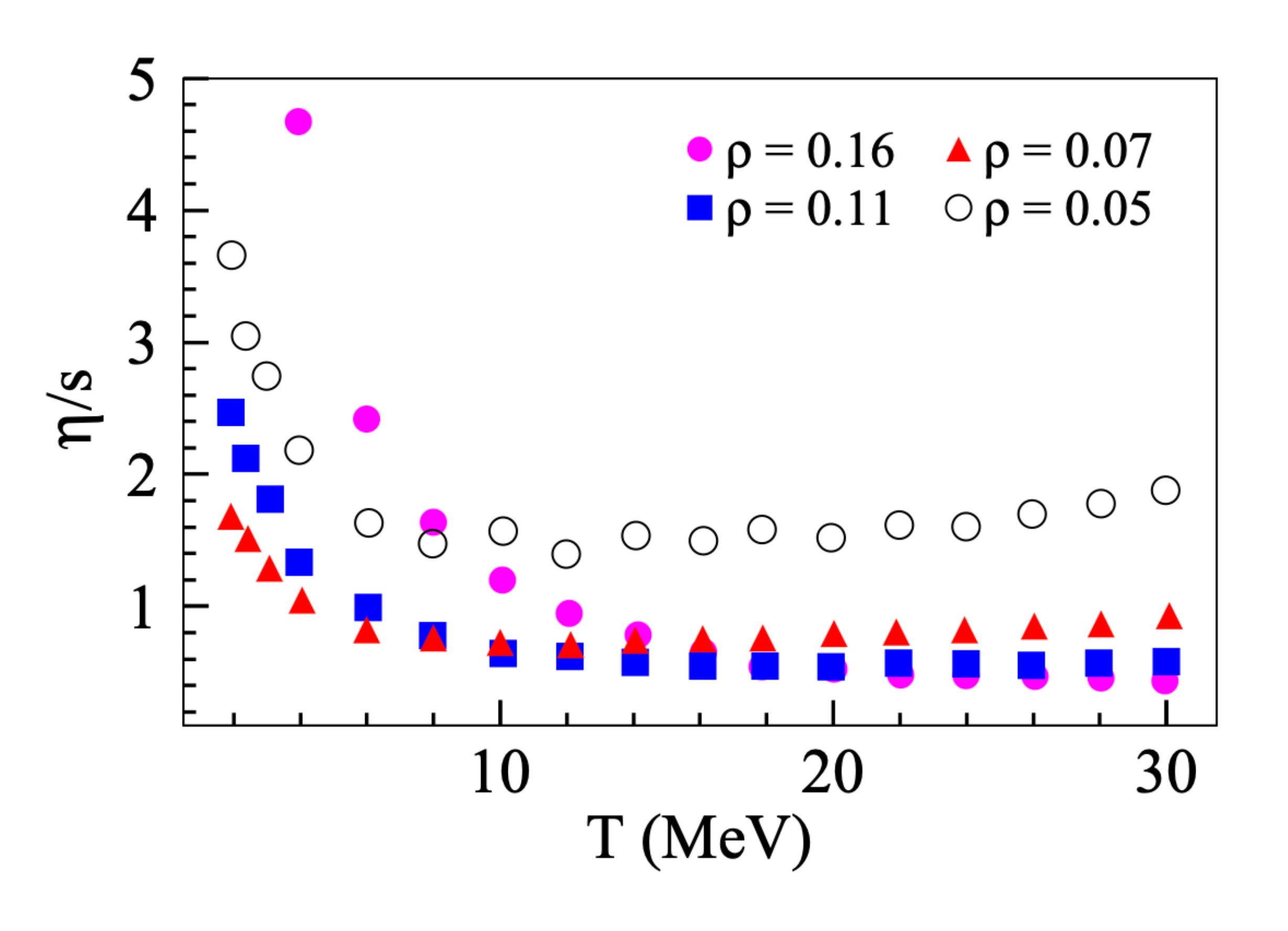}
\caption{(Color online) Specific shear viscosity as a function of temperature at different densities~\cite{Fang2014}.}
\label{fig:fig20}
\end{figure}
\begin{figure}
\setlength{\abovecaptionskip}{0pt}
\setlength{\belowcaptionskip}{8pt}
\centering\includegraphics[scale=0.38]{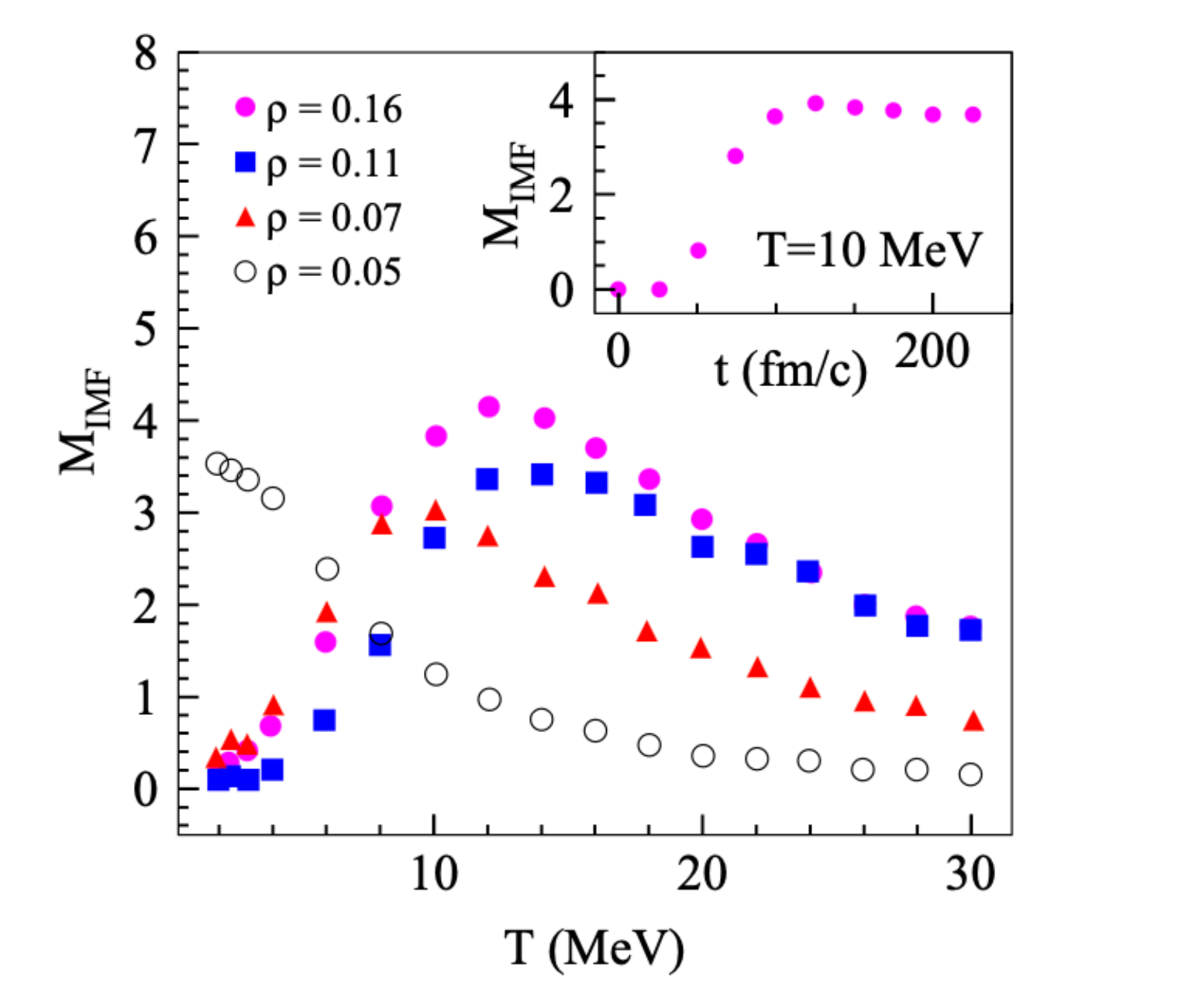}
\caption{(Color online) Temperature dependence of the multiplicity of intermediate mass fragment (IMF) after the freeze-out of the system for nuclear sources at different densities. Inset is the time evolution of multiplicity of  IMF at T$=$10 MeV for $\rho=0.16 $fm$^{-3}$~\cite{Fang2014}.}
\label{fig:fig21}
\end{figure}
In addition, Fang et al. ~\cite{Fang2014} extracted shear viscosities of finite-size nuclear sources by using the mean free path method within the framework of the isospin-dependent quantum molecular dynamics model，and the entropy density is from the Gibbs formula as Eq.~(\ref{FiniteNM-Eq3}). The classical kinetic theory relates $\eta$ of the system with the mean free path of the particle, as in Eq.~(\ref{chap2:Eqshear1}). From the results as shown in Fig.~\ref{fig:fig20}, $\eta/s$ is larger than 7 times KSS bound. And the authors found that a minimum of $\eta/s$ is seen at around $T$ = 10 MeV for $\rho\sim$1/2$\rho_{0}$. For higher densities, $\rho$ = 0.16,0.11 fm$^{-3}$, no minimum of $\eta/s$ is seen, but instead $\eta/s$ approaches an asymptotical value at higher temperatures. There may be a relation between the minimum of $\eta/s$ as a function of $T$ and liquid-gas phase transition of nuclear matter. And the authors have checked the multiplicity of the intermediate mass fragment (IMF), which is a widely studied observable for liquid-gas phase transition in heavy-ion collisions. As shown in Fig.~\ref{fig:fig21}, when the density is around 1/2 $\rho_{0}$ and at the same temperature, the multiplicity of the IMF of the system after freezing-out shows a maximum value which could be the signal of a liquid-gas phase transition~\cite{YGMa1995}. However, for the low density of 0.05 fm$^{-3}$, there is no maximum value for multiplicity of IMF as shown in Fig.~\ref{fig:fig21} but shows a minimum of $\eta/s$ in Fig.~\ref{fig:fig20}. It may indicate that the minimum of $\eta/s$ here is more than the liquid-gas phase transition. 

\begin{figure}
\setlength{\abovecaptionskip}{0pt}
\setlength{\belowcaptionskip}{8pt}
\centering\includegraphics[scale=1.3]{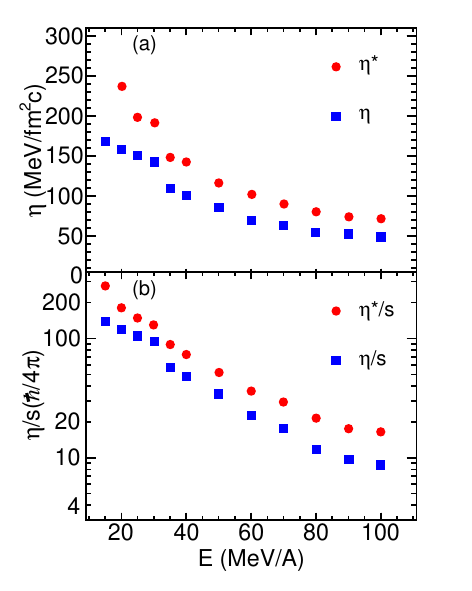}
\caption{(Color online) Shear viscosity and specific shear viscosity as functions of temperature with two different methods~\cite{Liu2017}. But the result for specific shear viscosity in (b) has been modified.}
\label{fig:fig22}
\end{figure}
Further, Liu et al. ~\cite{Liu2017} investigated the thermal and transport properties of central $^{129}$Xe + $^{119}$Sn collisions around the Fermi energy using the mean free path method and numerical expression~\cite{DPawel1984} in the framework of an isospin-dependent quantum molecular dynamical model. As shown in Fig.~\ref{fig:fig22} which has been corrected for the value of specific shear viscosity, it is seen that $\eta/s$ is larger than 8 times the KSS bound. And it decreases as the beam energy increases, which is consistent with that in Refs.\cite{LiSX2013,Zhou2013}. 

In this section, shear viscosity is extracted in QMD or BUU type model with numerical equations (\ref{FiniteNM-Eq2}), (\ref{FiniteNM-Eq4-0}), and (\ref{chap2:Eqshear1}). It can be seen that these results are consistent. However, these formulas can not be utilized from the experimental side and are hard to check with data.

To summarize this part of the discussion, the shear viscosity in infinite and finite nucleonic matter, Pauli blocking and nucleon-nucleon cross section play important roles. The values of specific shear viscosity by different methods or models are several times of $\hbar/4\pi$ and they are not very different from those of QGP matter.

\newpage
\section{Experimental aspects of shear viscosity in nucleonic matter}
\label{fouth}

\subsection{Analysis method by experiment}
\label{sub:fouth-1}

\begin{figure}
\setlength{\abovecaptionskip}{0pt}
\setlength{\belowcaptionskip}{8pt}
\centering\includegraphics[scale=0.3]{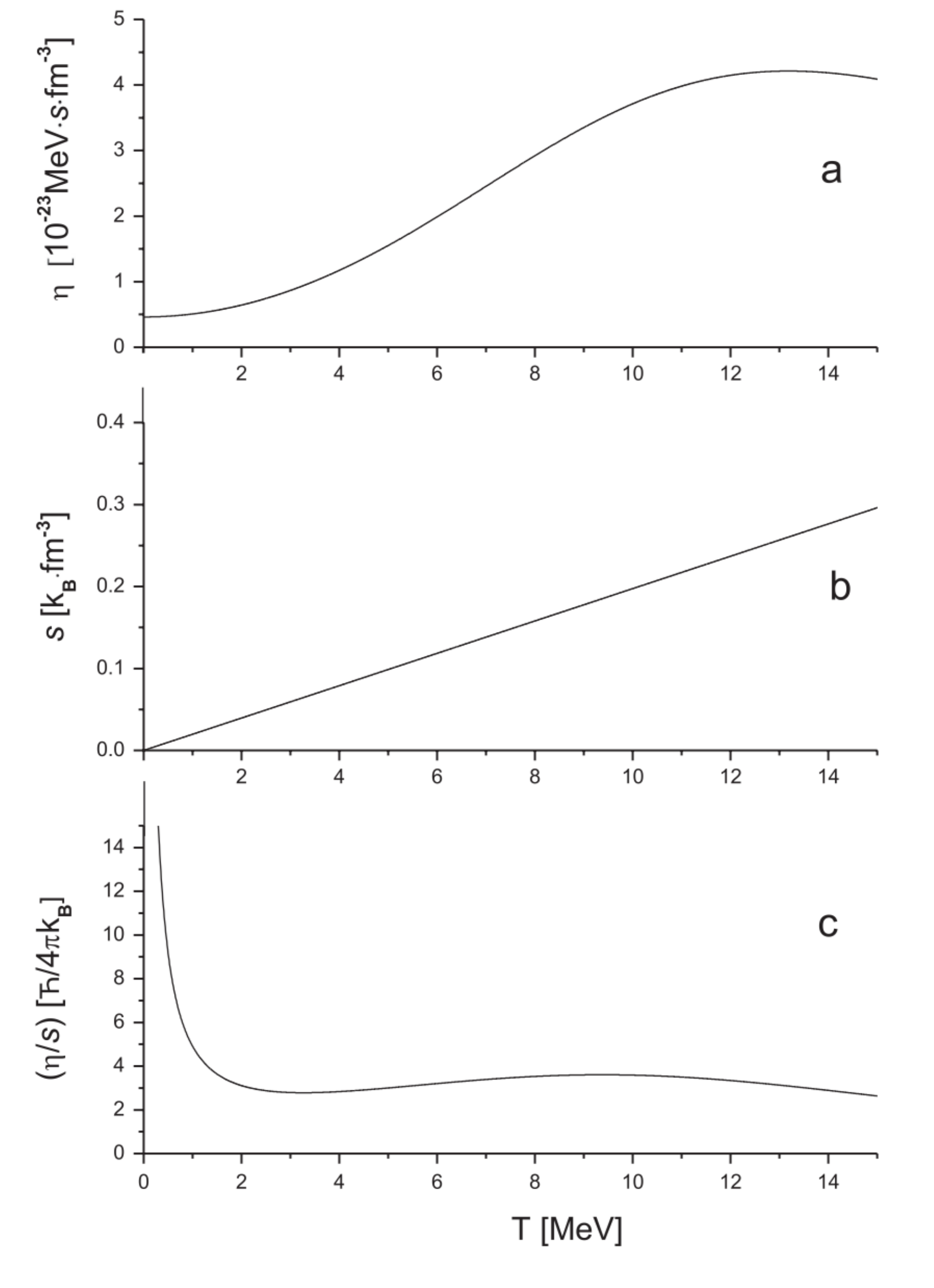}
\caption{(Color online) The nuclear shear viscosity $\eta$ (a), entropy density s (b) and their ratio, ($\eta/s$) (c), in units of $\hbar$/4$\pi{k_{B}}$, as functions of the temperature, T. The values of the parameters used in the calculations are $\epsilon_{F}$ = 40 MeV, $\rho$ = 0.16 fm$^{3}$, $\alpha$ = 9.2 MeV, and $\hbar\omega$ = 20 MeV~\cite{AuerbachN2009}.}
\label{fig:fig23}
\end{figure}

In the previous section, we reviewed some calculations for shear viscosity ($\eta$) and the ratio of shear viscosity to entropy density ($\eta/s$) in nucleonic matter. As mentioned above, the analysis of the state of matter produced by the ultra-relativistic heavy-ion collisions at the Relativistic Heavy Ion Collider (RHIC) shows the matter has very small values of $\eta/s$~\cite{TSDT2009}. How is it for finite nucleonic matter or infinite nuclear matter? The finite nucleonic matter like a finite nucleus, with the constituent nucleons being a many-body quantum system, is governed by strong interaction and shows highly correlated behavior~\cite{Mondal2017}. It can be an ideal system to search for near-perfect fluidity in matter. The search for the universal characteristic of the strong interaction of the many-body quantum systems with shear viscosity is significant.

Auerbach and Sholomo have done earlier theoretical work~\cite{AuerbachN2009} regarding the value of $\eta/s$ for finite nuclear matter in which the nucleons are governed by strong interaction. The authors have extracted shear viscosity within a generalized Fermi liquid  drop model by employing a collision kinetic equation, where the dissipative propagation of sound waves is taken into account in finite nuclei and nuclear matter~\cite{AuerbachN2009}. In the Fermi liquid drop model, shear viscosity is expressed as,
\begin{eqnarray}
\eta=\frac{2}{5} \rho \epsilon_{F} \frac{\tau_{coll}}{1+(\omega \tau_{coll})^{2}},
\label{ExperimentalShearEq-1}
\end{eqnarray}%
with
\begin{eqnarray}
\tau_{coll}= \frac{\tau_{0}}{1+\Big{(}\frac{\hbar\omega}{2{\pi}T}\Big{)}^{2}}.
\label{ExperimentalShearEq-2}
\end{eqnarray}%
where $\hbar\omega$ is the excitation energy of a sound wave and $\omega$-dependent terms are related to the dynamical distortion of the particle momentum distribution due to memory effects~\cite{AuerbachN2009}. In Eqs.~(\ref{ExperimentalShearEq-1}) and (\ref{ExperimentalShearEq-2}), $\tau_{coll}$ is the Landau approximation for the collision relaxation time. And $\tau_{0}=\hbar\alpha/T^{2}$ and the value of $\alpha$ is sensitive to the in-medium-nucleon-nucleon scattering cross section. $\alpha$ is taken to be 9.2 MeV in Ref.~\cite{Kolomietz2004}. As displayed in Fig.\ref{fig:fig23}, shear viscosity increases as temperature increases at density of $\rho$=0.16 fm$^{-3}$. Here the unit 10$^{-23}$ MeV$\cdot$s$\cdot$fm$^{-3}$ is equal to 3 MeV/(fm$^{2}$c), and c is speed of light. From Fig.\ref{fig:fig23}(c), the authors found that the specific shear viscosity is not drastically different from the RHIC result. So it could be a general characteristic feature in a many-body nuclear system with strong interaction, not just for the state created in the relativistic heavy-ion collisions~\cite{AuerbachN2009}. 

It raises a lot of interest in both theory and experiment. 
In 2011, Dang~\cite{DinhDang2011} proposed a formalism based on the Green-Kubo relation and the fluctuation dissipation theorem (FDT). With the Green-Kubo formula~\cite{R_Kubo_1966}, one can obtain the expression for the linear transport coefficient of any system at a given temperature T and density $\rho$ in terms of the time dependence of equilibrium fluctuations in the conjugate flux~\cite{DinhDang2011}. The Green-Kubo formula has been mentioned above. Here we write the shear viscosity $\eta(T)$ as,
\begin{eqnarray}
\eta(T)=\lim_{\omega\rightarrow0}  \frac{1}{2\omega} \int dt d^{3}x e^{i\omega} \langle T_{xy}(\mathbf{x},t)T_{xy}(\mathbf{x},t_{0})   \rangle
\label{ExperimentalShearEq-3-0}
\end{eqnarray}%
where $\langle...\rangle$ denotes average over an equilibrium ensemble of events. From the fluctuation dissipation theorem, the right-hand side of Eq.~(\ref{ExperimentalShearEq-3-0}) can be written as proportion to the absorption cross section $\sigma(\omega,T)$,
\begin{eqnarray}
\eta(T)&=&\lim_{\omega\rightarrow0} \frac{1}{2\omega i} [{\rm G}_{A}(\omega)-{\rm G}_{R}(\omega)] \notag \\
           &=&-\lim_{\omega\rightarrow0} \frac{{\rm Im} \, {\rm G}_{R}(\omega)}{\omega}  \notag \\
           &=&\lim_{\omega\rightarrow0} \frac{\sigma(\omega,T)}{C}
\label{ExperimentalShearEq-3-1}
\end{eqnarray}%
where ${\rm G}_{A}$ and ${\rm G}_{R}$ are the advanced and retarded Green functions, respectively, with 
\begin{eqnarray}
G_{R}&=&-i\int dt d^{3}x e^{i\omega} \langle T_{xy}(\mathbf{x},t)T_{xy}(\mathbf{x},t_{0})  \rangle 
\label{ExperimentalShearEq-3-2}
\end{eqnarray}%
and ${\rm G}_{A}(\omega)$=${\rm G}_{R}^{\ast}(\omega)$. The $C$ in Eq.~(\ref{ExperimentalShearEq-3-1}) is equal to 16$\pi$G with G being the ten-dimensional gravitational constant. As in Ref.~\cite{DinhDang2011}, one can obtain the final expression for the shear viscosity at temperature T,
\begin{eqnarray}
\eta(T)= \eta(0) \frac{\Gamma(T)}{\Gamma(0)} \frac{E_{GDR}(0)^{2}+[\Gamma(0)/2]^{2}}{E_{GDR}(T)^{2}+[\Gamma(T)/2]^{2}},
\label{ExperimentalShearEq-4}
\end{eqnarray}%
where $\Gamma$ is the full width at half maximum (FWHM) of the giant dipole resonance (GDR), and $E_{GDR}$ is its peak energy. The $\eta(0)$ is shear viscosity at T=0. It is worth to mentioning that in the calculations, an uncertainty is in extracting the value $\eta$(0) of the shear viscosity $\eta$(T) at T = 0 and it can be given different dissipation mechanisms~\cite{DinhDang2011}. By using Eq.~(\ref{ExperimentalShearEq-4}), one can calculate the shear viscosity $\eta$ of a finite hot nucleus directly from the width and energy of the GDR of this nucleus in the experimental data which gives model-independent estimation. The entropy density (entropy per volume V) is calculated,
\begin{eqnarray}
s=\frac{S}{V} =\rho \frac{S}{A}.
\label{ExperimentalShearEq-5}
\end{eqnarray}%
with the nuclear density ρ = 0.16 fm$^{-3}$. In Ref.~\cite{DinhDang2011}, two formulas in different methods were used to determine the entropy. One is from phonon-damping model (PDM)~\cite{Dang1998,Dang1998-1,Dang2003} and another is from the Fermi-gas formula,
\begin{eqnarray}
S=2a_{T}T
\label{ExperimentalShearEq-6}
\end{eqnarray}%
where a$_{T}$ the temperature-dependent parametrization of the density parameter as in Refs.~\cite{Drebi1995,Habior1987,TBaumann1998}. This formula (\ref{ExperimentalShearEq-6}) is used in the FLDM and the analysis of experimental data. Combined Eq.~(\ref{ExperimentalShearEq-5}) and Eq.~(\ref{ExperimentalShearEq-6}), one can obtain the entropy density $s=2\rho \frac{a_{T}}{A}T$. $T$ is the temperature corresponding to the final excitation energy $E^{\ast}$ and is given by $T=\sqrt{E^{\ast}/a_{T}}$, where $E^{\ast}=E_{0}^{\ast}-E_{rot}-E_{GDR}-\triangle P$, $E_{0}^{\ast}$, $E_{rot}$ and $\triangle P$ being average excitation energy, the average rotational energy and the pairing energy, respectively~\cite{Mondal2017}. For a given hot nucleus, one can extract the experimental temperature, the width, and the energy of the GDR. When substituted these quantities into Eq.~(\ref{ExperimentalShearEq-4}) and the equation for entropy density $s=2\rho \frac{a_{T}}{A}T$, it can be given a model-independent estimation of the ratio $\eta/s$. 
\begin{figure}
\setlength{\abovecaptionskip}{0pt}
\setlength{\belowcaptionskip}{8pt}
\centering\includegraphics[scale=0.3]{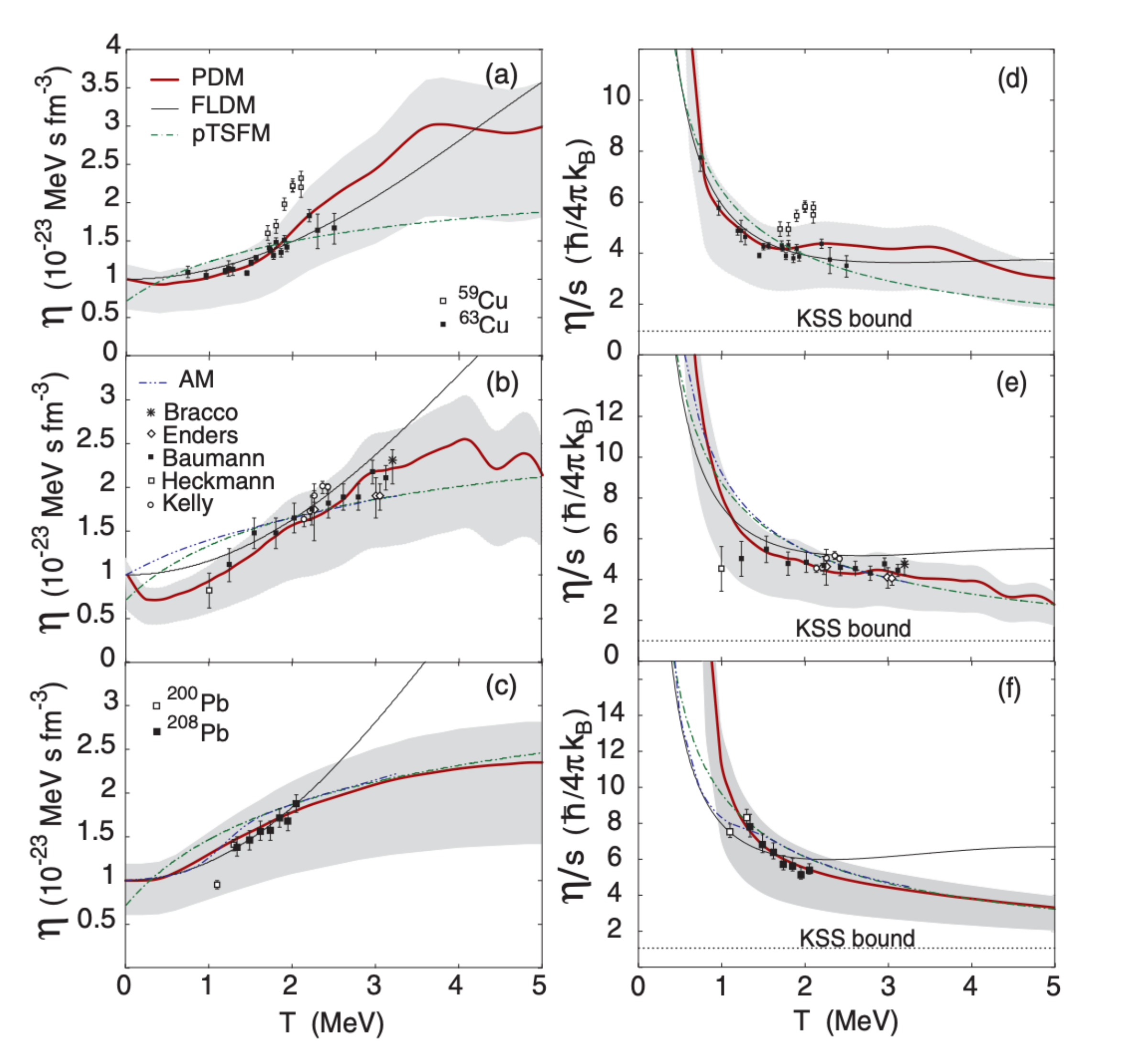}
\caption{(Color online)  Shear viscosity $\eta$(T) [(a)$-$(c)] and ratio of $\eta/s$ [(d)$-$(f)] as functions of T for nuclei in [(a) and (d)] copper, [(b) and (e)] tin, and [(c) and (f)] lead regions. The gray areas are the PDM predictions by using 0.6u$<$η(0)$<$1.2u~\cite{DinhDang2011}. Here u is an unit 10$^{-23}$ MeV s fm$^{3}$. The experimental data for the calculations of $\eta/s$ are from copper ($^{59}$Cu~\cite{Drebi1995} and $^{63}$Cu~\cite{Habior1987}), tin (by Bracco et al. ~\cite{Bracco1989}, Enders et al. ~\cite{Enders1992}, Baumann et al. ~\cite{TBaumann1998}, Heckmann et al. ~\cite{Heckman2003}, and Kelly et al. ~\cite{Kelly1999}), and lead ($^{208}$Pb ~\cite{TBaumann1998} and $^{200}$Pb ~\cite{Chakrabarty1987}).}
\label{fig:fig24}
\end{figure}

\subsection{Shear viscosities of finite nuclei in experiment}
\label{sub:fouth-2}

The predictions for the shear viscosity $\eta$ and the ratio $\eta/s$ by the phonon-damping model (PDM), the Fermi-liquid-drop model (FLDM), the two versions of TSFM, namely, the adiabatic model (AM), and the phenomenological parametrization of the two thermal-shape fluctuation models (pTSFM) for $^{63}$Cu, $^{120}$Sn, and $^{208}$Pb are plotted as functions of $T$ in comparison with the empirical results. The empirical values for $\eta$ in Figs.~\ref{fig:fig24}(a)$-$(c) are extracted from the experimental systematics for the GDR in copper, tin, and lead regions~\cite{Bracco1989,Drebi1995,Habior1987,Enders1992,TBaumann1998,Heckman2003,Kelly1999,Chakrabarty1987}, making use of Eq.~(\ref{ExperimentalShearEq-4}). The theoretical results show agreement with that of experiment. And shear viscosities for nuclei increase with temperature. The specific shear viscosity shows a mild decrease with temperature and is larger than 3 times the KSS bound.

\begin{figure}
\setlength{\abovecaptionskip}{0pt}
\setlength{\belowcaptionskip}{8pt}
\centering\includegraphics[scale=0.3]{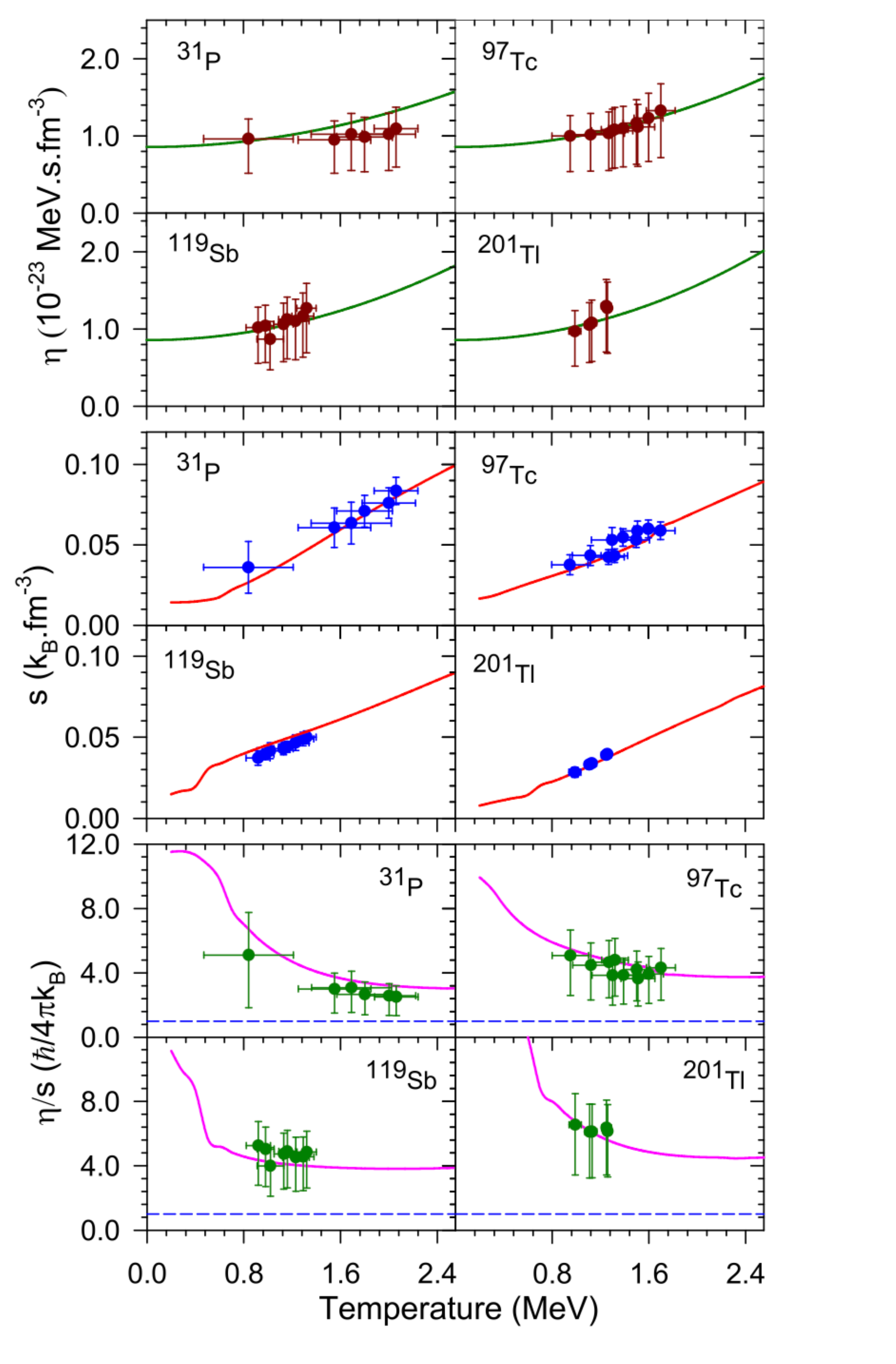}
\caption{(Color online)  Experimental data (symbols) along with the theoretical predictions (solid lines) for shear viscosity $\eta$ (upper panel), entropy density $s$ (middle panel), and specific shear viscosity ($\eta/s$) (lower panel). Blue short-dashed line (lower panel) is the KSS bound. The errors in $\eta$ and $\eta/s$ include the statistical errors as well as the systematic error due to the lower and upper bounds of $\eta$(0)~\cite{Mondal2017}.}
\label{fig:fig25}
\end{figure}
\begin{figure}
\setlength{\abovecaptionskip}{0pt}
\setlength{\belowcaptionskip}{8pt}
\centering\includegraphics[scale=0.35]{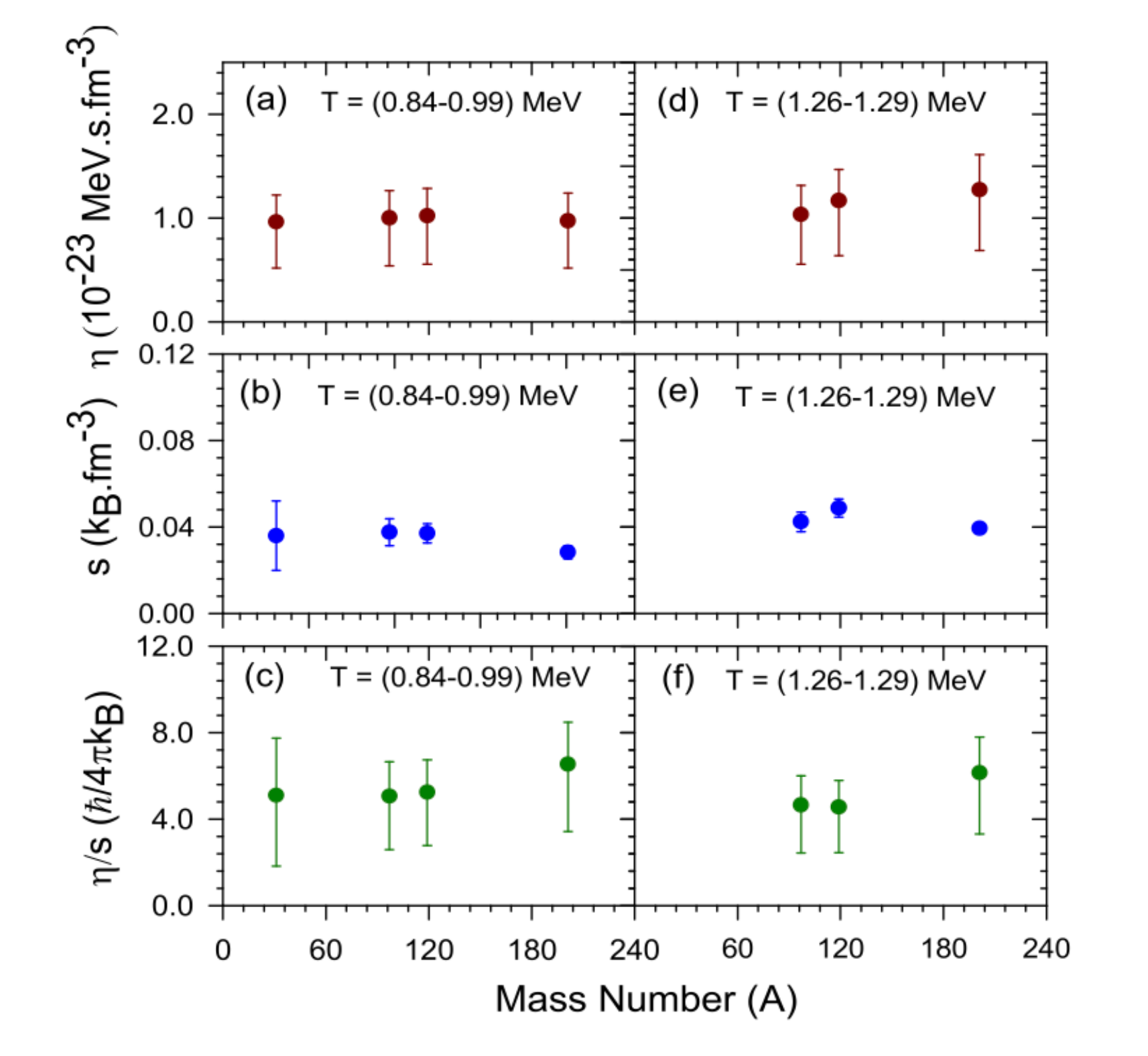}
\caption{(Color online)  Shear viscosity (a), entropy density (b) and specific shear viscosity (c) as functions of mass number ($A$) in the  different temperature regions~\cite{Mondal2017}.}
\label{fig:fig26}
\end{figure}

Heavier nuclear systems from A$\sim$30 to A$\sim$208 are investigated in Ref.~\cite{Mondal2017} for the shear viscosity and ratio of shear viscosity over entropy density by using the formalism as Eq.~(\ref{ExperimentalShearEq-4}). Shear viscosity, entropy density, and the ratio of shear viscosity to entropy density are well reproduced for the systems. Here, shear viscosity also shows an increase as temperature increases. It is interpreted by the kinetic theoretical calculations $\eta$ = $\frac{1}{3}$$\rho m v \lambda $ as Eq.~\ref{chap2:Eqshear1}. 
For an equilibrated nucleus, the mean free path $\lambda$ $\sim$ $v/N_{coll}$, where $N_{coll}$ is the collision number, and is not sensitive to temperature. Thus, shear viscosity increases with temperature for both the mean free path $\lambda$ and $v$ increasing with temperature with $v \sim \sqrt{T}$. One can notice that the finite nucleus system is not a dilute gas system but a fermion system with nucleon density reaching 0.16 fm$^{-3}$ and Pauli blocking effect. However, the results for shear viscosity of finite nucleus shown in Fig.~\ref{fig:fig25} and Fig.~\ref{fig:fig26} are different from those in Fig.~\ref{fig:fig3}, Fig.~\ref{fig:fig8} and Fig.~\ref{fig:fig10}. From the calculations with Green-Kubo, Chapman-Enskog and relaxation time approximation methods, shear viscosity at temperature $T$ = 0 is a large value. Here, $\eta$ at $T$ = 0 has a small finite value. Moreover, it is confined in the range (2.5$-$6.5) $\hbar/(4\pi{k_{B}}$) for the finite nuclear matter within the temperature range of $\sim$(0.8$-$2.1) MeV~\cite{Mondal2017}. For the dependence of mass number, it is seen that $\eta$ increases with the mass number at the highest available temperature for heavier nuclei, as in Fig.~\ref{fig:fig26} (d). However, $\eta/s$ is still within (5.1$-$6.5)$\hbar/(4\pi{k_{B}}$) and shows independence with the nuclear size and the neutron-proton asymmetry at a given temperature~\cite{Mondal2017}.

\begin{figure}
\setlength{\abovecaptionskip}{0pt}
\setlength{\belowcaptionskip}{8pt}
\centering\includegraphics[scale=0.5]{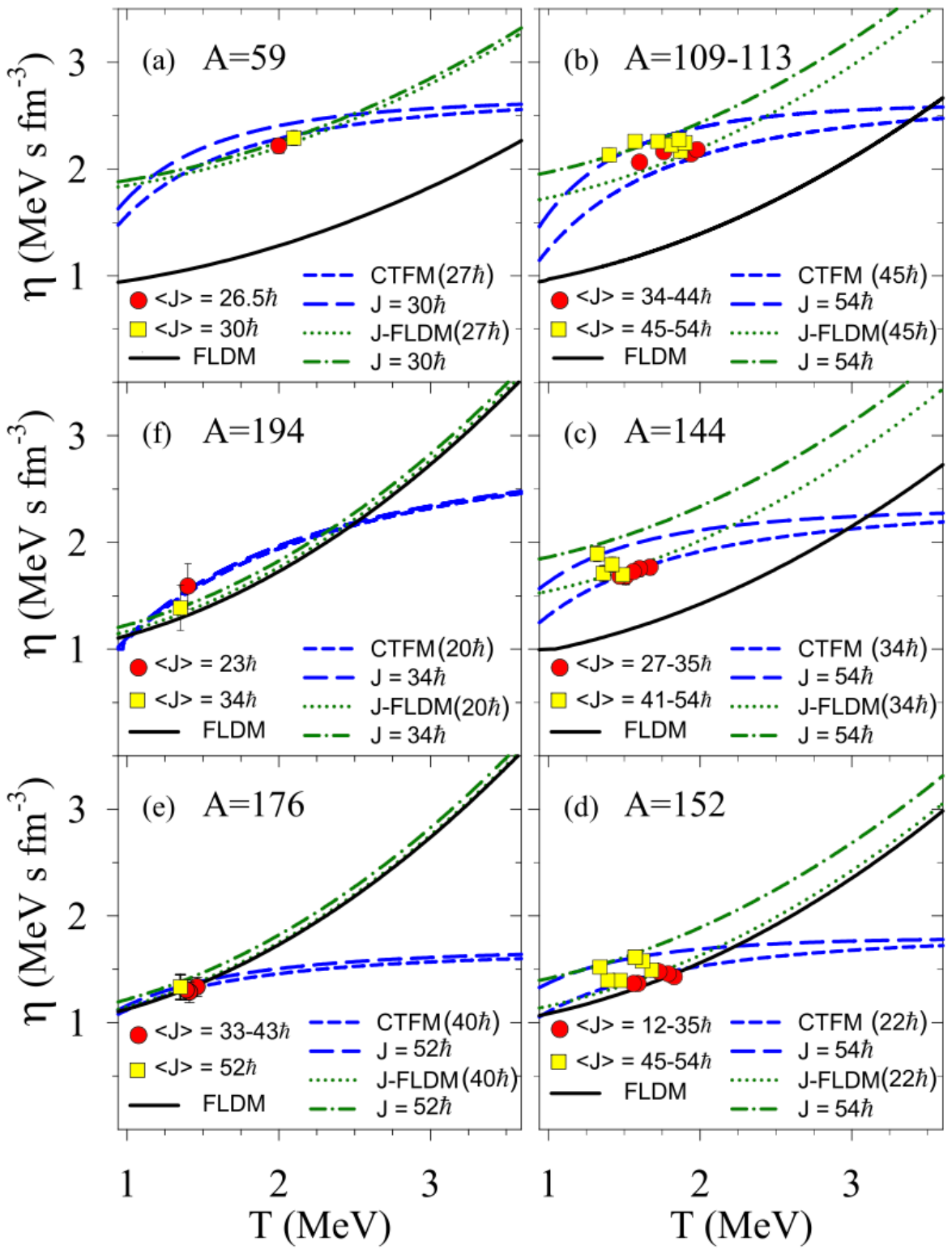}
\caption{(Color online)  Shear viscosity as a function of temperature for different mass nuclei at different angular momenta ($J$)~\cite{Bhattacharya2021}.}
\label{fig:fig26-1}
\end{figure}

In other aspects, the effect of high angular momentum on shear viscosity is discussed by Bhattacharya et al.~\cite{Bhattacharya2021}. With the help of the critical temperature included fluctuation model (CTFM) and FLDM, the predictions for shear viscosity at different angular momenta can be given, and comparison is done with the experimental data by using experimental giant dipole resonance (GDR) width ($\Gamma$) at high angular momenta ($J$ = 12$-$54 $\hbar$) and temperatures (T = 1.2$-$2.1 MeV) collected from the existing literature. As displayed in Fig.~\ref{fig:fig26-1}, it can be seen that the shear viscosity ($\eta$) increases with T and $J$. The results from the critical temperature included fluctuation model (CTFM) well describe J-induced $\eta$. However, the shear viscosity from FLDM, which is an extension of the liquid drop model (LDM) to quantum Fermi liquid, displays a large deviation in the higher angular momentum region and is failing to describe the experimental data. From Fig.~\ref{fig:fig26-1}, one could also learn that the effect of J in $\eta$ is important for lighter mass nuclei. The reason given by the author is due to smaller moment of inertia for the lighter mass nuclei in comparison to that of the heavier mass nuclei.

\begin{figure}
\setlength{\abovecaptionskip}{0pt}
\setlength{\belowcaptionskip}{8pt}
\centering\includegraphics[scale=0.35]{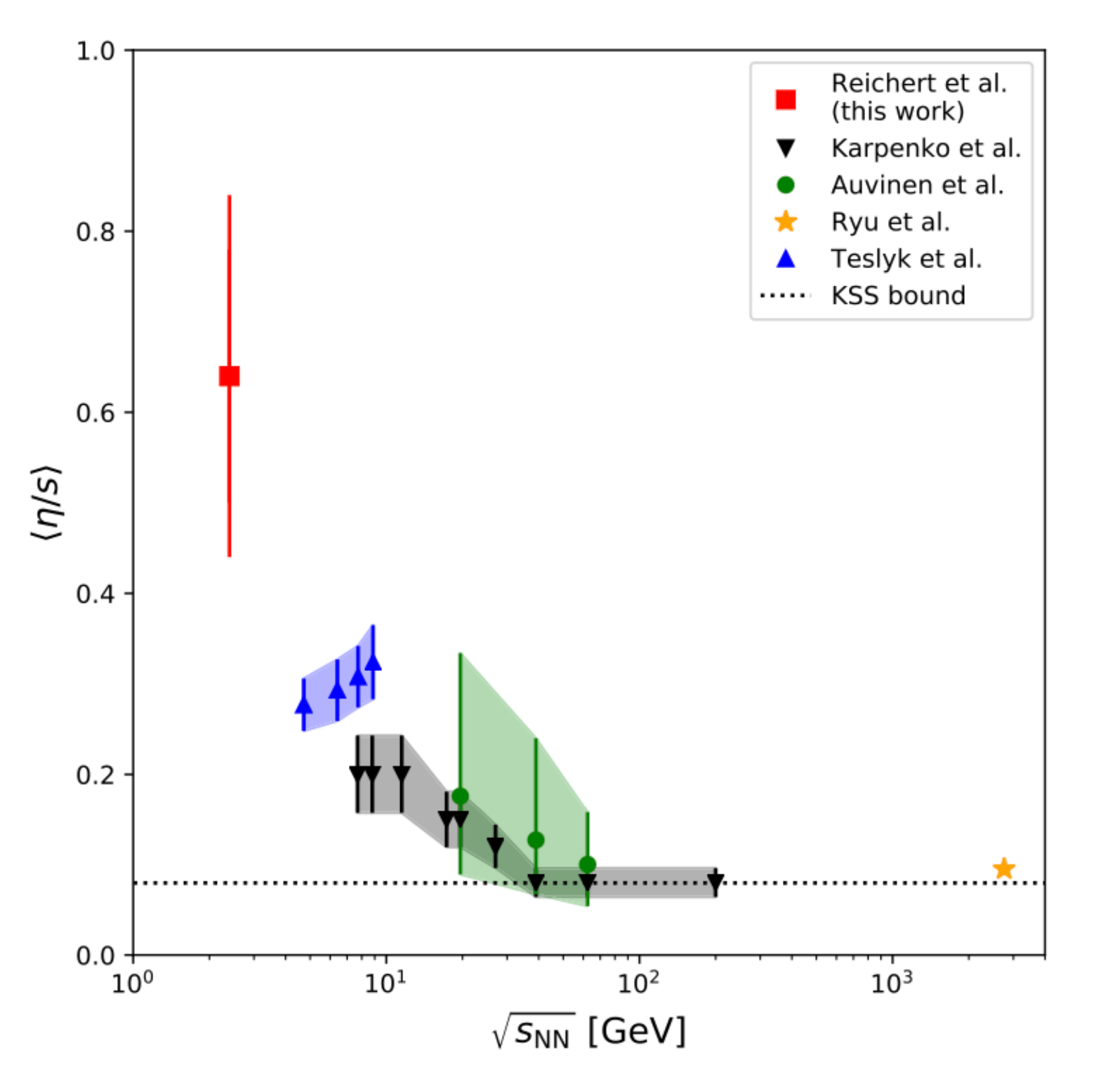}
\caption{(Color online)  Ratio of shear viscosity to entropy density $\eta/s$ as a function of collision energy $\sqrt{\rm s_{NN}}$~\cite{Reichert2021}. The blue triangles-up are extracted from~\cite{Teslyk2020} at full overlap, while the black triangles-down are from~\cite{Karpenko2015} and the green circles are from \cite{Auvinen2018}. The orange star denotes an estimate from~\cite{RyuS2015}, and the dotted black line shows the KSS bound~\cite{KSS2005}.}
\label{fig:fig27}
\end{figure}

Recently, Reichert {\it et al.}~\cite{Reichert2021} estimated $\eta/s$ from data provided by the HADES experiment at GSI on the flow coefficients $v_{1}$,$\cdots$,$v_{4}$ for protons in $^{197}$Au+$^{197}$Au reactions at E$_{\rm{lab}}$ = 1.23 AGeV (or $\sqrt{\rm s_{NN}}$ =2.4 GeV). Through the coarse graining analysis of the UrQMD~\cite{BASS1998,Bleicher1999,Petersen2008,Steinheimer2018} transport simulations of the flow harmonics in comparison to the experimental data, it was obtained that $\eta/s$$\approx$0.65$\pm$0.15 (or (8$\pm$2) $1/4\pi$) as shown by the red-square point in Fig.\ref{fig:fig27}. It is seen that even in the higher collision energy, $\eta/s$ shows a decrease as beam energy increases. Reichert {\it et al.} pointed out that one could explore $\Lambda$, $\bar{\Lambda}$ polarization from the spatial distribution of $\eta$/s in the transverse plane. Also, one can get information on the contributions of $\eta/s$ to the collective flow components since the flow components were created at different times of the collision. For instance, $v_{1}$ is mainly created during the maximal compression phase, where $\eta$/s is lowest and $v_{2}$ gains further contributions during the expansion phase, where $\eta$/s moderately increases~\cite{Reichert2021}. Moreover, one could investigate the effects of shear viscosity on the Fourier-decomposition of azimuthal particle distribution~\cite{Reichert2021}.

\subsection{Others}
\label{sub:fouth-3}

\begin{figure}
\setlength{\abovecaptionskip}{0pt}
\setlength{\belowcaptionskip}{8pt}
\centering\includegraphics[scale=0.4]{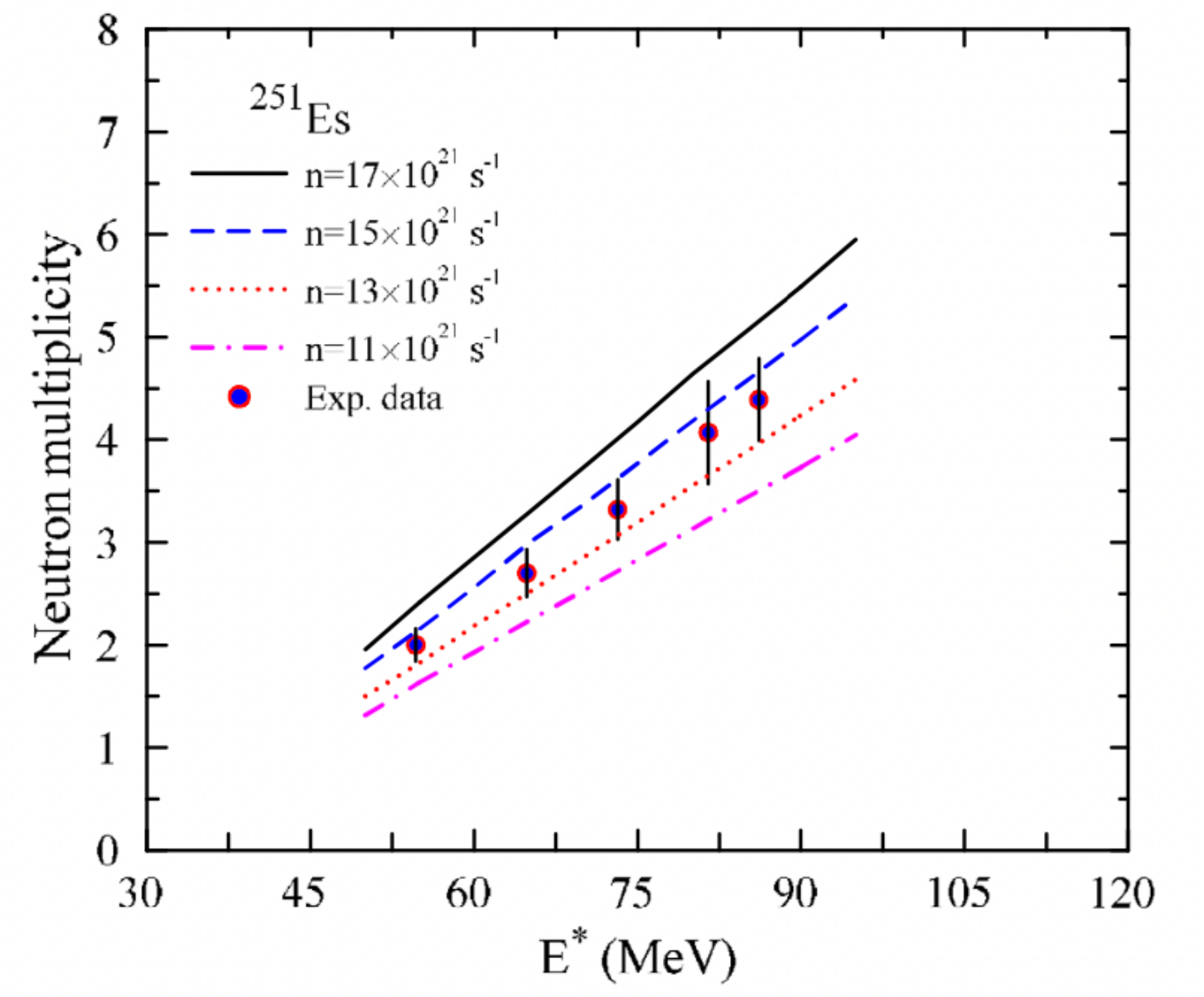}
\caption{(Color online)  The pre-scission fission neutron multiplicity for the compound nucleus $^{251}$Es as a function of energy calculated with different values of the post-saddle friction~\cite{ESLAMIZADEH2020527}.  The experimental data with filled symbols is from \cite{NEWTON_1988}.}
\label{fig:fig27-1}
\end{figure}

Fission is an important subject in the studies of nuclear physics, nuclear astrophysics, and nuclear applications~\cite{Holub_1983,Zank_1986,Chen_jw2023,Yi_jy2022}. Lots of work has attempted to reproduce experimental data on fission in heavy ion collisions by adjusting the properties of the viscosity of hot nuclear matter~\cite{Holub_1983,Zank_1986,GAVRON_1986,Lestone_2009,Kapoor_2019,ESLAMIZADEH2020527}. 
It is well known that the standard statistical model underestimates the number of measured pre-scission particles emitted in heavy-ion reactions ~\cite{Lestone_2009,Kapoor_2019,ESLAMIZADEH2020527}. The main reason could be due to the nuclear viscosity. For the fission, it has two main processes: one is the time needed for transition from equilibrium compound state to saddle point called the transient time ($\tau_{tr}$) and the other is the time of descent from saddle to scission point, called  $\tau_{ssc}$~\cite{Kapoor_2019}. And such processes have been investigated and it is indicated that dissipative in nature at high excitation energy is important~\cite{Kapoor_2019}. In Ref.~\cite{ESLAMIZADEH2020527}, Eslamizadeh and Pirpour simulated the fission process of the excited compound nucleus $^{251}$Es produced in $^{19}$F+$^{232}$Th reaction with a stochastic approach based on two-dimensional Langevin equations. Within the framework, 
it contains a friction coefficient ($n$) as a free parameter which depends on the nuclear viscosity. As shown in Fig.~\ref{fig:fig27-1}, it is shown that the results of calculations are in good agreement with the experimental data by applying values of the post-saddle friction equal to 13$-$15 $\times$ 10$^{21}$ s$^{-1}$~\cite{ESLAMIZADEH2020527}. In Ref.~\cite{ESLAMIZADEH2020527}, the author did not give the value of the nuclear viscosity but the post-saddle friction, which indicates that nuclear viscosity plays an important role in fission dynamics.

In this section, we have reviewed some results for shear viscosity $\eta$ and specific shear viscosity $\eta/s$. Shear viscosity in terms of the width and energy of GDR can help us extract it from experiments. Shear viscosity of finite nuclei shows an increase as temperature increases at low temperatures with a density $\rho$=0.16 fm$^{-3}$. However, for the shear viscosity of infinite nuclear matter as shown in Figs.~\ref{fig:fig9} and ~\ref{fig:fig10}, it can be seen that at density $\rho$=0.16 fm$^{-3}$, shear viscosity decreases as a function of temperature in the low temperature region since there is a competition between Pauli blocking and nucleon-nucleon collisions. Whether it is a finite-size effect or not should be checked in future investigations. And from the experimental results for specific shear viscosity $\eta/s$, it is not so different from that of QGP matter produced in relativistic heavy-ion collisions. Also, we have presented some previous results about the effects of mass and angular momentum on shear viscosity and the effect of nuclear viscosity in fission dynamics studies. 
	
\newpage
\section{$\eta$/s in relativistic heavy$-$ion collision}\label{fifth}

This review focuses mainly on the shear viscosity and ratio of shear viscosity to the entropy density $\eta/s$ of nucleonic matter. However, there is a strong connection between $\eta/s$ of nucleonic matter and QGP in relativistic heavy-ion collisions. For completion, we will shortly review $\eta/s$ in relativistic heavy-ion collisions. In heavy-ion collision, the exact shear viscosity and ratio of shear viscosity to the entropy density should be time dependence and depend on matter density, system temperature, the presence of magnetic fields, and possibly other system conditions since the collision system compression or expansion (in relativistic heavy-ion collisions as the QGP expands and goes through a transition into a hadron phase)~\cite{Gonzalez2021}. 

\subsection{Ratio of shear viscosity to entropy density across the QCD phase diagram}

\begin{figure}
\setlength{\abovecaptionskip}{0pt}
\setlength{\belowcaptionskip}{8pt}
\centering\includegraphics[scale=0.22]{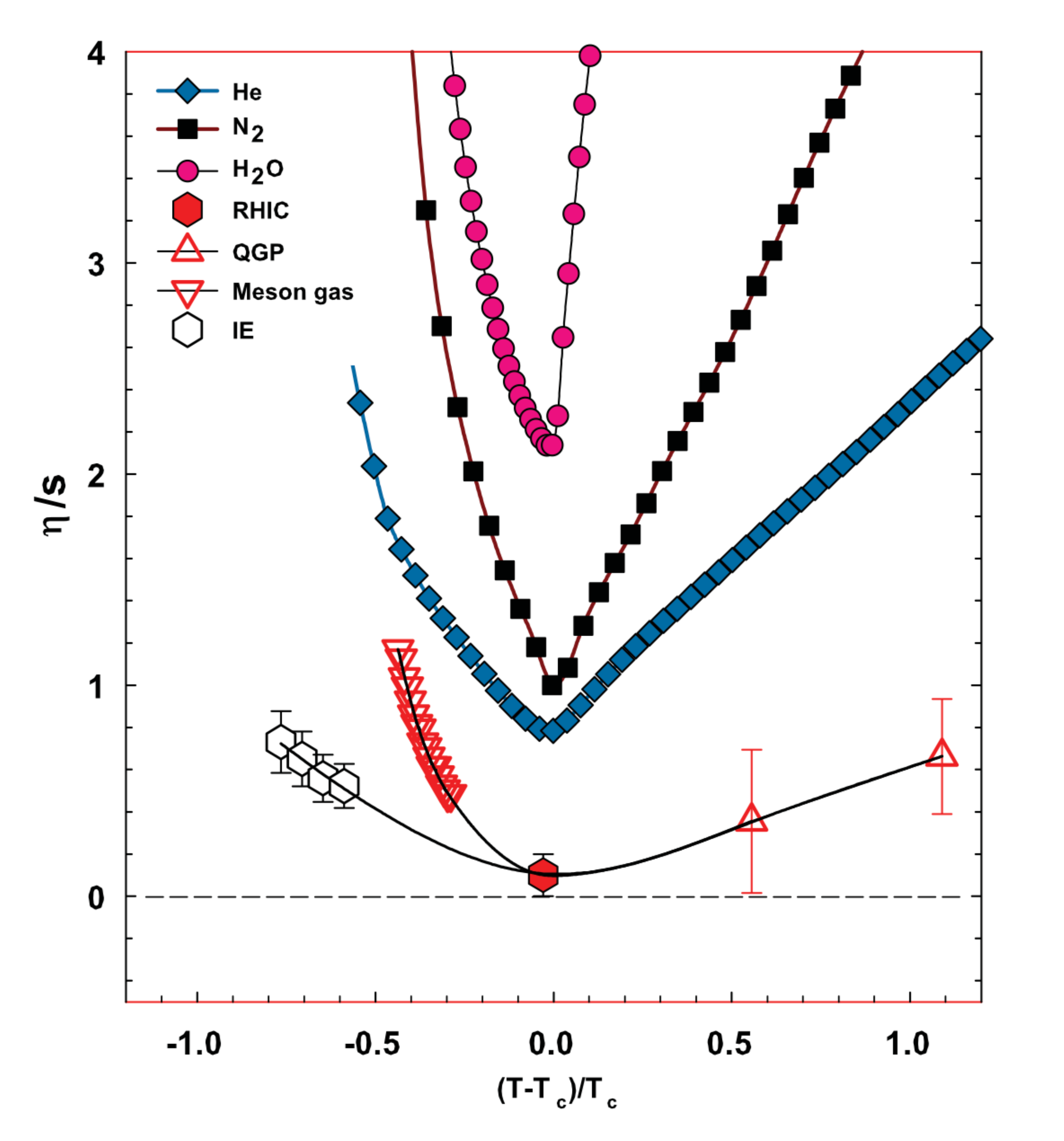}
\caption{(Color online)  Ratio of shear viscosity to entropy density ($\eta/s$) as a function of reduce temperature (T-$T_{c}$)/$T_{c}$ for several substances ~\cite{Lacey2007}. The lattice QCD value critical temperature $T_{c}$ for nuclear matter is set at 170 MeV~\cite{Karsch2001}.}
\label{fig:fig28}
\end{figure}

After the KSS bound proposed by Kovtun, Son, and Starinets~\cite{KSS2005}, lots of efforts were paid to the ratio of shear viscosity to entropy density $\eta/s$, the fluidity of QGP matter, the relation between $\eta/s$ and phase diagram, etc. According to the results in Ref.~\cite{Csernai2006}, the ratio of shear viscosity to entropy density near the phase transition from hadrons to quarks and gluons shows a similar behavior to helium, nitrogen, and water. As seen in Fig.~\ref{fig:fig28}, the experimental and calculated $\eta/s$ values for nuclear matter show a minimum in the vicinity of the critical temperature $T_{c}$~\cite{Lacey2007}. This value of $\eta/s$ for QGP and that in relativistic heavy-ion collision in RHIC is much lower than that obtained from helium, nitrogen, and water. It has been interpreted that the QGP created in the early phase of RHIC collisions is more strongly coupled than atomic and molecular substances. From Fig.~\ref{fig:fig28}, $\eta/s$ for RHIC collisions at $T_{c}$, is rather close to the conjectured lower bound of $1/4\pi$(generally, in relativistic heavy-ion collisions $\hbar$=1). The authors concluded that it is compatible with the minimum expected if the hot and dense QCD matter produced in RHIC collisions follows decay trajectories that are close to the critical end point (CEP)~\cite{Lacey2007}. 

\renewcommand\arraystretch{1.3}
\begin{table}
\setlength{\belowcaptionskip}{0.2cm}
\centering  
\caption{Compilation table ~\cite{Gonzalez2021} for measured charged particle densities, dN$_{ch}$/d$\eta$,
and longitudinal widths, $\sigma_{\bigtriangleup\eta}$, of the charge independent
correlator G$^{\rm CI}_{2}$, interpolated freezeout times, $\tau_{f}$, and computed values of $\eta$/s as a function of the centrality of Pb$+$Pb collisions at $\sqrt{s_{\rm NN}}$=2.76 TeV~\cite{ALICE2011_1,ALICE2011_2,ALICE2013_1,ALICE2020_1} and Au$+$Au at $\sqrt{s_{\rm NN}}$= 200 GeV~\cite{STAR2009_1,STAR2011_1}.} \label{G2shearTable}
\begin{tabular}{p{2cm} p{2.5cm} p{5.0cm} p{3.0cm} p{3.5cm} }
\hline
Centrality  &  dN$_{ch}$/d$\eta$ &  $\sigma_{\bigtriangleup\eta}$ of  G$^{\rm CI}_{2}$ & $\tau_{f}$ (fm/c) &  $\eta$/s  \\ %
\Xhline{1pt}
LHC   \\ %
0-5\%         & 1601.00$\pm$60.00   & 0.68$\pm$0.01$^{\rm sta}$+0.11$^{\rm sys}$-0.03$^{\rm sys}$ & 10.33$\pm$0.10$\pm$1.03$^{\rm fit}$ & 0.05$\pm$0.00$^{\rm sta}$$\pm$0.04$^{\rm sys}$ \\ %
5-10\%       & 1294.00$\pm$49.00   & 0.73$\pm$0.01$^{\rm sta}$+0.54$^{\rm sys}$-0.03$^{\rm sys}$ &  9.62$\pm$0.09$\pm$0.91$^{\rm fit}$ & 0.07$\pm$0.00$^{\rm sta}$$\pm$0.03$^{\rm sys}$\\ %
10-20\%     & 966.00$\pm$37.00     & 0.71$\pm$0.01$^{\rm sta}$+0.04$^{\rm sys}$-0.03$^{\rm sys}$ &  8.73$\pm$0.08$\pm$0.75$^{\rm fit}$ & 0.06$\pm$0.00$^{\rm sta}$$\pm$0.02$^{\rm sys}$\\ %
20-30\%     & 649.00$\pm$23.00     & 0.73$\pm$0.01$^{\rm sta}$+0.03$^{\rm sys}$-0.03$^{\rm sys}$ &  7.64$\pm$0.07$\pm$0.56$^{\rm fit}$ & 0.07$\pm$0.00$^{\rm sta}$$\pm$0.03$^{\rm sys}$\\ %
30-40\%     & 426.00$\pm$15.00     & 0.70$\pm$0.01$^{\rm sta}$+0.03$^{\rm sys}$-0.03$^{\rm sys}$ &  6.64$\pm$0.06$\pm$0.39$^{\rm fit}$ & 0.06$\pm$0.00$^{\rm sta}$$\pm$0.02$^{\rm sys}$\\ %
40-50\%     & 261.00$\pm$ 9.00      & 0.69$\pm$0.01$^{\rm sta}$+0.03$^{\rm sys}$-0.03$^{\rm sys}$ &  5.64$\pm$0.05$\pm$0.23$^{\rm fit}$ & 0.06$\pm$0.00$^{\rm sta}$$\pm$0.02$^{\rm sys}$\\ %
50-60\%     & 149.00$\pm$ 6.00      & 0.65$\pm$0.01$^{\rm sta}$+0.03$^{\rm sys}$-0.03$^{\rm sys}$ &  4.68$\pm$0.04$\pm$0.13$^{\rm fit}$ & 0.05$\pm$0.00$^{\rm sta}$$\pm$0.02$^{\rm sys}$\\ %
60-70\%     & 76.00$\pm$ 4.00        & 0.63$\pm$0.01$^{\rm sta}$+0.03$^{\rm sys}$-0.03$^{\rm sys}$ &  3.74$\pm$0.04$\pm$0.20$^{\rm fit}$ & 0.05$\pm$0.00$^{\rm sta}$$\pm$0.02$^{\rm sys}$\\ %
70-80\%     & 35.00$\pm$ 2.00        & 0.59$\pm$0.01$^{\rm sta}$+0.02$^{\rm sys}$-0.02$^{\rm sys}$ &  2.89$\pm$0.03$\pm$0.33$^{\rm fit}$ & 0.04$\pm$0.00$^{\rm sta}$$\pm$0.02$^{\rm sys}$\\ %
RHIC   \\ %
0-5\%        & 691.00$\pm$49.00  & 0.94$\pm$0.06$^{\rm sta}$$\pm$0.17$^{\rm sys}$ & 7.24$\pm$0.07$\pm$0.50$^{\rm fit}$ & 0.14$\pm$0.03$^{\rm sta}$$\pm$0.09$^{\rm sys}$\\ %
5-10\%      & 558.00$\pm$40.00  & 0.99$\pm$0.07$^{\rm sta}$$\pm$0.06$^{\rm sys}$ & 6.81$\pm$0.06$\pm$0.41$^{\rm fit}$ & 0.16$\pm$0.03$^{\rm sta}$$\pm$0.07$^{\rm sys}$\\ %
10-20\%    & 421.00$\pm$30.00  & 0.93$\pm$0.06$^{\rm sta}$$\pm$0.07$^{\rm sys}$ & 6.27$\pm$0.05$\pm$0.30$^{\rm fit}$ & 0.14$\pm$0.03$^{\rm sta}$$\pm$0.06$^{\rm sys}$\\ %
20-30\%    & 287.00$\pm$20.00  & 0.84$\pm$0.05$^{\rm sta}$$\pm$0.03$^{\rm sys}$ & 5.62$\pm$0.04$\pm$0.17$^{\rm fit}$ & 0.10$\pm$0.02$^{\rm sta}$$\pm$0.04$^{\rm sys}$\\ %
30-40\%    & 195.00$\pm$14.00  & 0.67$\pm$0.03$^{\rm sta}$$\pm$0.02$^{\rm sys}$ & 5.05$\pm$0.04$\pm$0.12$^{\rm fit}$ & 0.04$\pm$0.01$^{\rm sta}$$\pm$0.02$^{\rm sys}$\\ %
40-50\%    & 126.00$\pm$60.00  & 0.59$\pm$0.02$^{\rm sta}$$\pm$0.03$^{\rm sys}$ & 4.48$\pm$0.05$\pm$0.17$^{\rm fit}$ & 0.01$\pm$0.01$^{\rm sta}$$\pm$0.02$^{\rm sys}$ \\ %
50-60\%    & 78.00$\pm$ 9.00     & 0.57$\pm$0.02$^{\rm sta}$$\pm$0.02$^{\rm sys}$ & 3.95$\pm$0.06$\pm$0.27$^{\rm fit}$ & 0.01$\pm$0.01$^{\rm sta}$$\pm$0.02$^{\rm sys}$\\ %
60-70\%    & 45.00$\pm$ 6.00     & 0.55$\pm$0.02$^{\rm sta}$$\pm$0.04$^{\rm sys}$ & 3.43$\pm$0.08$\pm$0.38$^{\rm fit}$ & 0.003$\pm$0.009$^{\rm sta}$$\pm$0.022$^{\rm sys}$\\ %
\hline
\end{tabular}
\end{table}

\subsection{The dependence of ratio of shear viscosity to entropy density on collision centrality}

According to the work~\cite{Gonzalez2021}, the author gives the dependence of specific shear viscosity on collision centrality in ultra-relativistic heavy-ion collisions at RHIC and LHC by using the Gavin ansatz, which relates the longitudinal broadening of transverse momentum two-particle correlators, $G_{2}$. The $G_{2}$ correlator~\cite{Gavin2006,Sharma2009} is designed to be proportional to the covariance of momentum currents and is as such sensitive to dissipative viscous forces at play during the transverse and longitudinal expansion of the matter formed in A$+$A collisions. And these forces lead to a longitudinal broadening of $G_{2}$ measured as a function of the pseudorapidity difference of measured charged particles~\cite{Gonzalez2021}. Through this $G_{2}$ correlator, Gavin gave a clear image for relation between $G_{2}$ and system expansion. As the matter expands, neighboring fluid cells drag one another. Fast fluid cells tend to slow down whereas slow fluid cells accelerate. This has the effect of dampening the expansion and produces a progressive broadening of the $G_{2}$ correlator with time. The longer the system lives, the longer viscous effects play a role, and the broader the $G_{2}$ correlator becomes~\cite{Gonzalez2021}. It means that one can get the difference in the variance of the correlator among different collision centralities. For the Gavin ansatz, one can write it as,
\begin{eqnarray}
\sigma_{c}^{2}-\sigma_{0}^{2}=\frac{4}{T_{c}}\frac{\eta}{s}\Big{(}\frac{1}{{\tau}_{0}}-\frac{1}{\tau_{c,f}}\Big{)}
\label{ExperimentalG2-7}
\end{eqnarray}%
where $\sigma_{c}$ is the longitudinal width of the correlator measured in most central collisions whereas $\sigma_{0}$ is the longitudinal width of the correlator at formation time $\tau_{0}$ which is set 1.0$\pm$0.5 fm/c~\cite{Becattini_2014}. The critical temperature $T_{c}$ is used at 160 MeV$\pm$5 MeV~\cite{Becattini_2014}. $\tau_{c,f}$ is the freeze-out time in most central collisions which can be obtained from two pion Bose-Einstein measurements from AGS to LHC energies~\cite{Aamodt_ALICE_2011}. With Eq.~\ref{ExperimentalG2-7}, it gives one way to compare model simulations and experimental data. Based on RHIC and LHC data and with Gavin ansatz, the estimates for specific shear viscosity are given as in table~\ref{G2shearTable}. Seeing from table~\ref{G2shearTable}, $\eta$/s based on ALICE (A Large Ion Collider Experiment) data exhibits very weak or no dependence on collision centrality at LHC energy, while estimates obtained from STAR data hint that $\eta$/s might be a function of collision centrality at top RHIC energy. However, given the large systematic uncertainties of these data, one cannot draw a strong conclusion about the dependence on collision centrality. For the precision of estimates of the dependence of $\eta$/s, one needs to improve the accuracy of the STAR measurement of $G_{2}$. Additionally, it can be found that the mean values of $\eta$/s at some collision centralities are less than the KSS bound of 1/4$\pi$$\approx$0.08 and it is more obvious from LHC results than from STAR data. If one neglects the systematic uncertainties, it could raise the question of how the universal bound of 1/4$\pi$ is for QGP matter. 

\subsection{Magnetic field effect on ratio of shear viscosity to entropy density}

In noncentral relativistic heavy-ion collisions, an extremely strong magnetic field up to $\sim$$10^{18}$ G can be created~\cite{SKOKOV2009,Masayuki2010,Bzdak2012,DengWX2012}. As in Pb$+$Pb collisions at LHC, the computation shows that the strength of the magnetic field can reach eB$\sim$15$m_{\pi}^{2}$~\cite{SKOKOV2009}, where $m_{\pi}$=140 MeV is the mass of pion. With such a strong electromagnetic field, it could induce some anomalous transport phenomena in hot quantum chromodynamic (QCD) matter ~\cite{Dmitri2013,HuangXG2016}, such as the chiral magnetic effect (CME) ~\cite{Fukushima2008,Abelev2013}, chiral separation effect (CSE)~\cite{Son2004}, chiral electric separation effect (CESE) ~\cite{HuangXG2013-1}, chiral magnetic waves (CMW) ~\cite{Kharzeev2011}, chiral vortical effect (CVE)~\cite{Kharzeev2011-1} and particle polarization~\cite{adamczyk_global_2017}. The magnetic field created in noncentral collisions is perpendicular to the reaction plane and Lorentz force can vanish in the direction parallel to the field.
Lorentz force lay down in the reaction plane which makes azimuthal anisotropy for transverse flow of QGP~\cite{tuchin_2012}. And also, it would significantly influence on the properties of strange quark matter. With strong magnetic field, the condensates for light quarks/antiquarks (as u and d quark) degree of freedom are affected~\cite{Kurian_2019}. If the condensates increase with the magnetic field at low temperatures, it refers to magnetic catalysis (MC), and decrease for temperatures close to TPC which refers to inverse magnetic catalysis (IMC) \cite{Dmitri2013,andersen_2016}. These effects would change the masses of the $u$ and $d$ quarks which in turn play an important role to the transport feature~\cite{ZhuXQ2023}.

There is lots of work to investigate the influence of the magnetic field on the QCD phase diagram. And also, the impact of magnetic field on the shear viscosity of quark matter has been investigated in Refs.~\cite{tuchin_2012,ZhuXQ2023,Ghosh_2019,Rath_2021}. The simulations are in the framework of Nambu$-$Jona$-$Lasinio model (NJL) where magnetohydrodynamics is considered. Under the relaxation time approximation, one can obtain shear viscosity for the QGP system~\cite{ZhuXQ2023,Ghosh_2019}, 
\begin{eqnarray}
\eta=\frac{1}{15T} \sum_{a} {\rm g_{a}} \int \frac{d^{3}p}{(2\pi)^{3}} \frac{p^{4}}{E_{a}^{2}} \tau_{a}f_{a}^{0}(1-f_{a}^{0})
\label{Chap5-Eq-1}
\end{eqnarray}%
where `a' is index for quark and antiquark, $\tau_{a}^{c}$ is the relaxation time of the quark generated by the collision mechanism and $\rm g_{a}$ is degeneracy factor for quark and antiquark which is equal to $N_{s}\times N_{f}\times  N_{c} $=2$\times$2$\times$3 where \,$N_{s}$ is spin freedom; $N_{f}$ is flavor freedom [u and d, for SU(2)]; $N_{c}$ is color freedom~\cite{ZhuXQ2023,Ghosh_2019,Shaikh2021}. Here only two-flavor (isospin-symmetric) quark matter is considered for a system medium mainly with u and d quarks and their antiquarks. In Eq.~(\ref{Chap5-Eq-1}), $f^{0}$ is the equilibrium distribution function, which is written as covariant form~\cite{ZhuXQ2023},
\begin{eqnarray}
f^{0}(x,p)=\frac{1}{{\rm exp}\Big{[}\frac{u_{\mu}p^{\mu}\mp \mu}{T}\Big{]} +1}.
\label{Chap5-Eq-2}
\end{eqnarray}%
where $u_{\mu}(x)=\gamma_{\mu}(1, \textbf{u}(x))$; $u_{\mu}(x)$ is the four velocities of the fluid. The symbol $\mp$ is for quark (`$-$') and antiquark (`$+$'). With magnetic field, the shear viscosity coefficient of the dissipative fluid system can be decomposed into five different components (n=0,1,2,3,4)~\cite{ZhuXQ2023}. The $n=0$ component remains undisturbed by magnetic field. The other four components can be expressed~\cite{ZhuXQ2023},
\begin{eqnarray}
\eta_{2}=4\eta_{1}=\frac{1}{15T} \sum_{a} {\rm g_{a}} \int \frac{d^{3}p}{(2\pi)^{3}} \frac{p^{4}}{E_{a}^{2}} f_{a}^{0}(1-f_{a}^{0})\frac{(\tau_{a}^{B})^{2}}{\tau_{a}^{c}}
\label{Chap5-Eq-3}
\end{eqnarray}%
and
\begin{eqnarray}
\eta_{4}=2\eta_{3}=\frac{1}{15T} \sum_{a} {\rm g_{a}} \int \frac{d^{3}p}{(2\pi)^{3}} \frac{p^{4}}{E_{a}^{2}} f_{a}^{0}(1-f_{a}^{0})\tau_{a}^{B}
\label{Chap5-Eq-4}
\end{eqnarray}%
where $\tau_{a}^{B}$ is the relaxation time induced by the magnetic field.
\begin{figure}
\setlength{\abovecaptionskip}{0pt}
\setlength{\belowcaptionskip}{8pt}
\centering\includegraphics[scale=0.28]{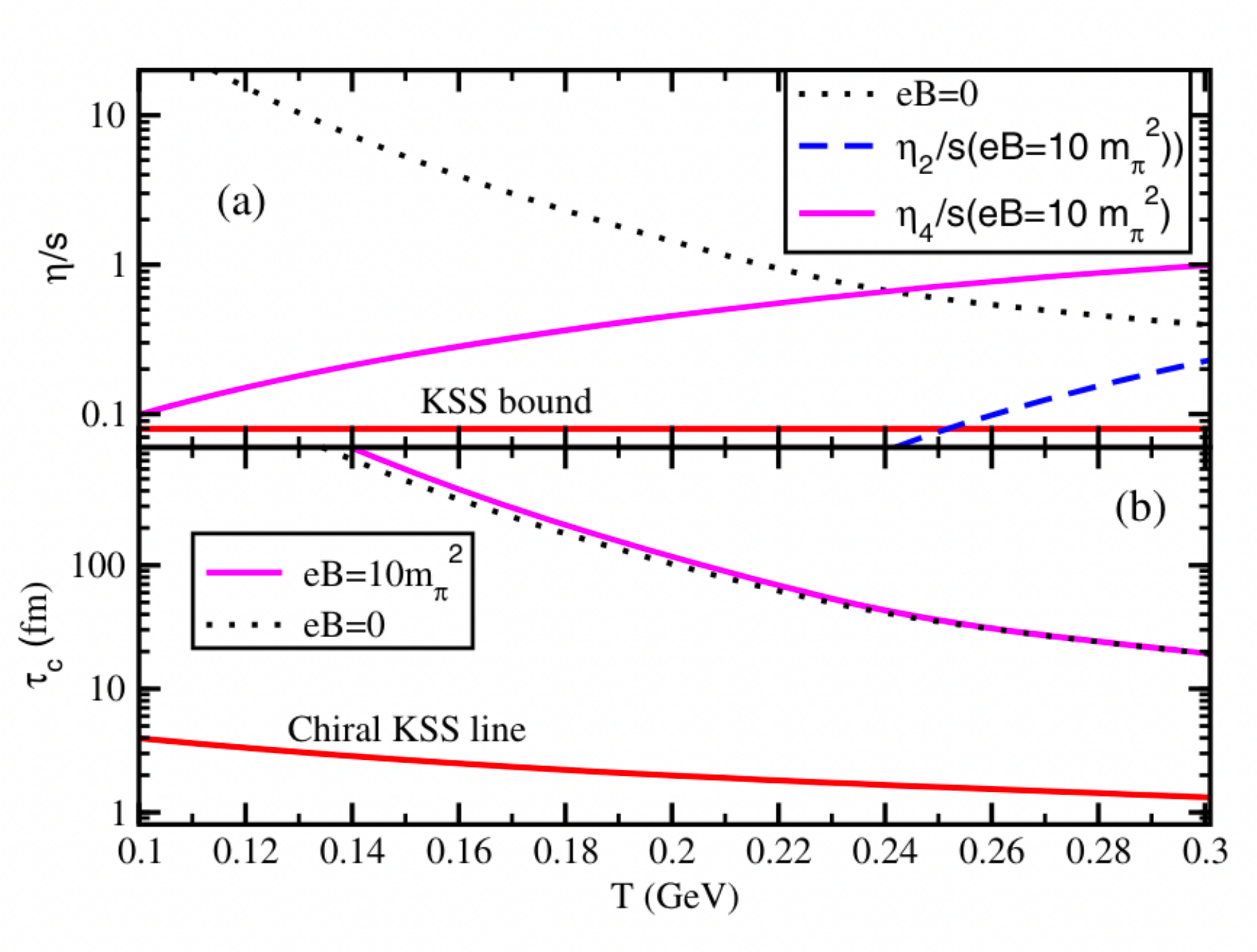}
\caption{(Color online)  The ratio of shear viscosity to entropy density $\eta/s$ (a) and relaxation time $\tau_{c}$ (b) as functions of temperature at eB=0 and eB=10 $m_{\pi}^{2}$~\cite{Ghosh_2019}.}
\label{fig:fig29}
\end{figure}

As shown in Fig.~\ref{fig:fig29}, without magnetic field, specific shear viscosity decreases as temperature increases in the range of 100$-$300 MeV. Here at eB=0, due to the increase in collisional frequency with temperature, relaxation time decreases with T. Being proportional with the decreasing function $\tau_{c}$ for eB=0, $\eta/s$ decreases with temperature~\cite{Ghosh_2019}. When the magnetic field is taken into account, the components of shear viscosity over entropy increase with the increasing of temperature. One can notice that in Eq.~(\ref{Chap5-Eq-3}), $\eta_{2}$ becomes inverse to the $\tau^{c}$ with the magnetic field. Thus, $\eta_{2}/s$ increases with temperature. For the Hall-type viscosity, $\eta_{3,4}\propto \tau^{B}$,  $\eta_{3,4}$ increases as temperature increases because of its phase space part~\cite{Ghosh_2019}. The $\eta_{4}/s$ in Fig.~\ref{fig:fig29} shows a value that crosses the KSS bound. It indicates that the effect of magnetic field could make the quark matter system a more perfect fluid.
 
\subsection{Bayesian estimation of ratio of shear viscosity to entropy density}

\begin{figure}
\setlength{\abovecaptionskip}{0pt}
\setlength{\belowcaptionskip}{8pt}
\centering\includegraphics[scale=0.26]{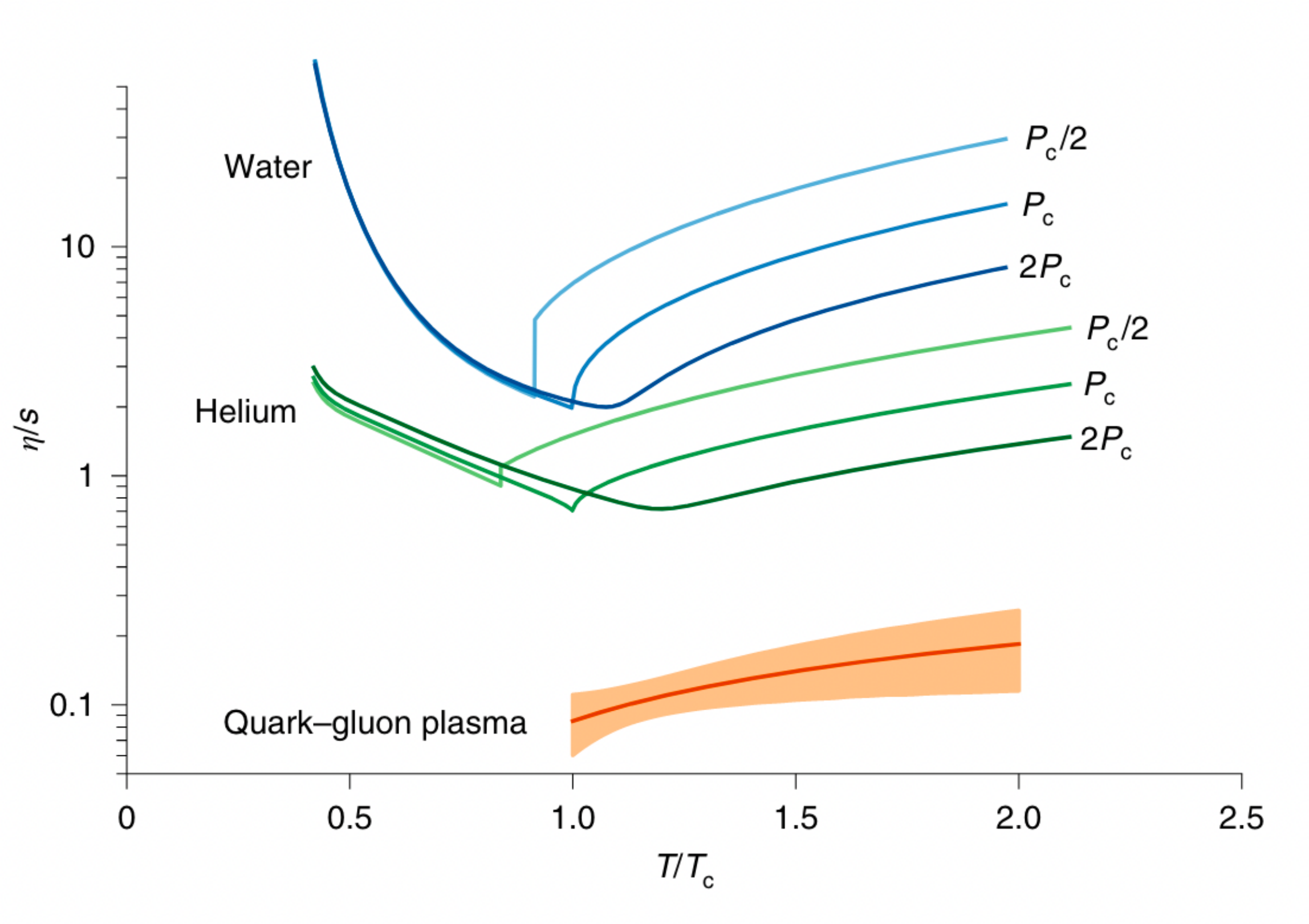}
\caption{(Color online) Specific shear viscosity as a function of reduce temperature T/$T_{c}$ for water, Helium and  QGP~\cite{Bernhard_2019}. The water and Helium are with different pressures which calculated from NIST data~\cite{NIST_2018}. The orange line and band show the posterior median and 90$\%$ credible region for the QGP $\eta/s(T)$ estimated from Pb–Pb collision data at $\sqrt{s_{\rm NN}}$=2.76 and 5.02 TeV.}
\label{fig:fig30}
\end{figure}

First principles for the transport properties of QGP matter in heavy-ion collisions remain challenging~\cite{Everett_2021,Everett_2021_PRL}. 
In recent years, machine learning and Bayesian analysis methods have attracted lots of attention on the study of nuclear physics~\cite{Alhassan_2022,He_2023,He_2023B,Bernhard_2019,Parkkila2021}. With the application of Bayesian analysis~\cite{Bernhard_2019,Parkkila2021}, one can give a meaningful credibility region for the ratio of shear viscosity to entropy density $\eta/s$ for the QGP matter~\cite{Bernhard_2019,Jussi_2020,Everett_2021,Everett_2021_PRL}.  In Ref.~\cite{Bernhard_2019}, the authors used Bayesian parameter estimation methods to constrain temperature-dependent specific shear and bulk viscosity $\eta/s$(T) and $\zeta/s $(T), which are fundamental properties of the QGP. In Bayes’ theorem, one has to give the posterior distributions as,
\begin{eqnarray}
P({\bf{x}}|{\bf{y}_{exp}}) = \frac{P({\bf{y}_{exp}}|{\bf{x}}) P({\bf{x}})}{P({\bf{y}_{exp}})}, 
\label{Chap5-Eq-5}
\end{eqnarray}%
where $P(\bf{y}_{exp}|\bf{x})$ is a likelihood function that quantifies how well the model describes the experimental measurement and $P(\bf{x})$ is a prior function. And the normalization $P({\bf{y}_{exp}})$ is called the `Bayesian evidence'~\cite{Everett_2021}. Then, one needs to build up a vector of model parameters $\bf{x}$ and do the simulations with parameters ($\bf{x}$). With the model calculations, we could do the comparison with a set of experimental data $\bf{y}$. The desired parameter estimates may be extracted from the posterior distribution. How to constraint the parameters in a Bayesian analysis process? In Ref.~\cite{Everett_2021_PRL}, the authors from JETSCAPE Collaboration have given some ways: first, this can be achieved with better external constraints on the model's parameters $\bf{x}$, reflected in a more realistic prior distribution $P({\bf{x}})$; second, new or more precise experimental data can tighten the likelihood $P({\bf{y}_{exp}}|{\bf{x}})$; third, theoretical progress on the model and better quantification of the model’s uncertainties lead to more reliable constraints through the likelihood as well.

In Ref.~\cite{Bernhard_2019}, the authors calibrate the model with a diverse set of experimental data measured by ALICE at LHC from Pb+Pb collisions at beam energies of both $\sqrt{s_{\rm NN}}$=2.76 and 5.02 TeV. As the orange band shows in Fig.~\ref{fig:fig30}, the $\eta/s$ is constrained in a 90$\%$ credible region. Here it can be seen that near the $T_{c}$, $\eta/s$ is very close to the KSS bound (1/4$\pi\approx$0.08) and even lower than it. With the help of hybrid model, denoted as T$_{\rm R}$ENTo + VISH (2 + 1) + UrQMD~\cite{Parkkila2021} which has successfully described the previous ALICE measurements~\cite{Everett_2021}, the authors give more results for transport properties in ultra-relativistic heavy-ion collisions using the Bayesian analysis method. They argued that the improved statistical uncertainties both on the experimental data and hydrodynamic calculations with additional observables do not help to reduce the final credibility ranges much, which indicates a need for improving the dynamical collision model before the hydrodynamic takes place~\cite{Parkkila2021}.

\begin{figure}
\setlength{\abovecaptionskip}{0pt}
\setlength{\belowcaptionskip}{8pt}
\centering\includegraphics[scale=0.5]{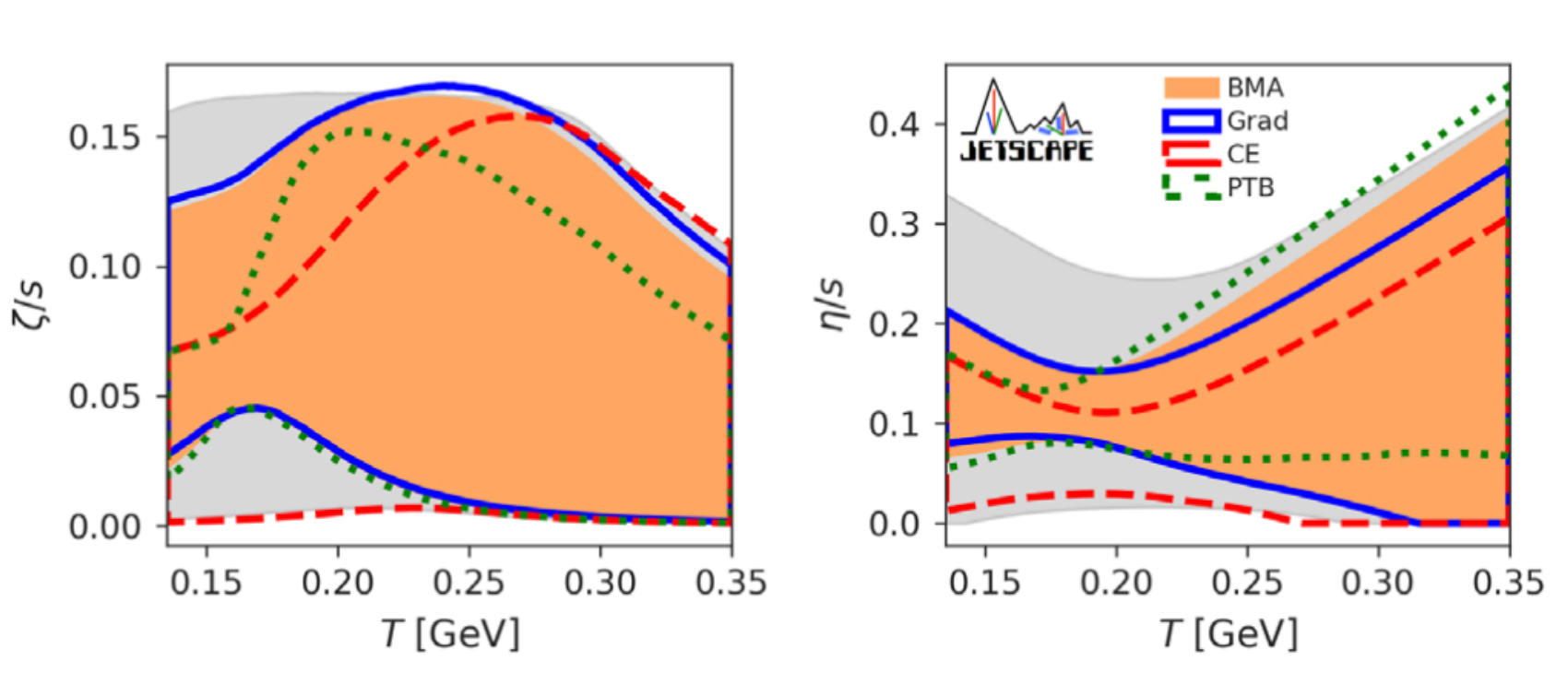}
\caption{(Color online) Specific bulk (left) and shear (right) viscosities of QGP as functions of temperature. The 90\% credible intervals for the prior (gray), the posteriors of the Grad (blue), Chapman-Enskog (red), and Pratt-Torrieri-Bernhard (green) models, and their Bayesian model average (orange) ~\cite{Everett_2021_PRL}.}
\label{fig:fig31}
\end{figure}

In the Bayesian analysis method, some uncertainties would come from different models. In Ref.~\cite{Everett_2021_PRL}, JETSCAPE Collaboration extended the Bayesian inference framework and proposed Bayesian model averaging method, in which the posterior distribution is defined by,
\begin{eqnarray}
P({\bf{x}}|{\bf{y}_{exp}}) \sim \sum_{i} P^{(i)}({\bf{y}_{exp}}) P^{(i)}({\bf{x}}|\bf{y}_{exp})
\label{Chap5-Eq-6}
\end{eqnarray}%
where $i$ is the index for different models. The authors with three kinds of hadronic phase-space distributions: the 14-moment Grad approximation, relativistic Chapman-Enskog series in the relaxation-time approximation (``RTA Chapman-Enskog''), and Pratt-Torrieri-Bernhard models, are for the particlization process which is the transition from hydrodynamics to hadronic kinetics and constrain the shear and bulk viscosities of quark-gluon plasma (QGP) at temperatures of 150$-$350 MeV by using combined data of Au+Au collisions at $\sqrt{s_{\rm NN}}$=200 GeV from the RHIC and Pb+Pb collisions at $\sqrt{s_{\rm NN}}$=2.76 TeV from the LHC. The results are shown in Fig~\ref{fig:fig31}.
It can be seen that different particlization models show similar qualitative features. In the deconfinement region, one can obtain the most likely values for $\eta/s$ of order 0.1 and $\zeta$/s of 0.05-0.1~\cite{Everett_2021_PRL}. For specific shear viscosity, the experimental data are significant constraints for 150 $\lesssim$ T $\lesssim$ 250 MeV. However, at higher or lower temperature region, the constraint becomes less. And the constraints on the specific bulk viscosity are weaker. Though the constrain from the orange region in Fig~\ref{fig:fig31} does not reach a very narrow band yet, the Bayesian model average method including both experimental and known theoretical uncertainties provides a better way to constrain the transport properties of the quark-gluon plasma. 
 
\newpage
\section{Summary and outlook}\label{sec:sum}

In this review, we have presented the shear viscosity and specific shear viscosity of nucleonic matter. Shear viscosity is an important transport property for classical and quantum systems. There are mainly two research directions on shear viscosity. One is the universality of the specific shear viscosity and the conjecture value $\hbar/4\pi$. Another is specific shear viscosity for the phase transition. By some methods, such as the mean free path, Green-Kubo method, shear strain rate method, Chapman-Enskog method and relaxation time approximation, one can extract shear viscosity for the infinite nuclear matter. Also, one can estimate shear viscosity and specific shear viscosity for finite nuclear matter or nuclear matter formed in heavy-ion collisions with the mean free path and Green-Kubo methods. From these results, it can be found that the viscosity of the nucleonic matter is mainly affected by four physical sources, i.e., density, temperature, isospin asymmetry, and nucleon-nucleon cross section. For lower-density systems, shear viscosity increases with temperature. For nucleonic matter at higher density, shear viscosity shows a minimum structure due to competition between temperature and collision. And lower isospin asymmetry or a higher nucleon-nucleon cross section leads to smaller shear viscosity. There are some extended effects on shear viscosity, such as the mean free path and Pauli blocking which can be fixed in a system with given density, temperature, isospin asymmetry, and nucleon-nucleon cross section. Actually, from a microscopic view, it rises by potential and nucleon-nucleon collision among particles. 

Moreover, it can be found that specific shear viscosity for nucleonic matter has no more difference from that of the state created in relativistic heavy-ion collisions. The value of specific shear viscosity for nucleonic matter is several times of $\hbar/4\pi$ which indicates the KSS bound postulate being an universal limit. For phase transition investigations, even though some substances show minimum values around critical temperature, the liquid state and pure neutron matter without phase transition can display minimum value from dependence on temperature due to the existence of the Pauli Block effect. Such a turning point is from the dynamic effect; thus, one should be careful when considering phase transition with specific shear viscosity. 

It may be easier to determine shear viscosity with an analytic formula. However, how to get the shear viscosity in heavy-ion collisions, which includes compression and expansion processes and is far from equilibrium, could be a challenge. In the heavy-ion collision process, shear viscosity and other transport properties have time and spatial dependence. From the experimental aspect, it is also hard to extract shear viscosity from heavy-ion collision, even though one can calculate shear viscosity with the width of giant dipole resonance or constrain shear viscosity within a hybrid model by comparing it with experimental data. For the later one, the accuracy of shear viscosity is dependent on the models. The estimation for shear viscosity with the width of giant dipole resonance from experimental data is only for the range of about several MeV. Moreover, it can be found that the dependence of shear viscosity with the width of giant dipole resonance on temperature for finite nuclear matter shows an inverse trend to the result of the theoretical calculations in the low temperature region. Therefore, the experimental investigations on shear viscosity for finite nuclear matter are expected. 

Shear viscosity and specific shear viscosity in relativistic heavy-ion collisions have also been briefly presented in this review. It is an important subject in relativistic heavy-ion collision, which is dependent on temperature as well as magnetic field effects etc. The strong magnetic field significantly affects the shear viscosity of QGP matter. It could be interesting to pay attention to that in intermediate-energy heavy-ion collisions. There are more issues for shear viscosity to be considered in the future for asymmetric nuclear matter which the fragment effect is taken into account. And it could be worth observing its correlations with dipole resonance or collective flow. As nucleons with spin degree, however, up to now, less work has been done considering how the spin and spin potential act on shear viscosity and specific shear viscosity. Especially in non-central heavy-ion collisions with large angular momentum, the effect of spin-orbit coupling on shear viscosity is also worthy of consideration.

\newpage
\section*{Acknowledgements}
 
This work is supported by the National Natural Science Foundation of China under Contracts Nos. 11890710, 11890714, 12205049，12147101, 11925502, 11961141003, and 11935001，the Strategic Priority Research Program of CAS under Grant No. XDB34000000, National Key R\&D Program of China under Grant No. 2018YFE0104600, and by Guangdong Major Project of Basic and Applied Basic Research No. 2020B0301030008.

\section*{Author's contributions \textit{(optional section)}}
Detailing here the contributions of the authors of the review.
	
\bibliography{mybibfile}
	
		
	
		
	\newpage
	\appendix
	\renewcommand*{\thesection}{\Alph{section}}
	
	\section{Appendices, if necessary}\label{appendix}
	
\end{CJK*}
\end{document}